# The impulse observations of random process generate information binding reversible micro and irreversible macro processes in Observer: regularities, limitations, and conditions of self-creation

*Vladimir S. Lerner,* USA, lernervs@gmail.com

*Abstract*

What is information originating in observation? Up to now it has not scientifically conclusive definition.

Numerous different interactive observations possess randomness whose probability is a source of its entropy-uncertainty. Observation under random impulses with opposite Yes-No probability events reveals hidden correlation, which connects Bayesian probabilities increasing each posterior correlation. That sequentially reduces relational entropy along the process, conveying probabilistic casualty with temporal memory collecting correlations which interactive impulse innately cuts.

Within the hidden correlation emerges reversible time–space *microprocess* with Yes-No conjugated entangled entropy flows, which the impulse dynamically cuts and memorizes the cutting entropy as information-certainty, allowing natural encoding during observation with interactive environment. Sequential interactive cuts along the process integrates the cutoff information in *information macroprocess* with irreversible time course. Each memorized information binds the reversible microprocess within impulse with the irreversible information macroprocess along the multi-dimensional process, which consecutively and automatically converts entropy to information and holds it.

The cutoff information models Data-Bits collected from observing frequencies-impulses by an Observer.

Each impulse No action cuts maximum of impulse minimal information while following Yes action transfers the maxmin between the impulses performing dual principle of converting the process entropy to information through uncertainty-certainly gap. Multiple Bits, moving the macroprocess, join in triplet minimax macro*units,* which logically organizes *information networks* (IN), encoding the units in the triplet's code information structures. The IN time-space distributed structure self-renews and cooperates information, decreasing the IN complexity. Integrating process entropy in an entropy functional and Bits information in information path integral' measures embraces the variation problem with the minimax law which determines all processes regularities. Solving the problem, mathematically describes the micro-macro processes, the IN, and invariant conditions of Observer's IN self-organization and self-replication. The integral maximum, integrating unlimited number of Bits' units, limits total information that process carries. The process regularities include the studied constraints, thresholds and gaps to overcome.

The considered random observations under interactive impulse is a constituent of any natural events in physical interactive processes, which originates information processes including brain logic, human thoughts, artificial intelligence.

*Definition: Information is memorized entropy cutting hidden correlation in observations.*

Random interactions of observing process, whose probability is source of process entropy, innately cut this entropy and memorize it as *information*. An interacting observer' intentional probing impulses reduce entropy hidden in the observing correlation, collect maximal entropy, cut and memorize is as information. Memory dynamically freezes the cutting process.

*Origin*: The repeated observations with opposite (No-Yes) probing impulses cutoff the observable random process' correlations. The probing impulses integrate and temporary memorizes the entropy of cutting correlation.

The impulse (0-1) actions covert maximal entropy of the cutting correlation to equivalent information Bit of information microprocess, which memorizes logic of the observed entropy probes' prehistory.

*Properties*: This quantum entangled Bit of the information microprocess, formed by the (0-1) actions, holds Yes-No logic of a primary information observer as Wheeler's Bit-Participator built without any priory physical law. The impulse, bringing the information microprocess and Bit, carries the information logical cost of getting the Bit. The sequential impulse cuts along the process convert its entropy to information, connect the Bits sequences, and integrate the cutting *micro processes* in information *macroprocess* and finally to Observer intelligence.

# Introduction

Suppose we are observing somewhat like a new Earth in unknown yet Galactic.

Is it the fact-certainty or is it complete uncertain for this observation?

What is scientific way to find it? How to uncover a path from uncertainty to certainty as the fact of realty, focusing not on physical process of the observation but on its information-theoretical essence?

Since at beginning of this process, there are no facts about reality, the beginning is uncertain regarding the facts.

Presenting an observation as multiple random events-a random process of interactive impulses with probabilities measure from near zero to one, allows us identifying reduction of uncertainty to certainty measures by related probabilities.

The studied problem addresses following questions:

What observation is? How does reality of facts-certainty could be observed or predicted in observation?

What drives the process of observation allowing to find the certainty- information about the fact and multiple facts?

This leads to fundamental question of origin of information in observation, specifically:

How does process of observation create elementary information unit-Bit?

Are there any underlying (microprocesses), connected uncertainly of randomness with certain (macro) dynamics?

What are relations between reversible micro-and irreversible macrodynamics?

What is a basic information macro unit assembling the Bits?

How do the multiple information units join in information process?

What is structure of information process revealing the fact and multiple facts?

These are fundamental questions essential for Science and applications.

The questions arise at observing multiple events-processes in physics, biology, cognition, cosmos, economy, sociology, acquiring knowledge, science experimenting, learning, reading, examining investigations, analyzing sequence of numbers, playing games, dealing with any other subjects of human activities, others.

Consequently, what is the common ground in all of these, and how to use only it for answering these essential questions?

Interactions build structure of Universe, and impulse is elementary discrete inter-actions $\downarrow\uparrow$ which models standard unit of information Bit originating in any natural interactive process independent of its physical nature. Such common impulse models variety of physical interactions including electrical, gravitational, elementary particles, and multiple measurements.

Hence, how the structure of Universe is actually built? What is the path from a primer uncertain forms to real structuration?

Studying the impulse observations independently of their specific substances allows focusing on the process converting uncertainty of randomness-entropy to certainty-information, its information regularities, while observing reduction of uncertainty-entropy's measure to information measure. An observer arises in these random observations, evolving from probabilistic microprocess within each interactive impulse up to creation the real information Observer from the information micro-macroprocesses, which finally acquires intellectuality. Such study is distinct from known basic results [A.1-A.10] by introducing more general formal approach to Information Observer rising on the path from uncertainty to reality and aimed on regularities of artificial design of human thoughts. Studying entropy-information specific of the observing process stages includes analytical descriptions, computer simulations, uncertainty-certainty gap, and thresholds with numerical constraints. The approach integrates all observation stages by unified mathematical formalism.

### A. The formalism general points

- Numerous different interactive observations possess randomness whose probability is a source of its entropy-uncertainty.

Uncertainty measures conditional entropy of relative a priori—a posteriory probabilities.

Maximal uncertainty measures non-correlating a priori-a posteriory probabilities, when their connection approaches zero.

Such theoretical uncertainty has infinite entropy measures, whose conditional entropy does not exist.

The *finite* uncertainty measure has a nonzero posteriory probability, connecting with its a priori probability through non-zero correlations holding the conditional entropy.

-Information, as notion of certainty–opposite to uncertainty, measures reduction of uncertainty to posteriory probability one (evaluating a probabilistic fact-truth).
-Since a Bit-the standard unit of information, is a discrete entity-an impulse of Yes-No actions models an inter-action ↓↑.
-The impulse inter-actions originates in any natural process generated in interactions, which include interactive observations generating a random process.
-The most common formal model of natural random non-stationary process is Markov diffusion process.
-The random process of interactive observation, possessing entropy and finite correlations, models series of random impulses with 0-1 probability measure satisfying Kolmogorov's (0-1) law of stochastic process.
-With no real physics affecting such virtual observation, it models virtual probing impulses which replicate the information impulses and start a path from maximal entropy (uncertainty) to certainty of real information.
-In the virtually observing process, a sequence of the probing impulse connects the sequential Bayesian a priori-a posteriori probabilities.
-Each impulse virtual No-0 action cuts maximum of impulse minimal entropy, while following Yes-1 action transfers this maxmin between the impulses decreasing the following process' entropy.
Thus impulse interactive actions express the impulse minimax principle.
-The Bayes posteriory probability measure 1 increases an initial finite correlation connecting it with the following correlation.
-The impulse minimax principle, imposed on sequential Bayesian correlations leads to reducing the process entropy at growing this correlation.
-That virtual probabilistic process under virtual probing Kolmogov's probabilities connected via sequential Bayesian correlations prognosis the decreasing process entropy by self-collecting the minimal virtual impulse entropy. A virtual observation potentially becomes real by obtaining Information.
-Hidden process correlations connect Bayesian probabilities, increasing each posterior correlation *of observing process*. Sequentially along the process, this reduces relational entropy, conveying probabilistic causality. Temporal memory collects correlations as interactive impulses are cut.
- The relational entropy conveys probabilistic casualty with temporal memory of correlations, which encloses a prehistory by integrating Bayesian probabilities on the path from the starting uncertain observation to current correlations.
-The cutting correlation reveals entropy and Bayes probability hidden within correlation.
-The interactive impulses cutting the correlation hidden entropy memorizes the cut.
-Cutting entropy memorizes the cutting time interval which freezes the probability of events with related Bayes probability hidden within correlation.
-The impulse virtual probability acquires both probabilistic one certainty and physical certainty after influx of energy of natural interactions compensating for appearance of physical impulse that brings irreversibility of minimax action, which creates this information unit by cutting the correlation.
-The probing minimax cut of the observing process checks the cutting probability and ability for the natural interactive compensation. Not any probability one entropy impulse produces information impulse.
-Collecting multiple information impulses creates real Information Observer, which through the probabilities of cutting correlation restores the virtual probe sequential logic.
 -From that follows definition: *information is memorized entropy (uncertainty) cutting from correlation in observations which process interactions on a path from reduction uncertainty to certainty.*
-The logical operations with information units achieve a goal, integrating discrete information hidden in the cutting correlations in time-space structure of information Observer.
-Multiple maxmin-minimax leads to the dual variation principle (VP), arising in natural interactions.  Using this VP allows synthesize an optimal information process minimizing observations time in Artificial designed information Observer with intellectual searching logic.

-For each initial observing conditions, solution of the VP brings particular extremal information process for each observation of information Observer, prognosis it along with Observer's multiple information process and sintering information structure.
-The prognosis extreme process for each particular starting observation identifies number of virtual impulse necessary to provide the actual cut producing the information impulse including interacting frequency.
This actual cut probes the projected probability through the cutting correlation which also memorizes the cutting current conditional probability hidden within correlation. The cutting information impulse possess invariant entropy measure which integrates information path functional applying the VP.
-This formal approach creates a path from the process uncertainty to certainty of real information impulses where an Observer interacts with an observing process (virtual-imaginable or real) via impulse searching observations.
-The approach enables generates the information Observer by computer synthesis an artificial designed information Observer with intellectual searching logic.
-The approach models any interacting process including observer's microprocess integrating its information in macroprocess and building it information physical structure.
-Since both interactions and its elementary *discrete* inter-actions ↓↑ define standard unit of information Bit, the impulse code is *universal* originates in any natural process generated in interactions.
-It allows natural encoding during observation with interactive environment. The specific interaction's energy quantity (power) limits the universal code length through its final encoding bit's information density; the energy quality, evaluated by the energy entropy, limits the related probability of the code length start.
- In a probabilistic measurement of the microprocess' quantum states, the approach can predict an optimal information rout between the quantum states and brings new understanding and formulation of quantum mechanics interactions.

***B. The essence of main stages of observations on the path from process uncertainty to certainty of real information***
-Randomness, as an abstraction of uncertainty, portrays a random field of random processes, which formally describes Kolmogorov's probabilities.
The experimental probability measure predicts axiomatic Kolmogorov probability if the experiment satisfies condition of a *symmetry* of the equal probable events in its axiomatic probability [1, 2].
In the theory of randomness, each events' probability is *virtual*, or, at every instant, prescribed to this imaginary event, many its potential probabilities might occur simultaneously, admitting multiple and concurrent measurement.
-The observing randomness begins with an elementary interaction of a random impulses of a random process, where each impulse consists of opposite Yes-No probability events according to Kolmogorov's 1-0 law [1] for random process.
Thus**,** the *observation is processing the interactions.* Impulse observations replicate frequencies of an observer.
-The observing sequence of Kolmogorov's probabilities measures Bayesian a priori-a posteriori probabilities of the impulses opposite actions ↓↑ (events). If each following a posteriori probability, in series of the Bayesian additive a priori-a posteriori probabilities grows, maximizing a final posteriori probability up to $P_p \to 1$, it reveals actual fact from its initial uncertainty.
-The probability transitions model an interactive random process, generated by an idealized (virtual) probability measurement of the finite uncertainty, as *observable process* of a potential (virtual) observer.
These probabilities connect a priori-a posteriori events, rising their correlation connections (from starting, weak to strong) that conveys *conditional* random entropy as negative logarithmic probability between the events along the process.
-Each impulse' opposite No-Yes interactive actions (0-1) carries a virtual impulse which potentially cuts off the random process correlation, whose conditional (relation) entropy decreases with growing the cutting correlations (Fig. A).

For example, each a priori to the cut process correlation $\vec{r}_{ai}$, in the impulse sequence $\uparrow \vec{r}_{ai} \downarrow \vec{r}_{pi} \uparrow$, may grow allowing then cutting more a posteriori correlation $r_{pi}$ from the process, which increases the Bayes probabilities connecting sequence $\uparrow \vec{r}_{pi} \downarrow \vec{r}_{ai+1} \uparrow$.

-If a preceding No action cuts a maximum of entropy (and a minimal probability), then following Yes action gains the maximal entropy reduction-its minimum (with a maximal probability) during the impulse cutoff. The impulse' maximal cutting No action minimizes absolute entropy that conveys Yes action (rising its probability), which leads to a maxmin of relational entropy between the impulse actions transferring the probabilities.

-This sequence of interacting impulses, transforming opposite No-Yes actions, increases each following Bayesian a posteriori probability and decreases the relative entropy. Consequently, initial uncertainty is gradually transforming to more probable (less uncertain) process and finally to certainty-as information about the reality.

-The probability-entropy measures a nearness (distance) of observable process' uncertainty to its information.

-Until reaching certainty, the observable process with the probabilities of uncertain events is not real–*virtual*, imaginable. With no real physics affecting such observation, the virtual observation starts with a maximal uncertainty or minimal finite probability of a virtual impulse.

-In the impulse's virtual Yes-No actions that are reversible, each second (No) through recursion [3] affects the predecessor (Yes) connecting them in a weak correlation, if there was not any of that.

-Arising the correlation connection memorizes this action indicating *start of observation* with following No-Yes impulse.

This correlation connects the Bayesian a priori-a posteriori probabilities in a temporal memory that does not store virtual connection, but renews, where any other virtual events (actions) are observed.

The Bayes connection increases each posterior correlation, sequentially reducing the entropy along the process.

-The starting observation *limits* a minimal entropy of virtual impulse, which depends on minimal increment of process correlation (Sec.1.1) overcoming a maximal admissible finite uncertainty.

-If the observing process is self–supporting through automatic renewal virtual inter-actions, it calls a *Virtual Observer*, which acts until these actions resume. Such virtual observer belongs to a self–observing process, whose Yes action virtually starts next impulse No action, and so on. Both process and observer are temporal, ending with stopping the observation.

Starting of virtual self-observation limits a *threshold* identified in Sec.I.2.

A longest virtual observer accumulates its rising temporal memory up to growing a priory probability $P_a \to 1$.

-The memory temporary holds the difference of the probabilities actions, as a virtual measure of an *adjacent distance* between the impulses' No-Yes actions and a probabilistic accuracy of measuring correlation (Sec.II.1).

-The measuring, beginning from the starting observation, identifies an interval from the start, which is also virtual, disappearing with each new connection that identifies a next interval memorized in that connection.

Thus, each new virtually observing event–action temporary memorizes a whole pre-history from the starting observation, including the summarized (integrated) maxmin-minimax entropy, which automatically holds in the memory a last of the current connection-correlation.

-The correlation indicates appearance of a *time interval* [4] of the impulse–observation.

The random process' impulses hold virtually observing random time intervals with hidden entropy.

-Collecting and measuring that uncertainty along the random process integrate entropy functional (EF), which is proportional to the running time intervals [5] (operating according to formulas (I.5- I.7))**.**

-With growing correlations, the intensity of entropy per the interval (as entropy density) increases on each following interval, indicating a shift between the virtual actions, which measures correlation of the related impulse in a *time interval's unit measure* $|1|_M$. The growing density *curves* an emerging ½ time units of the impulse time interval, and the action rotating curved time-jump initiates a *displacement* within the impulse on two space units-a counterpart to the curved time (Fig.1).

When two space units replace the curved ½ time units within the same impulses, such *transitional* time-space impulse preserves measure $|2 \times 1/2|_M = |1|_M$ of the initial time impulse.

-Within time interval impulse $|1|_M$, appearance of the space coordinate has the probability of growing this impulse correlation (Sec.II.2) for the observing random process ensemble. Probability of appearance this transitional time-space virtual impulse with its curved time interval requires more probing impulses compared to that for only time impulse.

An infinite impulse-jump on the impulse border, cutting the curving time, spots a "needle curve pleat" at transition to a finite space unit within the impulse. The Bayes probabilities may overcome this transitive gap. (Formally, the time-jump in opposite rotating Yes-No probabilistic actions curves the cutting correlation with curvature which originates a space shift measured by the entropy increment relative to that for curved time, Sec.II.1).

-The space displacement shifts the virtual observation from the source of random field to the self-observing process of a randomness and initiates probabilistic emergence of *time-space coordinate system* and gradient of entropy-an entropy force depending on the entropy density and space coordinates.

-The discrete displacement with the entropy' opposite jumps rans *anti-symmetric entropy* increments in a microprocess with the transitional impulse, starting its EF entropy measure along the impulse *discrete time-space* intervals.

-The jump-wise displacement preserves the Yes-No probability, and the continued time-space process' displacements also conserves these probabilities satisfying to Kolmogorov's (0-1) law of stochastic process. While virtual random observation of the stochastic process' Yes-No sequence evaluates their probability measure 1 or 0, the random impulse has no specific time-space shape and location. The emerging time-space movement conserves these probability measures in *discrete time-space form of the impulse* on these discrete intervals between the process probabilities. The curving impulse ↓↑ gets form (Fig.1a).

-The displaced self-observing process with the space-time priory-posteriori actions continues requesting virtual observation, which intends to preserve these probabilities. That initiates the observer's space–time entropy and correlations, starting self-collecting virtual space-time observation in a *shape* of correlation structure of the *virtual observer volume*.

-The observations under these impulses reduce entropy in space-time movement which encloses the volume. Reduction the process entropy under probing impulse, observing by Bayesian probability' links that increases each posterior correlation.

-The space-time entropy force rotates the curved time-space coordinate system (within the volume) with the rotating moment, depending on the gradient and velocity of running movement along the coordinate space trajectory.

-The gradient entropy along the rotating interval of the trajectory could engage next impulse in rotating action, which increases the correlation temporally memorizing the time-space observation.

-This action indicates appearance of a space-time *curved* shaping geometry structure of the virtual observer, which is concurrently memorized and developed under the impulses' generating minimax entropy.

-The memory temporary holds a difference of the starting space-time correlation as accuracy of its closeness, which determines the time-space observer location with its shape. The evolving shape gradually confines the running rotating movement which *self-supports* formation of both the shape and Observer.

-The virtual observer, being displaced from the initial virtual process, sends the discrete time-space impulses as virtual probes to test the preservation of Kolmogorov probability measure of the observer process with probes' frequencies. Such test checks this probability via a symmetry condition indicating the probability correctness and the time-space Observer' structural location. The increasing frequencies of the Observer self-supporting probes check the growing probabilities.

-The virtual Observer self-develops its space-time virtual geometrical structure during virtual observation, which gains its real form with transforming the integrated entropy to equivalent information.

-The impulse opposite actions at the displacement rotates their anti-symmetric entropy increments in the microprocess within the impulse, which connects the observation in correlation connections.

-Multiple impulses initiate a manifold of virtual Observers with random space-time shape in a collective probabilistic movement. For each random impulse, proceeding between a temporary fixed random No-Yes actions, such microprocess is multiple, whose manifold decreases with growing the probability measure. That also decreases the manifold of the virtual observers with the multiple probabilities. At approaching maximal a priori probability, minimum three simultaneous random impulse' cuts rise within space interval (Sec.II,(2.24)), which correlate in the rotating enfolding volume. At reaching this maximal probability, only a pair of additive entropy flows with symmetric probabilities (which contains symmetrical-exchangeable states) advances.

-At satisfaction of the symmetry condition, the impulses' axiomatic probability is transformed to the microprocess 'quantum' probability with pairs of conjugated entropies and their correlated movements.

-A maximal correlation adjoins these conjugated symmetric flows, uniting their entropies into a running *entanglement* which enables confine an entropy volume of the pair or triple superposition.

-The states of cutting correlations hold hidden process' inner connections, which initiate growing correlations up to running the superposition, entanglement and conversion of the cutting entropy in information. Both entangled anti-symmetric fractions appear simultaneously with starting space interval. The correlations binding this couple or triple with maximal probability are extremely strong. The correlated conjugated entropies of the entangled rotating virtual impulses are no separable and no real action *between* them is possible.

-The microprocess is different from that in quantum mechanics, since first, its interacting parts (anti-symmetrical entropy flows) do not carry energy, and the entanglement does not bind energy, just connects the entropy in joint correlation. The rotating anti-symmetric entropy flows have additive time–space *complex amplitudes* correlated in time-space entanglement with not limited distance.

-Arising correlated entanglement of the opposite rotating conjugated entropy increments starts entangling the entropy *volumes which* enclose the condense entropies of the microprocess complex amplitudes.

-The entangled increments with their volumes, captured in rotation, adjoin the entropy volumes in a stable entanglement, when the conjugated entropies reach equalization and anti-symmetric correlations cohere confining the coupling anti-symmetric entropy units in a minimal entangled real units.

The stable entanglement *minimizes* quantum uncertainty of the ongoing virtual impulses and increases their Bayesian probability. As a priory probability $P_a$ approaches it maximum, both the entropy volume and rotating moment grow. Still, between the maximal a priori probability of virtual process and a posteriori probability of real process $P_p = 1$ is a small microprocess' gap, associated with time-space probabilistic transitive movement, separating entropy and its information (at $P_a < 1$ throughout $P_p \to 1$).

-Estimation the gap transitive time interval (Sec.III) indicates that the stable entanglement, which joins the most a priory probable conjugated entropies inside the volume, finalizes *within* the gap.

In stable entanglement, both additivity and symmetry of probability for mutual exchangeable events vanish. That challenges pure probabilistic Kolmogorov productivity of the entanglement in quantum mechanics [6].

-It implies a distinction of statistical possibilities with the entropies-uncertainties in quantum virtual microprocess from information-certainty of reality. It is impossible to reach a reality in quantum world without overcoming the gap between uncertainty and certainty, located on edge of reality with probability $P_m \cong 0.985507502$ (Sec.II,(2.15)) that is limited by minimal uncertainty measure $h_\alpha^o = 1/137$ - physical structural parameter of energy, which includes the Plank constant's equivalent of energy. The gap holds a hidden real quantum locality which impulse cuts within the hidden correlation and transforms to information reality.

-The rotating moment, growing with increased volume, intensifies the time–space transition of the volume over the gap, acquiring physical property near the gap end, and, when last posterior probability $P_p = 1$ overcomes last prior virtual probability during the cutting moment, the entropy volume is transferred.

-The step-down (No) control's cut *kills* total entropy' volume during *finite* time-space rotation and *memorizes* dynamically this cutting entropy as the *equivalent information with its asymmetric geometry.*
-When the correlated entropy's conjugated pair being cut, this action transforms the adjoin entropy increments to real information which binds them in real Bit of *entangled asymmetrical couple* where changing one acts to other in *one direction.*
-Ability of an observer to overcome its gap depends on the entropy volume, collected during virtual probes, whose entropy force spins the rotating moment for transition over the gap, and the real control jump adds energy covering the transition. A shortage of the virtual probes collection and absence of this controlled energy will not create this ability and information. Cutting the curved volume spacing real needle pleat.
-Transition maximal probability of observation through the gap up to killing the resulting entropy runs a *physical* microprocess with both local and nonlocal entangled information units and real time-space, which preempt memorizing.
In physical terms, the sequence of opposite interactive actions models reversible micro-fluctuations, produced within observable irreversible macroprocess (like push-pull actions of piston moving gas in cylinder [7]).
More simple example, when a rubber ball hits ground, energy of this interaction partially dissipates that increases interaction's total entropy, while the ball's following reverse movement holds less entropy (as a part of the dissipated), leading to max-min entropy of the bouncing ball. Adding periodically small energy, compensating for the interactive dissipation, supports the continuing bouncing.
-Virtual process does not dissipates but its integral entropy decreases along No-Yes virtual probes.
-The real microprocess builds each information unit - Bit within the cutting impulse in real time (according to (I.6- I.8)).
-These processes, creating the Bit from an impulse, reveal structure of the Weller Bit [8] which memorizes the Yes-No logic of virtual actions, participating in getting the Bit information.
-The impulse carries logical cost of getting the Bit.
- Such Bit-Participator is a primary information Observer formed without any a priori physical law.
-The cutting action, killing the process entropy near $P_p \to 1,$ produces an interactive impact between the impulse No and Yes actions, which *requires the impulse access of energy* to overcome the gap.
The delivering equivalent information compensates for this impact, which needs only at real cutting control, while the virtual cuts avoid it. The impact emerges when virtual Yes action, ending the preceding imaginable microprocess, follows the real No action [9(2)].
-Information observer starts with real impulse cutting off the observing process and extracting hidden information Bit, which identifies such information observer as an extractor and holder this information.
-The killing physical action converts entropy of virtual Observer to equivalent information of real information Observer.
In multi-dimensional virtual process, correlations grow similarity in each dimension under manifold of the impulse observation. The correlations, accumulated sequentially in time, increase with growing number of currently observed process' dimensions. Hence, each next impulse with the same size cuts the increasing virtual correlation-entropy volume leading to rising density of the cutting entropy which increase its speed (as the volume related to the cutting impulse wide).
-Killing the distinct volumes densities converts them in the Bits distinguished by information density, while each Bit accumulates the observation of complimentary events in a *free information*, initiating the Bits' attraction.
-Between these different Bits rises information gradient of attraction minimizing the free information. That connects the Observer's collected information Bit in units of information process, which *builds Observer information structure.*
-Encoding each Bit extracts its hidden position by the cut, which erases the Bit of information at cost of energy. Each impulse time interval encodes invariant unit of information in physical process whose interactive time interval carries energy equivalent of the impulse information cutting from the correlation. That compensates for Maxwell Demon's cost while producing information during the interactive impulse.
-The information Bit from a cutting off random process includes the following specific:
information delivered by capturing external entropy during transition to the cut; information cut from the random process; information transferring to the nearest impulse that keeps persistence continuation of the impulse sequence via the attracting Bits; persistent Bits sequentially and automatically converts entropy to information, holding the cutoff information of random process, which connects the Bits sequences; the cutoff Bit has time–space geometry following from the geometrical form of discrete entropy impulse; information, memorized in the Bit, cuts the symmetry of virtual process; the Bit, generated in the attraction, is different from the primary Bit cut off from random process; the free information, rising between the primary Bits, is a part of its information spent on binding the attracted Bits.

-Both primary and attaching Bits keep persistence continuation in information process, which integrates the Bits time-space real impulses in the process elementary segments, joins in information units that composes a space-time information structure of Information Observer.

-The microprocess within a Bit's formation, connecting both imaginary entropy and information parts in rotating movement, also binds *multiple units* of Bits in *collective* movement of information macroprocess, which integrates information path functional (IPF).(Eqs. (1.9-1.10) connect EF-IPF with correlations).

The information marodynamics are reversible within each EF-IPF extremal segment by imposing the dynamic constraint on IMD Hamiltonian; irreversibility rises at each constraint termination between the segments. The IMD Lagrangian integrates both the impulses and constraint information on time space-intervals.

-A flow of the moving cutoff Bits forms a unit of the information macroprocess (UP), whose size limits the unit's starting maximal and ending minimal information speeds, attracting new UP by its free information. Each UP, selected automatically during the minimax attracting macro-movement, joins two cutoff Bits with third such Bit, which delivers information for next cutting Bit.

-Minimum three self-connected Bits assemble optimal UP-basic *triplet* whose free information requests and binds new UP triplet, which joins three basics in a knot that accumulates and memorizes triplets' information.

-During the macro-movement, the multiple UP triples adjoin in the time-space hierarchical network (IN), whose free information's request produces new UP at a higher level knot-node and encodes it in triple code logic. Each UP has unique position in the IN hierarchy, which defines exact location of each code logical structures. The IN node hierarchical level classifies *quality* of assembled information, while the currently ending IN node integrates information enfolding all IN's levels.

-New information for the IN delivers the requested node information's interactive impulse impact on the needed external information, which cutoff memorized entropy of a Data. Appearing new quality of information concurrently builds the IN temporary hierarchy, whose high level enfolds information logic that requests new information for the running observer's IN, extending the logic code.

-The emergence of current IN level indicates observer's information surprise thru the IN feedback's interaction with both external observations and internal IN's information, which renovate it information.

-The time-space information geometry shapes the Observer *asymmetrical structure* enclosing multiple INs.

-The IPF maximum, integrating unlimited number of Bits' units with finite distances, limits the total information carrying by the process' Bits and increases the Bit information density in a rising the process dimensions. While the Bit preserves its information, the growing information, condensed in the integrated Bit with a finite impulse geometrical size (limited by the impulse speed of light), intensifies the Bit information density, running up to finite IPF maximal information at infinite process dimension.

-The macroprocess integrates both imaginary entropy between impulses of the imaginary microprocesses and the cutoff information of real impulses, which sequentially convert the collected entropy in information physical process during the macro-movement.

-The observation process, its entropy-information and micro-macroprocesses are Observer-dependent, information of each particular Observers is distinct.

-However, the invariant information minimax law leads to equivalent information regularities for different Observers. By observing even the same process, each Observer gets information that needs its current IN during its optimal time-space information dynamics, creating specific information process.

-The equations of information macrodynamics [11-13], as the EF extremal trajectories, describe the IPF information macroprocess which averages all the microprocesses and holds regularity of observations under the maxmin-minimax impulses. At infinitive dimension of process $(n \to \infty)$, the information macroprocess is extremals of both EF and IPF, while EF theoretically limits IPF.

The limited number of the process macrounits that free information assembles leads to limited free information transforming the impulse microprocess to the macroprocess.

-When the information Observers unify-integrate their multiple facts in EFF-IPF variation problem, the VP solution determines regularities of the multiple observation, which become independent of each particular Observer. Such information-physical law establishes when the EF Hamiltonian changes maximum on minimum [86, Sec.III.8]. It minimizes the EF, maximizes probability on the EF trajectories, and reaches causal deterministic reality. The multiple randomly applied deterministic (real) impulses, cutting all process correlations, transforms the initial random process to a limited sequence of independent states.

-The macro-movement in rotating time–space coordinate system forms Observer's information structure confining its multiple INs that determine the *Observer time of inner communication* with *time scale* of accumulation information.

-Each Observer *owns the time of inner communication,* depending on the requested information, and *time scale,* depending on density of the accumulated information.

-The current information cooperative force, initiated by free information, evaluates the observer's *selective* actions attracting new high-quality information. *S*uch quality delivers a high density-frequency of related observing information through the selective mechanism. These actions engage acceleration of the observer's information processing, coordinated with the new selection, quick memorizing and encoding each node information with its logic and space-time structure, which minimizes the spending information and IN *cooperative complexity*.

-The observer optimal *multiple choices,* needed to implement the minimax self-directed strategy, evaluates the cooperative force emanated from the IN integrated node.

-The self-built information structure, under the self-synchronized feedback, drives self-organization of the IN and the *evolution macrodynamics* with ability of its self-creation.

-The free information, arising in each evolving IN, builds the Observer specific time–space information *logical* structure that conserves its "conscience" as intentional ability to request and integrate the explicit information in the observer IN highest level, which measures the Observer *Information Intelligence*.

-The coordinated selection, involving verification, synchronization, and concentration of the observed information, necessary to build its logical structure of growing maximum of accumulated information, *unites* the observer's *organized intelligence action*. The IN hierarchical level's amount quality of information evaluates *functional organization* of the intelligent actions spent on this action. Intelligence of different observer integrates the common IN's ended node.

-The IN node's hierarchical level cooperates the communicating observers' level of integrated knowledge.

-Increasing the IN enfolded information density accelerates grow the intelligence, which concurrently memorizes and transmits itself over the time course in an observing time scale.

-The intelligence, growing with its time interval, increases the observer life span.

-The self-organized, evolving IN's time-space distributed structure models *artificial intellect*.

## C. *Basic mathematical foundation*

The *integral measure* of the observing *process* trajectories are formalized by an *Entropy Functional* (EF), which is expressed through the regular and stochastic components of Markov diffusion process $\tilde{x}_t$ [5, 11]:

$$\Delta S[\tilde{x}_t]|_s^T = 1/2 E_{s,x}\{\int_s^T a^u(t,\tilde{x}_t)^T (2b(t,\tilde{x}_t))^{-1} a^u(t,\tilde{x}_t) dt\} = \int_{\tilde{x}(t)\in B} -\ln[p(\omega)] P_{s,x}(d\omega) = -E_{s,x}[\ln p(\omega)], \quad (I.1)$$

where $a^u(t,\tilde{x}_t) = a(t,\tilde{x}_t, u_t)$ is a drift function, depending on control $u_t$, and $b(t,\tilde{x}_t)$ is a covariation function, describing its diffusion component in Ito Eqs.[10]; $E_{s,x}$ is a conditional to the initial time and states $(s,x)$ mathematical expectation, taken along the $\tilde{x}_t = \tilde{x}(t)$ trajectories. The right side of (I.1) is the EF equivalent formula, expressed via probability density $p(\omega)$ of random events $\omega$, integrated with the probability measure $P_{s,x}(d\omega)$ along the process trajectories $\tilde{x}(t) \in B$, which are defined at the set $B$.

Generally, random process (as a continuous or discrete function $x(\omega,s)$ of random variable $\omega$ and time $s$), describes an elementary *changes* of its probabilities from one distribution (a *priory*) $P_{s,x}^a(d\omega)$ to another distribution (*a posteriori*) $P_{s,x}^p(d\omega)$ in the form of their transformation

$$p(\omega) = \frac{P_{s,x}^a(d\omega)}{P_{s,x}^p(d\omega)}. \quad (I.2)$$

Sequence of this probabilities ratios *generalizes* diverse forms of specific functional relations, represented by a series of different transformations.
The probability ratio in the form of natural logarithms:

$$-\ln p(\omega) = -\ln P_{s,x}^a(d\omega) - (-\ln P_{s,x}^p(d\omega)) = s_a - s_p = \Delta s_{ap}, \quad (I.2a)$$

describes the difference of a *priory* $s_a > 0$ and a *posteriori* $s_p > 0$ *random entropies, which* measure *uncertainty,* resulting from the transformation of probabilities for the process events, satisfying the entropy's additivity. A *change* brings a certainty or *information* if its uncertainty $\Delta s_{ap}$ is removed by some equivalent entity call information $\Delta i_{ap}: \Delta s_{ap} - \Delta i_{ap} = 0$. Thus, information is delivered if $\Delta s_{ap} = \Delta i_{ap} > 0$, which requires $s_p < s_a$ and a positive logarithmic measure with $0 < p(\omega) < 1$. Condition of zero information: $\Delta i_{ap} = 0$ corresponds to a redundant change, transforming a priory probability to equal a posteriori probability, or this transformation is identical–informational undistinguished. The removal of uncertainty $s_a$ by $i_a$: $s_a - i_a = 0$ brings an equivalent certainty or *information* $i_a$ about entropy $s_a$.

The logarithmic measure (I.2a) of Markov diffusion process' probabilities approximates the probability ratios for other random processes [14].
Mathematical expectation of random probabilities and entropies in (I.2a):

$$E_{s,x}\{-\ln[p(\omega)]\} = E_{s,x}[\Delta s_{ap}] = \Delta S_{ap} \Rightarrow I_{ap} \neq 0, \quad (I.3)$$

determines the mean entropy $\Delta S_{ap}$ as equivalent of nonrandom *information* $I_{ap}$ *of a random source.*

Being averaged by the source *events,* through a probability of multiple random variables-state, or by the source *processes* (through probabilities I.1), depending on what it is considered (a process, or an event), both (I.3) and (I.1) include Shannon's formula for relative entropy-information of a states (events).

For a continuous random variables, (1.3) brings also an equivalent of Kullback–Leibler's (KL) divergence measure, expressed through a nonsymmetrical logarithmic distance between the related entropies in (I.3, I.2).

The KL measure is connected to both Shannon's conditional information and Bayesian inference [12] of testing a priory hypothesis by observation a priory-a posteriori probability distributions.

A Markov diffusion process, with its statistical interconnections of states, represents the most adequate formal model of the information process, we are considering.

The EF functions (I.1) are *partially observable,* measuring only current covariation function on the process' trajectories.

Let us have a single dimensional EF (I.1) with drift function $a = c\tilde{x}(t)$ at given nonrandom function $c = c(t)$ and diffusion $\sigma = \sigma(t)$. Then, the EF acquires form

$$S[\tilde{x}_t / \varsigma_t] = 1/2 \int_s^T E[c^2(t)\tilde{x}^2(t)\sigma^{-2}(t)]dt, \tag{I.4}$$

from which, at current $\sigma(t)$ and nonrandom function $c(t)$, we get

$$S[\tilde{x}_t / \varsigma_t] = 1/2 \int_s^T [c^2(t)\sigma^{-2}(t)E_{s,x}[x^2(t)]dt = 1/2 \int_s^T c^2[2b(t)]^{-1}r_s dt, \tag{I.4a}$$

where for the diffusion process, the following relations hold:

$$2b(t) = \sigma(t)^2 = dr/dt = \dot{r}_t, E_{s,x}[x^2(t)] = r_s. \tag{I.5}$$

These relations *identify* entropy functional (I.1) on observed Markov process $\tilde{x}_t = \tilde{x}(t)$ by measuring the covariation (correlation) functions at applying positive function $c^2(t) = u(t)$, and *represent* the EF functional through a regular integral of functions

$$A(s,t) = r_s[2b(t)]^{-1} = r_s \dot{r}_t^{-1} \text{ (I.6a) and } u(t):$$

$$S[\tilde{x}_t / \varsigma_t] = 1/2 \int_s^T u(t)A(t,s)dt. \tag{I.6}$$

The *n*-dimensional functional integrant (I.6) follows directly from the related *n*-dimensional covariations (I.5), dispersion matrix, applying *n*-dimensional function $u(t)$.

Correlation function in (I.5) on a small interval $o(s)$ in form:

$$r(s) = \int_s^{s+o(s)} 2b(t)dt = 2b(s)o(s) \tag{I.6a}$$

leads to $A(s,s) = [2b(s)]^{-1}2b(s)o(s) = o(s)$, and to function

$$A(s,t) = o(s)b(s)/b(t) = o(\text{s},t), \tag{I.6b}$$

which brings integral (I.6) to form

$$S[\tilde{x}_t / \varsigma_t] = 1/2 \int_s^T u(t)o(\text{s},t)dt. \tag{I.7}$$

If function $u(t)$ cuts off the diffusion process on small interval $\delta_o = o(\text{s},t)$, it cuts correlation function (I.6a) of function (I.6b), which bring entropy of cutting correlation (I.7).

The impulse $\delta$-cutoff of $u(t)$ evaluates the quantity of information which the functional EF conceals, when the correlations between the non-cut process states being bound. The cut-off leads to dissolving the correlation between the process cut-off points, losing the functional connections at these discrete points.

Applying delta-function $c^2(t,\tau_k) = \delta u_t(t - \tau_k)$ to integral (1.7) determines the cutting functions:

$$\Delta S[\tilde{x}_t / \varsigma_t]\Big|_{t=\tau_k^{-o}}^{t=\tau_k^{+o}} = \begin{cases} 0, t < \tau_k^{-o} \\ 1/4o(\tau_k^{-o}), t = \tau_k^{-o} \\ 1/4o(\tau_k^{+o}), t = \tau_k^{+o} \\ 1/2o(\tau_k), t = \tau_k, \tau_k^{-o} < \tau_k < \tau_k^{+o} \end{cases}, \tau_k^{-o} < \tau_k < \tau_k^{+o}. \tag{1.8}$$

In such impulse, represented through opposite No-Yes (0-1) actions, each No action carries the cutting impulse part with a maximum of cutting entropy, while Yes action following the impulse cutting part gains the maximal entropy reduction. The $\delta_o = o(\text{s},t)$ impulse' cut of correlation $r(s)$ at moments $s$ maximizes this entropy part; the correlation maximal jump at following moment $t$ dissolves mutual correlation $r(\text{s},t) \to 0$ that maximizes the derivation, minimizing part $\dot{r}_t^{-1}$ of the entropy. That leads to a max-min principle of relational entropy between the impulse parts $(s,t)$ transferring the probabilities (1.2). The max-min variation principle implies the invariant functional (I.6), (1.7) under $u(t)$. Sequential cuts transform the entropy contributions from each maximum though minimum to the next maximal information contributions, where each next maximum decreases at the following cutoff moments.

Each $\delta$-cutoff at these points loses the amount of 0.5 Nats [10] minimizing integral (1.7).

The equations of max-min variation principle for the EF describes trajectories of information process, which the optimal EF integrates [10,13]. Information path functional (IPF) unites information cutoff contributions $\Delta I[\tilde{x}_t / \varsigma_t]_{\delta_k}$ taking along $n$ dimensional Markov process:

$$I[\tilde{x}_t / \varsigma_t]|_s^{t \to T} = \lim_{k=n \to \infty} \sum_{k=1}^{k=n} \Delta I[\tilde{x}_t / \varsigma_t]_{\delta_k} \to S[\tilde{x}_t / \varsigma_t], \tag{I.9}$$

where the IPF along the cutting time correlations on optimal trajectory $x_t$ in a limit determines

$$I_{x_t}^p = -1/8 \int_s^T Tr[\dot{r}r^{-1}]dt = -1/8 Tr[\ln(r(T)/\ln r(s)], (s = \tau_o, \tau_1, ..., \tau_n = T). \tag{1.10}$$

### D. Basis of the observing regularities and their specifics

The regularities of the observing cutoff correlation start from two basic concepts: integral measuring of the impulse' cut random process Bayesian entropy and reaching symmetry of the testing probabilities, while alternating Bayesian probabilities of the impulses run the observing process.

Applying these concepts brings minimax principle of maximal extraction entropy-uncertainty from each probing impulse's minimal cut and its conversion to maximal certainty–information, producing maximum information from this minimum. It leads to multiple maxmin –minimax steps of sequential growing both minimal integral contributions of entropy and probability along the process trajectory' time, which satisfies the microlevel reversibility and macrolevel irreversibility. This dual complementary principle establishes a variation information minimax law whose equations determines the sustainable structure of the Observer and functionally unifies its specific regularities. That implies the following.

-The observing process impulses which cut the process correlation, releases entropy of the prior hidden correlation;

-Cutting (killing) the entropy converts it to Bits (from qubits) of information quantum (micro) process;

-Each Bit, losing the correlation connection to random process, holds memory that comprises logic of its time-space prehistory;

-Such Bit is a self-organized elementary unit of information which the microprocess creates within the cutting impulse though ordering the observing probabilities;

-In a sequence of the Bits logics, each Bit conveys information about previous Bit, enfolds and connects Bits in information process;

-The observed Bit integrates its time–space prehistory in new Bit, which holds its explicit information for each Observer allowing restoration of the prehistory logic in the memorized dynamics and geometry;

-The micro- and macro information processes establish information structure of any physical process;

-Information process self-organizes its logic in space-time shaping structures of information networks, enable natural encoding and producing new information, needed for its extension, adaptation, and integrate the INs in new self-forming structure modeling a human thought;

-The network encodes the logic in a triplet code holding spiral space-time information structure that is running self-restoration and a genetic self-creation. The logic encloses a human cognition an intelligence.

The time-space path from entropy of uncertainty and irregularity to information of regularity and the following asymmetry of information macroprocess models a path from ignorance to knowledge.

**Summarizing the stages determines *what information is*.**

*Definition: Information is memorized entropy cutting hidden correlation in observations.*

Random interactions of observing process whose probability is source of process entropy, innately cut this entropy and memorize it as *information*. An interacting observer' intentional probing impulses reduce entropy of observation, cut it, and memorize the collected information. Memory dynamically freezes the cutting process.

*Origin*: The repeated observations, acting by probing impulses on an observable random process, cutoff the random process' correlation by the opposite (No-Yes) actions. Growing correlations connect the observing process Bayes probabilities increasing each posterior correlation by probabilistic casualty. The probing impulses integrate and temporary memorizes the entropy of cutting correlation. The 0-1 action coverts maximal entropy of the cutting correlation to equivalent information Bit of information microprocess within the impulse, which memorizes logic of the observed entropy probes' prehistory.

*Properties*: The quantum entangled Bit of the information microprocess, formed by (0-1) actions, holds Yes-No Bit's logic of a primary information observer as a Wheeler a Bit-Participator [8] built without any a priory physical law. Forming structure of this Bit includes: logic information of collected 0-1 actions, information of cutting entropy, and free information enables information attraction.

The Bit memorizes a time-space path from getting entropy (via random –virtual cuts) to its entangling within the microprocess, the path of transferring this entropy up to its killing in a finite interval, and a potential path to new Bit. The observer Bit-Participator holds geometry of its prehistory.

The impulse, bringing the information Bit, carries both real (information) microprocess and the information cost of getting the Bit [15]. Sequential impulse cuts along the process convert its entropy to information, connecting the Bits sequences and integrating them in information macroprocess.

**Significance of main finding:**

**1.** *Reduction the process entropy under probing impulse, observing by Bayesian probability' links that increases each posterior correlation; the impulse cutoff correlation sequentially converts the cutting entropy to information that memorizes the probes logic in Bit, participating in next probe-conversions;*

**2**. *Identifying this process stages at the information micro-and macrolevels, which govern the minimax information law;*

**3**. *Finding self-organizing information triplet as a macrounit of self-forming information time-space cooperative distributed network enables self-scaling, self-renovation, and adaptive self-organization.*

**4**. *Creating a path from the process uncertainty to certainty of real information Observer interacting with an observing process (virtual-imaginable or real) via impulse searching observations. The Observer self-creates its conscience and intelligence.*

*The results' analytical and computer validations and illustrations on simulations and experimental applications.*

F. *Numerical attributes of the main stages*

Implementation of the regularities requires to *discriminate numerically the considered main stages, their conditions and constraints* as the followings.
-Starting virtual observation with minimal probability and maximal uncertainty;
-Arising correlation and temporal memory;
-Minimal entropy increments of the interactive impulse;
-Emerging an elementary virtual observer with virtual cutoff the observing process;
-Estimation of the emerging curvature of an impulse carrying the definite quantity of entropy, which leads to the curved impulse.
-Emerging a time-space coordinate system, starting with time-shift displacement, its curving in rotating coordinate system's angular velocity;
-Appearance of virtual observer's geometrical shape in the rotating space-time impulse that encloses a microprocess;
-Evaluation of maximal angular velocity rotating the starting entropy increments in the microprocess' correlation connections and the maximal angular velocity of the entangling entropy increments;
-Estimation of frequency of the Observer probing impulses, verifying the *symmetry* condition and preservation 0-1 law, when the impulses' axiomatic probability is transformed to the microprocess probability with pairs of anti-symmetric (conjugated) entropies;

-Minimal time and space intervals from the starting microprocess and their correlated movements on the path to running a pair entanglement of the space-time conjugated entropies;
-Condition of stable entanglement and the time interval limiting it formation;
-Minimal entropy of the volume entangled in the rotating conjugated entropies;
-Potential entropy per volume, forming by the cutoff impulse;
-Entropy spent on transfer this volume during the impulse cut's wide;
-Entropy measures of the speed killing the entropy volume;
-Transitional speed, transferring the entangled entropy volume to the uncertain zone (gap), where the step-down action enables killing this entropy volume with the above speed;
-Condition of overcoming the gap, depending on the entangled entropy volume rotating with the constraint speed of the entropy rotating moment; probability condition on the edge of reality within the gap;
-Entropy measuring a potential cost for converting the entangle volume entropy to information Bit by the end of Bayesian posteriori action (that killing the impulse entropy);
-Time-space intervals of conversion during the *gap,* needed for both the transition and killing during a finite curved impulse jump, estimating the process of memorizing the moving entangled volume;
-Condition of curving information impulse initiates curvature of gravitation and appearance of mass-energy;
-Condition of rising a cooperative attraction in the rotation with the requiring entropic force, the speed of rotating entropy moment that depends on entropy volume and the gap distance; the probabilistic edge of reality within the gap;
-Constraint, imposed on an information potential function, required by the minimax variation conditions, which defines both the extreme maximal and minimal trajectories of information macrodynamics (IMD);
-Condition of forming a triple consolidation of information macrounits (UP) during cooperative rotation in the information dynamics, requiring synchronization of the forming UP in time-space movement;
-Limitations on binding information Bits in a stable triplet UP;
-Information invariants defining the UP;
-Information frequencies identifying both the UP and information network (IN) invariants;
-Space parameters defining the space trajectory of cooperating UP movement according to minimax;
-Information Observer geometry's shape and volumes generated by the IMD rotating space-time trajectories;
-Necessary condition of joining each next information units in a sequence of triplet's information structures forming the IN, which progressively increases information bound in each triplet knot; forming the IN node on the unit knot and the IN ending node' triplet conserving all IN information;
-Condition of renewal the IN node information through interaction of its requested information with the currently delivering external information holding the required information frequency, which satisfies the needed density of the node information;
-Condition of progressive increasing the IN nodes' information density through shortening the control impulse delivering its information with increasing speed of communication within the IN;
-Iimitations on the IN's parameters of cooperative dynamics building a single cooperation-a doublet and triplet's boundary;
-Space-time structures of various self-cooperating INs;
-Limitation of minimal process' dimension for the cooperating elementary doublets and triplets;
-Minimal elementary uncertainty threshold, separating the process nearest dimensions for double cooperation;
-The cooperative limitations on creating self-connecting sequence of the UP in the IN hierarchical cooperative complexity;
-The observer *time of inner communication* depending on requested information and the *time scale* depending on density of the IN accumulated information;
-The observer conditions of logical encoding-decoding of the information units in the IN doublet-triplet code for quantum and classical computations;
-The invariant information condition for self-structuring, self-replicating observer and cyclic functioning with feedback;
-Limitations on the evolution macrodynamics including the constraints on the feedback invariants and an information potential of evolution. Creation cooperative information mass and space curvature holding the triplet energy.
-Information condition extending the observer information intelligence and getting needed information.
***The paper organizes in Sections divided on Paragraphs:***

Sec.I introduces limitations on starting observable process, evaluates its maximal posteriori and priory probabilities, conditional entropy, frequency of testing impulses' accuracy, and the cutting invariant entropy of a maxmin impulse.

Sec.II presents information conditions of rising information and its observer, emerging microprocess in time-space rotating movement, evaluation of maximal angular velocity and forming entropy volumes in correlation entanglement, minimal time interval on a path to entanglement, conditions of self-stability and converting virtual impulse's cutting entropy in real impulse information.

Curving impulse's time-interval measure as equal probable equivalent of impulse's space-time interval measure.

Sec.III estimates minimal entropy by finding elementary phase volume, developing in rotating movement, which initiates the entropy volume formation in the microprocess' three-dimensional space coordinate system. Finding the entropy per volume, time and speed of the volume transition, and the speed of killing the entropy volume. That confirms existence of an inter-step during the transition to information-uncertain zone (gap), where the step-down action, with the killing speed, cuts this entropy volume. The minimal impulse interval, approaching zero borders at conversion of the impulse entropy to information, which separates Virtual and Information Observer forever. The impulse finite curved jump evaluates the time interval of conversion during the gap, which includes both transition and killing that time. The entropy measure of the information cost for overcoming the gap estimates logical equivalent of Maxwell Demon's energy spent on this conversion and encoding the cutting correlations.

Sec.IV focuses on conditions of rising cooperative attraction in collective rotating movement of information units. Found entropic force determines speed of rotating moment, which numerically evaluates intensity of rotation and collectivization of the entangled grouping ensembles of the units and forming a triple consolidation of information macrounits during cooperative rotation. The information invariants, following from the minimax, numerically evaluate attracting speed and conditions of forming optimal stable triplet's cooperative information structures including the needed frequency of observation. The IMD equations, as solutions of minimax variation problem, describe the averaged attracting microprocesses, which persistently integrates the EF and assembles Bits in triplets' UP units and the UPs in information macroprocess. The dynamic trajectory sequentially orders and composes the UP in series of new triplets. The IPF path functional integrates and measures the information process trajectories. The found minimax numerical restrictions on the space-time rotating trajectory, triple collective spiral movement and forming joint geometrical shape.

Sec.V identifies the cooperative conditions and restrictions on the information parameters assembling the UN triplets in the hierarchical IN. At satisfaction of the necessary condition, each next information units joins in sequence of triplet's information structures forming the IN. Satisfaction of the sufficient condition progressively increases both quantity and quality information bound in each following triplet, while the IN ending triplet's node conserves all IN information and requests the requires quantity and quality of information for extending the IN. The attracting information is enclosed in the IN if it overcomes the minimal threshold for building the IN next triplet's node. The observer ability of selecting and sending the needed frequency of the probing impulses to build such IN defines a selective observer. The restricted information invariants limit minimal process' dimension of cooperative dynamics starting elementary doublet and triplet for the IN formation.

The invariants determine a border of the minimal elementary uncertainty threshold, separating the process nearest dimensions, needed for the double cooperation. The cooperative limitation define the options of creating the INs self-connecting sequence of information units in the IN hierarachy.

Sec.VI introduces operations for encoding information units in the IN code-logic from the observing process, chosen by the observer's (0-1) probes and the logic integration. Performing the observer's task (as an information program) implements logical computation using the doublet-triplet code of encoding information minimizing the IN cooperative complexity.

Sec. VII presents information invariants for self-structuring and self-replicating observer, including the IN feedback with information quality messenger structuring an intelligent observer, limitations on the evolution macrodynamics and cyclic functioning. The information cooperative information mass and space curvature holding the triplet energy.

Sec.VIII illustrates various applications on the selected experimental examples and reviews the results in natural sciences.

Conclusion summarizes the information formalism of observer micro-macrodynamics, including reduction the entropy, origination of the information dynamics, creation of information observer, and conditions of its self-replication.

# I. The observable process of interactive impulses

## 1. Evaluation starting virtual observation, and minimal entropy of virtual impulse.

**1.** Stochastic process with 0-1 probabilities holds a virtual reversible Yes-No actions, where each second (No) through recursion affects the predecessor if there is any their correlation.

Until that, the current random process is complete uncertain and virtually is not observable. The emerging correlation connects initial Bayes a priory-a posteriori probabilities $P_{aoo} \to P_{poo}$ increasing a posterior correlation. The current correlation temporary memorizes this connection, indicating start of Yes- No virtual recursive actions which reveal the virtual observations. Continuing the actions bring sequence of correlations $\uparrow \vec{r}_{ai} \downarrow \vec{r}_{pi} \uparrow$ when each a priori process correlation $\vec{r}_{ai}$ connects to a posteriori process correlation $r_{pi}$ reducing the relative entropy and increasing cumulative Bayes probabilities along the process.

The memory temporary holds the difference of the probabilities actions, as a virtual measure of an adjacent distance between the actions and a probabilistic accuracy of measuring correlation.

The correlation indicates appearance of a time interval [4] of the impulse–observation. Random process' impulses hold virtually observing random time intervals measuring the minimal adjacent distance, Fig.**A**.

Each opposite No-Yes interactive actions (0-1) potentially carries an impulse which might cutoff the random process correlation whose conditional (relation) entropy decreases with growing the cutting correlations.

What is a minimal entropy which an impulse's delta-function cuts off?

**2**. Impulse model of Dirac's delta-function represents a difference of step-down $u_-^t = u_-(t)/\delta_o$ and step-up $u_+^t = u_+(t+\delta_o)/\delta_o$ functions on interval $\delta_o$:

$$u_t^{\delta_o} = [u_-(t) - u_+(t+\delta_o)]/\delta_o = \delta^o u_\pm(t) \tag{1.1}$$

at switching moment $t$ of the step-functions.

**2a.** Entropy functional (I.1) under impulse (1.1) at real switching moments $t = \tau_k$ takes value

$$\Delta S[\tilde{x}_t(\delta^o u_\pm(\tau_k)] = 1/2. \tag{1.2}$$

At locality of $t = \tau_k$: at $\tau_{k-o} \to \tau_k$, where the impulse step-down control's acts, it evaluates

$$\Delta S[\tilde{x}_t(u_-(\tau_{k-o}))] = 1/4. \tag{1.2a}$$

At $\tau_k \to \tau_{k+o}$, where the impulse step-up control's acts, it evaluates

$$\Delta S[\tilde{x}_t(u_+(\tau_{k+1}))] = 1/4, \; u_+ = u_+(\tau_k), \; \tau_k \to \tau_{k+o}. \tag{1.2b}$$

Each cutoff interval $\delta_k = \tau_k^{+o} - \tau_k^{-o} = 1/2o(\tau_k)$ (1.2c) brings amount of entropy integral (1/2), becomes proportional, or these intervals in the process time course run the integration.

Proof [10] also validates that impulse function $\delta u_{\tau_k}^{\mp}$ switches the entropy functional from its initial minimum to cutting maximum and then back from the maximum to a next minimum delivering *maximal amount* of information $\ln 2 \cong 0.7 Nat \approx 1 Bit$ from these max-min actions.

**3**. The real impulse cut delivers that maximal information, while minimal entropy (1.2) evaluates a virtual impulse cutoff, which cannot bring the real step-control with delivering information (1.2a,b).

If a virtual impulse delivers minimal entropy(1.2) to the following virtual impulses, reaching this threshold starts self–observing process, whose posteriori action virtually coveys a next impulse cutting action, enables self-supporting the process continuation, and a virtual observer rises as a part of the primary random process with interactive impulses.

**3a.** An interactive impulse of the virtual observer cuts observable process when minimal discrete increment of process correlation $\Delta r_i$, related to a priory process correlation $r_i$, reaches discrete threshold

$$\Delta r_i / r_i = 4, \tag{1.2c}$$

which determines the *condition of getting minimal entropy* increment $s_{eo} = 0.5 Nat$.
The minimal process' correlation emerges from the connections of Bayes priory-posteriori probabilities measure, which starts virtual observer. The self–observing virtual observer sends virtual probes to continue self-observation intendent to preserve initial (0-1) probabilities measure of the stochastic process.

**2. *Evaluation the probability path from virtual probes to probability approaching a real cutoff***
1. The real cut, delivering information $\ln 2$, starts with minimal posteriori probability
$$P_{po}^{t-} = \exp(-\ln 2) \approx 0.86, \qquad (1.1)$$
and ends with maximal posteriori probability $P_{po}^{t+} \to 1$.

**2.** Relational probability increment $\Delta p_k$ measures ratio of Bayesian a priori probability $P_{ao}$, starting virtual probes, to a posteriori $P_{po} \to P_{po}^{t-}$ probabilities *approaching a real cutoff*
$$\Delta p_k = \Delta p_{ap} = P_{ao} / P_{po}. \qquad (1.2)$$
A lower limit of maximal increment $\Delta p_k$ with accuracy $\varepsilon_k \in (0,1), i = 1,2,....n$ estimates [16,10]:
$$\Delta p_k = \max \varepsilon_k \exp-[2S_{ki} /(1-\varepsilon_k)]^{1/2}, i = 1,2,4,..,N, N = n \times m_o, \qquad (1.3)$$
Extremal solution of (1.3) in the form
$$(1-\varepsilon_k)^3 / \varepsilon_k^2 = 1/2NS_{ki}, \qquad (1.3a)$$
where $S_{ki}$ is a potential entropy corresponding $\Delta p_k$, $m_o$ is number of virtual probes within each of $n$ process' dimensions, and $i \to N = n \times m_o$ evaluates a number of total probes.
For $m_o$ virtual probes in each dimension at $n = 1$, relation (1.3a) leads to
$$(1-\varepsilon_k)^3 / \varepsilon_k^2 = 1/2m_o S_{ki}. \qquad (1.3b)$$
Relations (1.3, 1.3a,b) evaluate the lower limit of maximal probabilistic distance (path) from probability of real cut (1.2) down to some starting probability of $N$ or $m_o$ virtual probes which a virtual observer can send. Increasing accuracy $\varepsilon_k$ of probability $P_{ao}$ will brings more correct evaluation of this probability distance, or $P_{ao}$ at given $P_{po}$ in (1.2). For each impulse with minimal entropy
$$S_{ki} = s_{eo} = 0.5, \qquad (1.4)$$
minimal realistic $\varepsilon_k = 4.5 \times 10^{-4}$ evaluates relative probability increment
$$\Delta p_k \cong 4.5 \times 10^{-4} \exp(-1) \approx 1.65 \times 10^{-4}, \qquad (1.4a)$$
which estimates number of virtual impulses $m_o = 8800$, with frequency $f_o = 8800 Hz$ enables to send.
At accuracy $\varepsilon_k = 0.1$, $S_{ki} = 0.5$ we get
$$m_o \cong 29, \Delta p_k \cong 0.03499, f_o = 8800Hz, P_{ao} = 0.86 \times 0.03499 \cong 0.03. \qquad (1.5)$$
Choice of accuracy depends on a virtual observer that starts probing impulses with a priori probability $P_{ao}$ determined by $\varepsilon_k \in (0,1), i = 1,2,....n$, which for the impulse with $S_{ki} = 0.5$ settles the probes frequency.
When the frequency ratio of probing (virtual) impulses reaches 0.15, which corresponds to Bayes entropy 0.5 Nat (~0.6 Bit), the microprocess starts (by overcoming automatically this threshold).
Eq.(1.3b), at fulfilment (1.4), in particularly, satisfies $\varepsilon_k = 1/2^{m_o}$, which at $m_o = 1$ corresponds $N = n$, or
$$\varepsilon_{kN} = 1/2^N = 2^{-n}. \qquad (1.5b)$$
Thus, increasing dimensions of virtual process minimizes primary error (1.5b) (between primary (0-1) dual actions, which start minimal relative probability increment that initiates decrease of uncertainty:
$$\Delta p_{kN} = 2^{-n} \exp[-1(1-2^{-n})^{-1}]^{1/2}. \qquad (1.5c)$$

At hudge $n \to \infty, \Delta p_{kN} \to 2^{-n}, \exp[-1(1-2^{-n})^{-1}]^{1/2} \to 0$. (1.5d)

The entropy of error $S_k = 2(1-\varepsilon_k)^3 / N\varepsilon_k^2$ at $\varepsilon_{kN} = 1/2^N$ and $N \to \infty$ gets $S_{kN} = 2^{N+1}/N$. (1.5e)

A virtual observer tests the observable process by virtual impulses starting with a minimal $P_{ao}$ and $\varepsilon_k$, specifically using Kolmogorov-Smirnov test [17,18].)
A chosen accuracy (1.4a),(1.5b) estimates potential start of observation with a posteriori probability
$$P_{poo} = P_{po}^{t-}/m_o \cong 0.977 \times 10^{-4}. \quad (1.6)$$

Thus, if increasing correlation brings impulse with entropy $S_{ki} = 0.5$, such impulse temporary holds the probabilities difference (closeness) consistent with accuracy $\varepsilon_{ko}$ of starting correlation and minimal a posteriori probability $P_{poo} < P_{ao}$. The recursive action, overcoming a threshold of a maximal uncertainty with minimal a priory probability $P_{aoo} < P_{poo}$, automatically starts virtual observation that connects impulses in potential virtual test.

The start activates summarizing the current max-min entropies of the impulses cutting random process (which include a previous cuts) in entropy integral measure (1.4) of the process.
The entropy measures beginning of the virtual observer to virtually cut impulses along the observation. When the measured entropy integral reaches its maximum (among previous local max-min), the real cut arise, which minimizes that entropy maximum creating information Bits. At proximity to real cut, priori probability $P_{ao}$ probes grows, reaching for the impulse with entropy $s_k = 1/2$ value

$$P_{ao} = \exp(-s_k) = \exp(-0.5) = 0.6015. \quad (1.6a)$$

If probing action with this $P_{ao}$ precedes a posteriori probability $P_{po} \approx 0.86$, then ratio of a priori – posteriori probability $\Delta p_{ap} = 0.6/0.86 \cong 0.69767$ brings relational entropy $S_{ko} = 0.36$.

It contributes approximately $\cong 1/2\ln 2 \cong 0.35 Nat \approx 0.5 Bit$ of real information in the actual cut.
Therefore, the preceding virtual probe, that corresponds No action of the last impulse, brings ~half of its Bit, while the following Yes action with probability $P_{po} \approx 0.86$ bring another half.

Comments1.1. The Bayes probabilities logic grasps how the Brain does Plausible Reasoning [19], which evaluate the probability relations between entropy and correlations in virtual observing random process. Sociology uses an "Illusory correlation" [20] as the phenomenon of perceiving a relationship between variables (typically people, events, or behaviors) even when no such real relationship exists.

## II. Emerging of the space-time observer, constrains and a microproces
### 1. The correlations, impulse time, and step-wise actions
1. The process correlation emerges from the connections of Bayes priory-posteriori probabilities measure.

A priory $r_{ia}$ or a posteriori correlations $r_{ip}$ implies related time interval $r_{ia} = c^i \tau_i$ [4] with coefficient $c^i$ depending on a time unit mesure $M_p$, which for impulse $(0-1) = \downarrow\uparrow$ satisfies

$$|c^i \tau_i|_M \xrightarrow{P_{ao}} M_p = [1], \quad (2.1)$$

with probability $P_{ao}$ of appearance of the correlation and the related impulse.

For an impulse's size square measure, impulse (1.1) has time unit $1/2o(\tau_k)$ (1.2c) and measure $M = [1/2o(\tau_k)]^2$.

Such a time impulse, preserving measure (2.1), extends initial time unit to $o(\tau_k) = 2$ that holds measure $M_p = [1]$.

2. With growing correlations an adjacent distance between the virtual actions increases, and a shift between the impulse actions moves the observing process from the initial random field to an emerging virtual observer enables renews the virtual actions. The impulse stepwise action displaces the origin of space-time coordinate system of the virtual observer relatively to the origin of space-time coordinate system of virtual observing process. The displacement within the impulse changes the discrete time-space form of the impulse that requires *preserving* its measure (2.1) in the emerging time-space coordinate system. The measure should be conserved in following time-space movement.

**3**. Impulse, modeling step-down $u_-^t = u_-(t)/\delta_o$ and step-up $u_+^t = u_+(t+\delta_o)/\delta_o$ functions on discrete interval $\delta_o = \Delta$, preserves its measure $U_{am}$, if opposite functions $u_-(t), u_+(t+\Delta)$ on this interval satisfy the limitations:

$$[u_+(t+\Delta) - u_-(t)] = U_a(\Delta), \quad (2.2a) \text{ or } [u_+(t+\Delta) + u_-(t)] = U_a(\Delta), \quad (2.2a1)$$

$$u_+(t+\Delta) \times u_-(t) = U_m(\Delta), \quad (2.2b)$$

$$U_a(\Delta) = U_m(\Delta) = U_{am} = c^2 > 0. \quad (2.2c)$$

Condition (2.2a1) or (2.2a) satisfies additivity of these functions (carrying additive probabilities, or entropies); (2.2b) fulfills condition of the functions' multiplicativity; and (2.2c) fulfills condition of preservation both these properties along a Markov process for an elementary impulse with two instance-jumps ↓↑ on $\Delta$ and appearance of a space coordinates in observer's time movement.

Specifically, at $t = \tau_k^{-o}$, $\Delta = \tau_k^{+o} - \tau_k^{-o}$, functions $u_-(t) = u_-(\tau_k^{-o}), u_+(t+\Delta) = u_+(\tau_k^{+o})$ in form

$$u_-(\tau_k^{-o}) = \downarrow_{\tau_k^{-o}} \bar{u}_-, u_+(\tau_k^{+o}) = \uparrow_{\tau_k^{+o}} \bar{u}_+, \quad (2.3)$$

where instance-jump $\downarrow_{\tau_k^{-o}}$ has time interval $\bar{u}_-$ and opposite instance jump $\uparrow_{\tau_k^{+o}}$ which supposedly emerges as a space coordinate, has high $\bar{u}_+$ during $\tau_k^{+o}$, satisfy (2.2a-c) at

$$\bar{u}_- = 0.5, \bar{u}_+ = 1, \bar{u}_+ = 2\bar{u}_-. \quad (2.4)$$

Indeed. Substitution (2.4) leads to $U_a(\tau_k^{+o} - \tau_k^{-o}) = U_m(\tau_k^{+o} - \tau_k^{-o}) = U_{am} = 0.5$. (2.5)

Since $U_m(\Delta) = M$, such impulse, preserving measure (2.1), needs duplication: $U_{amk} = |1|_k$.

It requires additional third instance jump $\uparrow_{\tau_{k+}^{+o}} \bar{u}_\pm$ mounting space coordinate $\bar{u}_\pm$, which forms an emerging space–time impulse $[\downarrow_{\tau_{k+1}^{-o}} \uparrow_{\tau_{k+1}^{+o}}]$ made after transforming function $\uparrow_{\tau_{k+}^{+o}} \bar{u}_\pm$ to function $\downarrow_{\tau_{k+o}^{\pm}} \bar{u}_-^{k=1}$ at $\bar{u}_-^{k=1} = -\bar{u}_+$. (2.5a)

First part $[\downarrow_{\tau_k^{-o}} \uparrow_{\tau_k^{+o}}]$, with time unit $\bar{u}_- = 0.5$, is a transitional impulse starting space coordinate $h = \bar{u}_+$ which at $h(\tau_k) = 2/o(\tau_k)$ preserves measure (2.1) for such potential time-space impulse:

$$M_p = [1/2o(\tau_k) \times 2/o(\tau_k)] \xrightarrow{P_i} M_h = [1/2o(\tau_k) \times h] \to [1]_M, M = [1/2 \times 2] \to [1]. \quad (2.6)$$

If starting function $u_-(t) = u_-(\tau_k^{-o})$ applies to an impulse with only time interval unit measure $[1]$ having probability of appearance $P_i$, than *this* time-space impulse has the same probability of appearance $P_i$.

In equal time measures $M_p$ (2.1) and time space $M_h$ (2.6), time unit $1/2o(\tau_k)$ is extending up to time unit $o(\tau_k) = 2$ which replaces the equal space unit $h = 2$. Thus both impulses have equal measures and equal probability equivalent, which preserves each Yes-No action of these functions.

That also preserves the impulse entropy measure, when virtual Observer is cutting the correlations.

Comments 2.3. The transitive virtual impulse, while preserving measure (2.6), may not satisfy (2.2c). Each step-function (2.3) cuts observable process having particular probability, associated with the cutting function. Satisfaction condition (2.2a-2.2c) for the cutting functions keeps both additivity and multiplicativity within such impulse, whose preservation holds Markov diffusion process with both additive and multiplicative functionals [5] under these functions.

**3a.** Let us implement transition of time unit impulse to time-space impulse applying time jumping function $u_- = \downarrow_{\tau_k^{-o}} \bar{u}_-$ until instant $\tau \to \tau_k^{+o}$ with function $u_+(\tau \to \tau_k^{+o}) = \uparrow_{\tau_k^{+o}} \bar{u}_+$ jumps space coordinate $\bar{u}_+ = h$.

Jumping function $u_-$ replicates an elementary discrete impulse $\downarrow\uparrow$ with both opposite functions $u_-, u_+$ acting within interval $0[t_o^{\mp}] = \delta_o[t_o^-, t_o^+]$ where $u_-(t_o^-) = \downarrow_{\tau_k^{-o}} \bar{u}_-$ starts at $t_o^- \in \tau_k^{-o}, \bar{u}_- = 1/2o(\tau_k)$.

Functions $u_+, u_-$ acting within transitional impulse satisfy relations (I.C) for
$$c^2 = |\bar{u}_+ \bar{u}_-| = c_+ c_- = \bar{u}^2, c_+ = \bar{u}_+, c_- = \bar{u}_-, \tag{2.7}$$
where the impulse time unit measure is $M = [1/2o(\tau_k)]^2 = c^2$.
These functions, satisfying (2.7), hold conjugated imaginary discrete complex:
$$u_+ = j\bar{u}_+, u_- = -j\bar{u}_- \quad j = \sqrt{-1} \tag{2.7a}$$
Assuming both functions (2.7a) starts within instant $\tau_k^{-o}$ instantaneously, functions hold
$$u_{\mp} = \mp j\bar{u}_{\mp}, \bar{u}_{\mp} = \bar{u}, \tag{2.7b}$$
which preserve measure (2.6a) of that impulse.

Function $u_+(\tau \to \tau_k^{+o}) = \uparrow_{\tau_k^{+o}} \bar{u}_+$ ending with instant $\tau \to \tau_k^{+o}$ creates space coordinate high $\bar{u}_+ = h = \bar{u}$.

<u>Comments 2.3a</u>. Functions $u_+ = (j-1), u_- = (j+1)$, satisfying both (2.2a, 2.2b) in forms $u_+ - u_- = (j-1) - (j+1) = -2$, $u_+ \times u_- = (j-1) \times (j+1) = (j^2 - 1) = -2$, do not preserve positive measure (2.2c).

## *1. Starting the microprocess*

When the frequency ratio of probing (virtual) impulses reaches 0.15, which corresponds to Bayes entropy 0.5 Nat (~0.6 Bit), the microprocess starts overcoming automatically this threshold.

Instant time-jump of the time measured entropy within impulse under opposite functions (2.7a) curves the time measure, which creates a space shift within this impulse.

**1.** The functions jumps provide the increments the relative entropy:
$$\frac{\delta S}{S} / \delta t = \mp j\bar{u} \tag{2.8}$$
which in a limit leads to differential Eqs. for opposite entropies
$$\dot{S}_+(t) = -j\bar{u} S_+(t), \dot{S}_-(t) = j\bar{u} S_-(t) \tag{2.8a}$$
starting processes-solutions of (2.8a) at initial conditions $S_+(t_o^+), S_-(t_o^+)$:
$$S_+(t) = S_+(t_o^+)[\operatorname{Cos}(\bar{u}t) - j\operatorname{Sin}(\bar{u}t)]\big|_{t_o^+}^{t=\tau}, S_-(t) = S_-(t_o^+)[\operatorname{Cos}(\bar{u}t) + j\operatorname{Sin}(\bar{u}t)]\big|_{t_o^-}^{t=\tau}, \tag{2.8b}$$
on the process time interval $\delta_o[t_o^-, t_o^+]$ within this virtual impulse with time measure $|1|_M$.

Trajectories (2.8b) describe potential anti-symmetric conjugated wave dynamics of a microprocess within the impulse.

**2.** *The relation between the curved time and equivalent space length within an impulse*

Let us have a two-dimensional rectangle impulse with wide $p$ measured in $[\tau]$ unit and high $h$ measured in $[l]$ unit, with the rectangle measure $M_i = p \times h$. (2.9a)

The problem: Having a measure of wide part of the impulse $M_p$ to *find* high $h$ at equal measures of both parts:
$$M_p = M_h \text{ and } M_p + M_h = M_i. \tag{2.9b}$$
Since that, $M_h = 1/2 M_i = 1/2 p \times h$, and measure $M_p$ of the impulse wide $1/2 p$ equals to measure $M_h$ of the rectangle with high $h$ and the same wide as measured by $M_p$.

Assuming that first part of the impulse $M_p$ has only wide part $1/2p$, it equals $M_p = (1/2p)^2$.

Then from $M_p = (1/2p)^2 = M_h = 1/2 p \times h$ follows

$h/p = 1/2$. (2.9)

Let us find a length unit $[l]$ of the curved wide time unit $[\tau]$ using relations

$2\pi h[l]/4 = 1/2 p[\tau], [\tau]/[l] = \pi h/p$. (2.10)

Substitution (2.9) leads to relation of the measured units:

$[\tau]/[l] = \pi/2$. (2.11)

Ratio (2.11) sustains orthogonality of these units in time-space coordinate system, but since initial relations (2.9b),(2.10) are linear, ratio (2.11) represents a linear connection of the time-space units, received through the curving the time unit in the impulse-jumps. The microproces, built in rotation movement curving the impulse time, adjoins the initial orthogonal axis of time and space coordinates (**Fig.1**). The impulses, preserving measures (2.9a), (2.9), have common ratio of $h/p = 1/2$, whose curving wide part $p = 1/2$ brings universal ratio (2.11), which concurs with (2.4).

At above assumption, measure $M_h$ does not exist until the impulse-jump curves its only time wide $1/2p$ at transition of the impulse, measured only in time, to the impulse, measured both in time $1/2p$ and space coordinate $h$.

According to (2.9a), measure $M_h$ emerges only on a half of that impulse' total measure $M_i$.

The transitional impulse could start on border of the virtual impulses $\downarrow\uparrow$, where transition curving time $\delta t_p = 1/2p$ under impulse-jump during $\delta t_p \to 0$ leads to

$M_h \Rightarrow M_i = p \times h, M_p \to 0$. (2.11a)

If a virtual impulses $\downarrow\uparrow$ has equal opposite functions $u_-(t), u_+(t+\Delta)$, at $\bar{u}_+ = \bar{u}_-$, additive (2.2a): $U_a(\Delta) = 0$ and the impulse has only multiplicative $U_m(\Delta) \neq 0$, its relation (2.2c): $U_m(\Delta) = U_{am}$ is finite only at $\bar{u}_+ = \bar{u}_- \neq 0$. If any of $\bar{u}_+ = 0$, or $\bar{u}_- = 0$, both multiplicative $U_m(\Delta) = 0$ and additive $U_a(\Delta) = 0$. At $\bar{u}_- \neq 0$, $U_a(\Delta)$ is a finite and positive, specifically, at $\bar{u}_- = 1$ leads to $U_a(\Delta) = 1$ preserving measure $U_{amk} = |U_a|_k$.

An impulse-jump at $o[t_o^{\mp}] \to \delta t_p \to 0$ curves a "needle pleat" space at transition to the finite form impulse. The Bayes probabilities measure may overcome this transitive gap.

**2a.** Potential time-space impulse (2.5a) satisfies (2.2a-b) in forms $U_a(\Delta) = \bar{u}_+^{k+1} - \bar{u}_-^{k+1} = \bar{u}_+^{k+1} + \bar{u}_+$,
$U_m(\Delta) = \bar{u}_+^{k+1} \times \bar{u}_-^{k+1} = \bar{u}_+^{k+1} \times \bar{u}_+$, (2.11b) which fulfils (2.2c) only at

$\bar{u}_+^{k+1} = \bar{u}_+ = 2$, $U_{am} = 4$. (2.11c)

That also confirms $\bar{u}_+ = h = 2$ for transitional impulse jump $u_+(\tau \to \tau_k^{+o}) = \uparrow_{\tau_k^{+o}} \bar{u}_+$ changing sign to

$\bar{u}_-^{k+1} = \downarrow_{\tau_{k+1}^o} \bar{u}_+$. (2.11d)

Such new a posteriory impulse preserves its both additive and multiplicative measures at the renoved measure (2.11a), for example entropies $S_-^{k+1} = S_+^{k+1} = 2$. (2.11e)

Complex functions $\bar{u}_+^{k+1} = (1+j), \bar{u}_-^{k+1} = (1-j)$, satisfying (2.11b) at $(1+j)+(1-j) = (1+j)\times(1-j) = 2$, hold a microprocess within a posterior impulse. These functions, jumping with $\bar{u}_-^{k+1}$ on $o[t_o^{\mp}]$, develops impulse interval $|\bar{u}_+^{k+1} \times \bar{u}_-^{k+1}| = \bar{u}_+ = 2$, which preserves time-space positive measure (2.11c).

**2b.** Let us find the impulse length unit $[l]$ and time unit $[\tau]$ for the transitional impulse' time-space measure $|M_{io}| = p[\tau] \times h[l]$ preserving the impulse measure $M[\bar{u}_k] = |2 \times 1/2| \xrightarrow{p[\bar{u}_k]} |1|_M$ at $h = 2, p = 1/2$ and

$|M_{io}| = |1|_M [\tau] \times [l]$, $M = M[\bar{u}_k] = h \times p$. (2.12)

Substituting (2.11) leads to

$[l] = \pm[(|M_{io}|/|1|_M)(2/\pi)]^{1/2}, [\tau] = \mp[(|M_{io}|/|1|_M)(\pi/2)]^{1/2}$, (2.12a)

and to $|M_{io}| = M[\bar{u}_k]\pi/2 \times [l]^2$, which at $p = 1/2h, M[\bar{u}_k] = 1/2h^2, 1/2h^2[l]^2 \pi/2 = |M_{io}|$ and $h[l] = 2$ holds

$|M_{io}| = \pi$. (2.12b)

Then
$$[l] = \pm(2\pi)^{1/2}, [\tau] = \mp 2^{1/2}(\pi)^{3/2}, \quad |M_{io}| = [\tau] \times [l] = \pi \qquad (2.12c)$$

**3.** The entropy microprocess rises when the cutting action of the impulse curves its time $\tau = 1/2 o(t)$ up to appearance of space unit $\delta l \sim h(l)$ during that time.

Probabilities of transfer ½ time unit to two space units equivalents within the transitional impulse.
Both increments of entropy (2.8b) are equivalent for these finite time and space units, their time and space location is not distinctive: both have the equivalent time and space shift.
Entropy increments (2.8b) at moment $\tau = 1/2 o(t)$, $\bar{u} = h$ are

$$S_+(t = 1/2 o(t))) = S_+(t_o^+)[Cos(\bar{u}1/2 o(t)) - jSin(\bar{u}1/2 o(t)),$$
$$S_-(t = 1/2 o(t))) = S_-(t_o^-)[Cos(\bar{u}1/2 o(t)) + jSin(\bar{u}1/2 o(t))]$$

with initial conditions at beginning of the impulse:
$$S_+(t = t_o^+) = S_+(t_o^+), S_-(t = t_o^-) = S_-(t_o^-) \qquad (2.13a)$$

At $\bar{u} \to \bar{u}(\tau) = 1/2 o(\tau) \to h(\tau) = 2/o(\tau)$, the relative values of these increments:

$$S_+^* = S_+(t = 1/2 o(t))) / S_+(t_o^+) = [Cos(\bar{u}1/2 o(t)) - jSin(\bar{u}1/2 o(t))]\big|_{t_o^+}^{t=1/2 o(\tau)} = Cos\pi = 1$$
$$S_-^* = S_-(t = 1/2 o(t)) / S_-(t_o^-) = [Cos(\bar{u}1/2 o(t)) + jSin(\bar{u}1/2 o(t))]\big|_{t_o^-}^{t=1/2 o(\tau)} = Cos\pi = 1 \qquad (2.13)$$

have equal amplitudes $|S_+^*| = |S_-^*| = 1$ of these trigonometrical functions and the absolute value of these complexes. Since multiplication $\bar{u}(\tau)1/2o(\tau)$ is equivalent to space-length interval unit $|M_{io}| = [\tau] \times [l] = \pi$.
Both entropy functions (2.13) during these time or space movement become equal, while each function has a relative probability measure $S_+^* = -\ln p_+, S_-^* = -\ln p_-$ which are antisymmetric.
Potential sum of these functions:

$$S_\mp^* = S_+^* + S_-^* = -\ln p_+ + -\ln p_- = -\ln[p_+ p_-)], p_+ \times p_- = exp(-S_\mp^*) \qquad (2.14)$$

holds opposite multiplicative probabilities in this microprocess.

**3a.** Complex functions $\bar{u}_+^{k+1} = (1+j), \bar{u}_-^{k+1} = (1-j)$, starting at $o[t_{ok}^\mp] = 1/2$ time unit, develops on interval $\bar{u}_+ = 2$ space units the entropy function in form (2.13), which preserves measure $|M_{ko}| = 2\pi$ at $|M_{ko}| = M[\bar{u}_k]|M_{io}|, M[\bar{u}_k] = \bar{u}_+^{k+1} \times \bar{u}_-^{k+1} = 2, |M_{io}| = [\tau] \times [l]$, and brings $S_+^* = S_-^* = Cos 2\pi = -1$.
Changing the sign, corresponding (2.11d), preserves the equal entropy amplitudes.
For $\bar{u}_-^{k+1} = \bar{u}_+^{k+1} = 2$, measure $|M_{ko1}| = 4\pi$ will bring $S_+^* = S_-^* = Cos 4\pi = 1$ coinciding with (2.13).

**3b.** The amplitudes of these probability functions, at $|S_+^*| = |S_-^*| = 1$, can be equal for
$$p_{+a} = 0.3679, p_{-a} = 0.3679. \qquad (2.14a)$$

That leads to
$$p_{+a} p_{-a} = p_{\pm a}^2 = 0.1353, p_{\pm a}^2 = p_{+a}^2 = p_{-a}^2, S_{\mp a}^* = -\ln p_{a\pm}^2 = 2, \qquad (2.14b)$$
or at $S_{\mp a}^* = 2, p_{a\pm} = exp(-2) = 0.1353$, \qquad (2.14c)
it corresponds to entanglement of entropy fractions(2.13) during space-length interval
$$\pi = 2[\tau]/[l]. \qquad (2.14d)$$

Functions $u_+ = (j-1), u_- = (j+1)$, satisfying (2.2a-c) at the impulse starting interval $o[t_o^\mp]$, extend it to $h(\tau) = 2/o(\tau)$ running the antisymmetric entropy fractions along the extension up to $\uparrow_{\tau_k^{+o}} \bar{u}_+$, while opposite functions $u_+ = (1+j), u_- = (1-j)$, satisfying (2.2a-c) by the impulse end at $\uparrow_{\tau_{k+}^{+o}} \bar{u}_\pm$, mount entanglement of these entropy fractions on impulse space interval $\bar{u}_\pm = \pm 2$ (relative to $o(\tau)$ to hold the impulse probability).
*That indicates appearance of both entangled antisymmetric fractions simultaneously with starting space interval.*

The entangle probabilities amplitude $p_{\pm a}$ of the interacting probability amplitudes $p_{+a}, p_{-a}$ satisfies $p_{\pm a} = \sqrt{p_{+a} p_{-a}}$, and sum of the non-interacting probabilities $p_+ + p_- = \exp(-S_+^*) + \exp(-S_-^*) = p_\pm \neq p_{a\pm}$ nonequals to both interacting probability $p_{\pm a}$ and the summary probability $p_{\pm am} = 0.7358$ of the non-interacting entropy modules $\|S_+^*\| = \|S_-^*\| = 1$.

That violates additivity of these interacting probabilities in transitional impulse according to (2.11a), but preserves additivity of the entropy increments.

The posterior impulse' microprocess preserves both additivity and multiplicativity only for the entropies. These basic results are the impulse' entropy equivalents of the related quantum mechanics (QM) relations.

However $p_+, p_-$, as well as probability amplitudes $p_{+a}, p_{-a}$, are probability of random events in the hidden correlations that the impulse cuts, which distinguishes the considered microprocess from the related QM relations. Along the observations, the entropy microprocess integrates the entropy of multiple impulses. With minimal in impulse entropy is ½ Nat, each initial entropy $S_\pm(t_o) = 0.25 Nat$ in such impulse self-generates entropy $S_{\mp a}^* = 0.5 Nat$, starting the virtual observer' time–space microprocess with probability $p_{a\pm} = \exp(-0.5) = 0.6015$. Probability $p_{a\pm} = 0.1353$, relational to initial conditions, evaluates appearance of time–space impulse that *decreases* an initial entropy on $S_{\mp a}^* = 2 Nat$ (satisfying (2.11c)). The impulse's invariant measure preserves $p_{a\pm}$ along the time-space microprocess for multiple time-space impulses.

**5.** Reaching probability of appearance the time-space impulse needs $m_p = 0.6015/0.1353 \cong 4.4457 \approx 5$ multiplications of invariant $p_{a\pm} = 0.1353$.

Hence, each reversible microprocess within the impulse generates invariant increment of entropy, which enables sequentially minimize the starting uncertainty of the observation.

Assigning initial entropies (2.13a) a minimal uncertainty measure $h_\alpha^o = 1/137$ - physical structural parameter of energy, which includes the Plank constant equivalent of energy,-leads to Eqs:

$S_+(t) = h_\alpha^o [Cos(\bar{u}t) - jSin(\bar{u}t)]|_{t_o}^{t=\tau}$, $S_-(t) = h_\alpha^o [Cos(\bar{u}t) + jSin(\bar{u}t)]|_{t_o}^{t=\tau}$,

$S_{\mp a}^* = 2h_\alpha^o$, $p_{\pm a} = \exp(-2h_\alpha^o) = 0.992727305 = inv$,  (2.15)

which evaluates probability of real impulse'physical strength of coupling independently chosen units.

The initially orthogonal non-rotating entropy fractions $S_{+a}^*(t_o^+)$, $S_{-a}^*(t_o^-)$ satisfy Eqs

$S_{\mp a}^* = (h_\alpha^o)^2 [Cos^2(\bar{u}t) + Sin^2(\bar{u}t)]|_{t_o}^{t=\tau} = (h_\alpha^o)^2 = inv$  (2.15a)

which at $S_{\mp a}^* = (h_\alpha^o)^2 = 0.000053279 \to 0$ leads to $p_{\pm a}^* = \exp[-(h_\alpha^o)^2] = 0.999946722 \to 1$.

The rotation adjoins the orthogonal entropy fractions' geometrical sum in linear sum $2h_\alpha^o = e_\alpha^o$ indicating the fractions entanglement, coupling two anti-symmetric entropy units in minimal entangled real unit $e_\alpha^o$.

### 3. Evaluation the information constraints and conditions of the starting observer

**1.** Estimation of a priori relative $\Delta p_{apei} \cong 0.3679$ in (2.14a) at impulse $S_{ki}$ evaluates physically real minimal $\varepsilon_k = \varepsilon_{ei} = 0.039$ that determines minimal a posteriori probability $P_{ao} = 0.8437$ preceding the real cut. This $\varepsilon_{ei}$, according to (1.3b), evaluates probing $m_{oei} \cong 1248$ up to reaching this probability.

**2**. At real cut of posteriory probability $P_{po} \to 1$, ratio of their priory-a posteriori probabilities

$P_{ao}/P_{po} \cong 0.8437$  (2.16)

determines minimal entropy shift between interactive probabilities $P_{ao} \to P_{po}$ during real cut:

$$\Delta s_{apo} = -\ln(0.8437) \cong 0.117 Nat, \qquad (2.16a)$$

**3**. The information analog of Plank constant $\hat{h}$, at maximal frequency of energy spectrum of information wave in its absolute temperature, evaluates maximal information speed of the observing process:

$$c_{mi} = \hat{h}^{-1} \cong (0.536 \times 10^{-15})^{-1} Nat/\sec \cong 1.86567 \times 10^{15} Nat/\sec, \qquad (2.17)$$

which also estimates a minimal time interval corresponding the shift of entropy (2.16a):

$$\delta t_e \cong 1.59459 \times 10^{-14} \sec \approx 1.6 \times 10^{-14} \sec. \qquad (2.17a)$$

Time shift (2.17a) at maximal light speed $c_o = 3 \times 10^9 m/\sec$ allows estimate a minimal space shift:

$$\delta_{lo} \approx 4.8 \times 10^{-5} m. \qquad (2.17b)$$

**4**. Relations (2.17), (2.17a) lead to estimating maximal information extracted from a real interaction:

$$i_{\max} = c_{mi} \delta t_e \cong 3 \times 10^{29} Nat \cong 4.328 \times 10^{29} Bit, \qquad (2.18)$$

and (2.18) and (2.17b) estimate maximal information density on $\delta_{lo}$:

$$i_{\max}^d = i_{\max} / \delta t_{lo} \cong 6.25 \times 10^{33} Nat/m. \qquad (2.18a)$$

Both (2.18) and (2.18a) limit total information and its density extracted by a minimal impulse.
Since the information path functional integrates the collected information in its final Bit, this Bit's maximal information (2.18) condenses all information path of the extracted bits.

Because sequence of impulses $\overline{u}_k, k=1,...,\infty$ is limited by

$$\lim_{k \to \infty} |1| \overline{u}_k \leq 1|_{k \to \infty} \overline{u}_k = 3^k \overline{u}_1, k = 1,...,n \to \infty, \qquad (2.18b)$$

the IPF integrates the sequential information contributions into a finite integral's impulse [10].

Each Bit $|1|_{k \to \infty}$ acquires maximal information density (2.18a). Therefore maximal information (2.18) should be multiplied on number of impulses which had condensed this information:

$$I_{\max} = i_{\max} \times 3^n, n \to \infty,$$

where $n \to n^{\max}$ is limited by $n^{\max} \cong 14^{14} \cong 1.1112 \times 10^9$, Sec.V(6). That estimates

$$I_{\max} = i_{\max} \times 3^{14^{14}} Bit. \qquad (2.18c)$$

**5**. The estimated $c_{mi}$, $\varepsilon_{ei}$ determine a relative closeness the nearest entropy speed $c_{me}$ to $c_{mi}$ by ratio

$$\varepsilon_{ei} = c_{me} / c_{mi} = 0.039. \qquad (2.19)$$

Entropy (2.16a) and interval (2.17a) estimate the entropy speed on this interval interval:

$$c_{me} = 0.117 / 1.6 \times 10^{-14} = 0.073 \times 10^{15} Nat/\sec. \qquad (2.19a)$$

**6**. Each opposite impulse interval $\delta_{lo}$, carrying anti-symmetrical entropies $\delta_{ei}^o = 0.0585 Nat$, develops an entropic force

$$X_{\delta e}^i = \delta_{ei}^o / \delta_{lo} \approx 0.12 \times 10^4 Nat/m. \qquad (2.20)$$

The space displacement and growing intensity of the entropy force per this interval curves this interval, starting rotation of the observer opposite coordinate systems.

**7**. Angular velocity of the emerged rotating opposite coordinate' eigenvectors defines the eigenvectors inverse matrixes $\overline{A}(t)$ and $(\overline{A}(t))^{-1}$:

$$w(t) = \dot{\overline{A}}(t)(\overline{A}(t))^{-1}. \qquad (2.21)$$

**8**. The angular velocity, emerging with maximal linear speed $c_o$, curves length $\delta_{lo}$ to the length

$$\delta_{low} = \pi \delta_{lo} [m], \quad \delta_{low} = 15 \times 10^{-5} m \qquad (2.22)$$

Ratio $c_o / \delta_{low} \cong w_o$ approximates a maximal angular velocity for the curved length $\delta_{low}$:
$$w_o \approx 0.1989 \times 10^{14} \sec^{-1}. \quad (2.22a)$$
Starting maximal entropy speed (2.17) can rotate initial entropy increments (2.16a) with maximal entropy angular velocity
$$w_{oe} = 0.73 \times 10^{15} Nat/\sec. \quad (2.22b)$$
Rotation with angular velocity (2.22a) under entropy force (2.20) possesses speed of rotating moment
$$\delta M^i_{\delta e} / \delta t = X^i_{\delta e} w_o = 0.2387 \times 10^{17} \text{ Nat/m}\sec^{-1} \quad (2.23)$$
which characterizes intensively of the rotation.

**9**. The covariant dynamics, emerging from the cutting fractions of Markovian diffusion, "is a type of curved manifold known as a Riemannian symmetric space" [21], where the space of diffusion tensor represents the covariance matrix in a Brownian motion.

**10**. The opposite rotating conjugated entropy increments in the microdynamic process start forming imaginary entropy volumes during a rising correlated entanglement. The entangled increments with their volumes, captured in rotation, adjoin in the entropy volume of a stable entanglement.

When the conjugated entropies reaches equalization, the entanglement *minimizes* quantum uncertainty, which initially belongs to probability of interaction within each dimension of random process observing by its probing virtual impulse. (It is assumed that a virtual random process, even before interactive impact, holds quantum uncertainty in a probability field).

While the uncertainty-entropy satisfies symmetry and max-min principle, the observer retains *minimal quantum uncertainty* kept in the entangled volumes as a source of an information unit.

**11**. Satisfaction of the minimax variation principle (created automatically by each impulse in a sequence) guarantees *invariance* of the rotating movement in *n*-dimensional space, which brings the *rotation symmetry*, preserving both *isometry* (orientations) and *translation symmetry*, forming a group of *orthogonal matrices* of rotation (2.21). These properties might not be preserved after entanglement.

**12**. Minimal space interval from the starting microprocess to entanglement evaluates relation:
$$\Delta_{lo} / \delta_{lo} \cong \pi \approx 3, \Delta_{lo} \approx 14.4 \times 10^{-5} m, \quad (2.24)$$
where step-action's time interval $\delta t_e$ (2.17a) starts space interval $\delta_{lo}$ (2.17b); (2.24) concurs with (2.14d). The angular velocity for the curved length (2.24) is $w_{ol} \approx 0.2 \times 10^{13} \sec^{-1}$.

At a fixed speed, space shift (2.17b) evaluates a related time interval assuming $\delta_{te} \cong \delta t_e$:
$$\Delta_{te} \approx 4.8 \times 10^{-14} \sec. \quad (2.24a)$$
During this time-space transition arises the rotation movement, which provides transfer the entropy volume to a gap separated entropy and its information. The transition associates with transferring from the entropy of virtual observer to the information of actual observer with transferring imaginary time interval to the observer real time interval. This time transition starts after the conjugated entanglement becomes *self –stabilized*. That ends the conjugated process with an imaginary time and starts a real time course, which initiates transitional space-time movement emerging within the gap.

The end finalizes quantum interaction, when both conjugated entropy increments adjoin in the stable entanglement. Relation (2.24) identifies minimum *three simultaneous random impulse cuts within that space interval, which correlate in the rotating enfolding volume until reaching the posterior probability, where real cut confines them in the entangling triple.*

**13**. The stable entanglement requires first equalization of entropy speeds for the conjugated entropy movements with subsequent equalization of the entropy increments during the rotation process [11,14]. The first step follows from maximizing entropy speeds directed to their equalization, while the second step follows from minimizing entropy increment at the entanglement.

Thus, moment $\tau_k^{-o}$ preceding the stable entanglement needs a pair $i,m$ of entropy increments $\Delta s_{ei}, \Delta s_{em}$ (analogous to (2.13)) with equal maximal entropy speeds:

$$\Delta \dot{s}_{ei}(\tau_k^{-o}) = \Delta \dot{s}_{em}(\tau_k^{-o}), \tag{2.25}$$

and moment $\tau_k^o$ of the entanglement requires the equal minimal entropy increments satisfying (2.14c):

$$\Delta s_{ei}(\tau_k^o) = \Delta s_{em}(\tau_k^o). \tag{2.25a}$$

Condition (2.25) holds equalization of maximal entropy phase speeds measured by the $i,m$ eigenvalues

$$\lambda_{ei}(\tau_k^{-o}) = \lambda_{em}(\tau_k^{-o}). \tag{2.25b}$$

**14**. The time interval limiting formation of stable entanglements determines [22] ratio

$$\Delta t_d / \Delta t_e = 1/8, \tag{2.26}$$

where $\Delta t_e$ is time interval of the potential non-decaying entanglement, which evaluates:

$$\Delta t_e \leq 0.1 \sec, \tag{2.26a}$$

and $\Delta t_d$ is time interval of self-disentanglement -"sudden death" [23], possible in a transitive impulse.

**15**. Under the minimax principle, verification of *optimal* sequence of probing impulses checks maximal frequency of their occurrences for a minimal number of total checked samples.

This increases the probability of both a maximum (extracted by Yes action) and minimum (extracted by No action) that verifies the observation. End of the observable process indicates a last observable Yes action of the virtual observer, defined by the related maximal posteriori probabilities of the observable process, which assumes reality of this action.

Between the last observable uncertain Yes and a first observing certain No-action locates the gap with zone of uncertainty, following the start of observing (information) process.

Information observer cuts off the observing process extracting information by this real impulse holding space-time geometry. Thus, information observer is an extractor and holder this information.

**16**. The virtual impulses satisfy the relations for both unstable and stable entanglements, except crossing the gap by real Yes action. That precedes the last virtual probing No action within the forming impulses microprocess. The microprocess with entropy fractions within all virtual probing impulses is imaginary. The real microprocess is between that No and real Yes actions with imaginary fractions (2.8a,b) prior to entanglement. This microprocess is forming the Bit which encloses the probes of Weller's predicted Bit.

**III. Converting the virtual impulse's cutting entropy in real impulse information**
*Specific, constrains, and limitations for the microprocess*

**1.** The rotating increments of a space-time coordinated system develop an elementary phase volume of entanglement $v_{eo}$ with entropy per a volume $s_{eo}$ of the entropy fractions pair at entanglement

$$v_{eo} = 2^{s_{eo}}. \quad (3.1)$$

When minimal entropy (1.2) $s_{eo} = 1/2\ln 2$, overcoming threshold (2.16a), initiates the microprocess of the volume formation, the volume entangled by that entropy is:

$$v_{eo} = 2^{1/2\ln 2} \cong 1.272. \quad (3.1a)$$

This volume evaluates a base of orthogonal coordinated system with increments $\Delta \vec{l}_j^i$ needed to starts rotation of the entropy fractions. Span length of the increments' three-dimensional vector approximates relation

$$l^{io} = |\Delta \vec{l}_3^i| = \sqrt{3}[(\sqrt[3]{v_e})]^2 \cong 1.846, \quad (3.2)$$

where $\Delta \vec{l}_3^i = \sum_{j=1}^{3} \Delta \vec{l}_j^i$ is the vector sum of the phase coordinate (at $\sqrt[3]{v_e} \cong 1.066189$) per increment of entropy $s_{eo}$, or the equivalent probability $\Delta p_{eo} \cong 0.7$.

**2.** An external step-down control carries entropy-information evaluated by

$$\delta_{ue}^i = 1/4(u_{io} - u_i), \quad (3.3)$$

where $u_{io} = \ln 2 \cong 0.7 Nat$ is total entropy of the impulse and $u_i \cong 0.5 Nat$ is its cutting part.
The same entropy-information carries the impulse step-up control, while both cutting controls carry $\delta_{ueo}^i \cong 0.1 Nat$.
That evaluates information wide of each single impulse control's cut which the impulse carries:

$$\delta_{ue}^i \cong 0.05 Nat. \quad (3.3a)$$

Actually according to more precise estimation [86], starting the step-down part, the step-up part transferring for creating information, and the final cutting part generating information are accordingly

$$\delta_{ue1}^i \cong 0.025 Nat, \delta_{ue2}^i \cong 0.02895, \delta_{ue3}^i \cong 0.01847 Nat$$

These relations allow estimate both initial curvature $K_{e1}$ of the impulse step-down part, currying entropy $0.25 Nat$, and its relative increment $\delta K_{e1}$ to that part:

$$K_{e1} = (r_{e1})^{-1}, r_{e1} = \sqrt{1 + (0.025/0.25)^2} = \pm 1.0049875, K_{e1} \simeq +0.995037, \delta K_{e1} \simeq +0.004963. \quad (3.3a1)$$

The cutting part's curvature estimates relations for $K_{eo}$ and its relative increment $\delta K_{eo}$:

$$K_{eo} = (r_{eo})^{-1}, r_{eo} = \pm\sqrt{1 + (0.1/0.5)^2} = 1.0198, K_{eo} \simeq +0.98058, \delta K_{eo} \simeq +0.01942. \quad (3.3a2)$$

The transferred part's curvature estimates analogous relations

$$K_{e2} = (r_{e2})^{-1}, r_{e2} = \sqrt{1 + (0.02895/0.25)^2} \cong \mp 1.0066825, K_{e2} \simeq -0.993362, \delta K_{e2} \simeq -0.006638 \quad (3.3a3)$$

with opposite signs to the curvature off step-down part.
The final part cutting all impulse entropy estimates curvatures

$$K_{e3} = (r_{e2})^{-1}, r_{e3} = \sqrt{1 + (0.01847/0.7)^2} \cong \pm 1.000348, K_{e3} \simeq +0.999652, \delta K_{e3} \simeq +0.000348 \quad (3.3a4)$$

whose sign is same as the step-down part.
Thus, the entropy impulse is curved with three different curvature values (**Fig.1a**). These values estimates each impulse' curvatures holding the invariant entropies, which emerge in minimax cutoff impulse carrying entropy $S_{ki} = 0.5$ and a priori probability (1.6a) after getting multiple of probing impulse (1.5).

Since impulse measure $M = |1|_M$ is for cutting time correlation, its cutting curving correlation measure defines ratio

$$r_{iM} = M \times K_{ei}, \quad (3.3a5)$$

where time correlations which have not being cut possess Euclid's curvature $K_{iM} = 1$.

Accordingly, the impulse with both time and space measure $|M_{io}|=\pi$, which could appear with curvature of cutting part $K_{eo}$, define correlation measures

$$r_{icM} = M_{io} \times K_{eo}. \tag{3.3a6}$$

Ratio $r_{icM}/r_{iM} = \pi/|1| K_{ei}/K_{eio}$ (3.3a7) measures increment of the curved impulse correlations at appearance the impulse with emerging space coordinate, relative to the curved correlation of the impulse with only time correlation measure (3.34). Counting (3.3a7) determines $r_{icM}/r_{iM} = (\pi/|1|)K_{eo}/K_{e3} \cong 3.08$. The relative increment of the correlation $\Delta r_{iM}/r_{iM} = (r_{iM}+r_{icM})/r_{iM} = 1 + r_{icM}/r_{iM} \simeq 4$ leads to result (1.2), which in a limit: $\lim_{\Delta r(\Delta t), \Delta t \to 0}[\Delta r_{iM}/r_{iM}) = \dot{r}_{icM}/r_{iM}$ brings the equivalent contribution to integral functional IPF (1.10). The shortening of cutting time intervals [10] triples density of each minimax cut impulse (2.18b) preserving it invariant curving correlation measure (3.3a5). Since any impulse with virtual 0-1 Kolmogorov probability preserve its virtual measure (3.3a5), the related time virtual correlations be able to create the space should triple its measure. That compresses the impulse rising curvature of the invariant impulse which increases probability of cutting its time and emerging space coordinate.

Probability of appearance such virtual impulse compare to only time impulse $p_{a\pm}=0.6015$ requires more probing impulses (Sec.11 (3b)).

3. The impulse of $\ln 2 Nat$ cuts entropy during minimal time interval

$$\delta_{ei}^t = \ln 2 \times \hat{h} \cong 0.4 \times 10^{-15} \sec \tag{3.3b}$$

which measures the impulse wide time.

Time interval $\delta_{eio}^t = 1/2 \ln 2 \times \hat{h} \cong 0.2 \times 10^{-15}$ sec measures the wide of each cutting impulse part.

4. Assuming that cutting entropy $u_i \cong 0.5 Nat$ needs to process formation of volume (3.1a), allows us evaluate a potential entropy per forming volume

$$s_{ev} = 1/2 \ln 2 \times 1.272 \cong 0.4452 Nat. \tag{3.3c}$$

5. Each opposite directional impulse wide (3.3a), carrying anti-symmetrical entropies $\delta_{ue}^i$ per interval $\delta_{lo}$ (2. 17a), starts entropic force

$$X_{\delta ue}^i = \delta_{ue}^i / \delta_{lo} = 0.05/4.8 \times 10^{-5} \approx 0.1 \times 10^4 Nat/m, \tag{3.4}$$

which rotates these entropies with angular velocity (2.22a) and the speed of rotating moment, characterizing intensively of each starting rotation:

$$\delta M_{\delta ue}^i / \delta t = X_{\delta ue}^i w_o = 0.1989 \times 10^{17} \text{ Nat/ m sec}^{-1}. \tag{3.4a}$$

Entropy force, currying entropy $s_{ev}$ (3.3c) per curved interval $\delta_{low}$ (2.22):

$$X_{\delta i}^i = s_{ev}/\delta_{low} \approx 0.296 \times 10^4 Nat/m, \tag{3.4b}$$

rotates $s_{ev}$ with angular velocity $w_o$ and develops speed of rotating moment

$$\delta M_{\delta i}^i / \delta t = X_{\delta i}^i w_o = 0.588 \times 10^{17} \text{ Nat/ m sec}^{-1}. \tag{3.4c}$$

The speed of rotating moment of volume's $s_{ev}$ starts with speed (2.23) which (3.4a) accelerates up to (3.4c).

6. The correlated opposite rotating pairs near entanglement catch the cutting anti-symmetric entropy parts $u_i \cong 2 \times 0.25 Nat$ with speed (3.4c) of synchronous rotation, the related angular frequency $f_{oi} \sim \omega_{oi}$ and transfer it along the created rotary channel, enfolding all correlated entropy flows in the entropy volume at the entanglement.

These spinning couple self-moves with a phase speed which evaluates
$$c^{io} = l^{io}/\delta^{t}_{eio} \cong 1.846/0.2\times 10^{15} = 9.23\times 10^{15}\ \text{sec}^{-1}\ . \tag{3.4d}$$
The rotaing channel's edges, spinning with that speed, form a tunnel movement, which contains the entangled volume by analogy with classical spin Hall Effect in quantum mechanics [24].

As the impulse time interval decreases in the moves, information of the impulse approaches zero when no interaction could be observed. Such control information cannot cover the entropy volume.

A virtual Observer, during a preceding observation, may capture an entropy volume, while the observer impulse's opposite rotating pairs transfer the volume to entanglement. Until next (last) real impulse did not appear, the uncertain gap could not be overcomed, killing did not occur, and the virtual Observer, identifies by the moving entangled entropy volume, would move forever. Its entropy will never convert to information, and Information Observer will never emerge. The observation remains uncertain. But the coupling entropy units can emerge within the gap confining the *entangled units as a part of potential Bit*.

<u>Comments 3.1.</u> In Physical analogy, when impulse time interval equals to light wavelength, an observer moving with light speed misses relative observation, losing information, and becomes virtual observer, which cannot observe any certainty. Additionally, according to Relativity Theory, a real observer, moving within a light wave, such shortens his size that it cannot hold information, and moves nowhere.

Processes within such Observer are reversible. With only probabilistic entropy, such processes are free of material substance. Recent work [25] reveals light's intrinsic quantum spin Hall Effect.

**7**. Transfer volume (3.1a) during the impulse real cut' wide (3.3a) requires spending entropy
$$v_{eo}\delta^{i}_{ue} = s_{ev} \cong 0.0636 Nat\ , \tag{3.5}$$
where sum
$$\delta_{e} + s_{ev} \cong 0.5 Nat\ , \tag{3.5a}$$
is equivalent of the impulse cutting cutting part $\text{u}_{i} \cong 0.5 Nat$, or (1.2).

An external step-down control, which is not a part of this impulse, spends the same entropy $\text{u}_{i} \cong 0.5 Nat$ killing entropy volume (3.5a).

**8**. Speed of killing entropy (and dissolving the hiden correlation) $s_{cev} = s_{ev}/\delta^{t}_{ei}$ measures
$$s_{cev} \cong 1.59\times 10^{15} \approx 1.6\times 10^{15}\ Nat/\sec\ . \tag{3.5b}$$

**9**. The *phase* speed, required for transition the entropy volume with $s_{ve} = 0.0636 Nat$ during transition time $\delta^{t}_{eio} \cong 0.2\times 10^{-15}\ \sec$, evaluates (3.4d), which allows estimate the needed *entropy* minimal speed
$$c_{ev} = s_{ve}c^{io} = 0.587\times 10^{15}\ Nat/\sec\ . \tag{3.6}$$
This transitional speed is only $0.587/1.59 = 0.3692$ of (3.5b) needed for birth of information Bit.

That minimal speed transfers the entangled entropy volume to the uncertain zone (gap), where the step-down action with speed (3.5b) cuts this entropy volume $s_{ev}$. This estimation confirms existence of an inter-step of the transition before generation of information anticipating memory at the gap end $P_{po} \to 1$.

Comparing estimation (2.26a) with the gap transitive time interval indicates that the stable entanglement, which joins the most a priory probable conjugated entropies inside the volume, finalizes *within* the gap. The time –space transition arises during a rotation movement, which transfers of the entropy volume to the gap separating end of uncertainty and start of certainty- reality. The gap holds a hidden real locality whose *edge evaluates probability* (2.15); impulse cuts the locality within the hidden correlation, which confines entangling *qubits*, limited by that minimal uncertainty measure (of a joint couple). According to realtions (3.3b), (3.5b), the entangled entropy volume is killed during the impulse time interval which holds its dynamics and geometry.

Since killing associates with freezing, it memorizes the geometrical form and dynamics of the entangled moving spinning anti-symmetric couples in the starting real space of information units.

The correlation binding this 'quantum' information couples with maximal probability is extremely strong and real.

**10**. The transition time from the Bayesian a priori to a posteriory inferring action that delivers entropy $s_{ve}$ is a cost for converting that entropy to information Bit ($\ln 2$ Nat) by the end of Bayesian posteriori action.

Killing the impulse entropy estimates the cost coefficient [15]:
$$k_e = s_{ev} / \ln 2 \cong 0.0917554. \tag{3.7}$$

**11**. The real posteriori cutting part of the impulse that kills entropy $u_i \cong 0.5 Nat$ brings its information equivalent which includes impulse real step-up control with $\delta_{ui}^i \cong 0.05 Nat$. That impulse covers entropy cost $s_{ev}$ and evaluates $k_e$ value:
$$k_{eu} = s_{ev} / u_i \approx 0.1272. \tag{3.8}$$

An external step-down control spends equal information by killing entropy $u_i$, which ealso evaluates $k_{eu}$.

**12**. The impulse step-up action launches the information unit, while its certain control step-down action finishes the unit formation bringing memory and energy from the interactive jump that de-correlates the entangled entropy. This finite jump transits from uncertain Yes-logic transferring the entangled entropy of observation to certain information No-logic with elementary information of quantum qubit. Bit $\mathbf{a}_{io} \cong \ln 2 = 0.7 Nat$ includes information, which delivers total step-down controls (1.2a): $0.25 Nat$ minus information of the control cutting wide (0.35a): $0.05 Nat$, plus killed entropy $0.5 Nat$.

**13**. Uncertain step-up logic does *not* require energy like the probes of observable-virtual process, or a media whose information is not observed yet. This potential (uncertain) logic belongs to sequential test-cuts before appearance in a priori certain logic, which becomes a part of forming the elementary information unit.

Both transfer the impulse along the virtual process and the inter-entropy transition within each impulse have none predecessor, satisfying reversible logic computation [3] on the microlevel.

**14.** The observing last Yes-action opens a path from entanglement to the certainty when the last cutting No action eliminates both random and quantum uncertainties by producing information.

According to (3.5,3.5a), the quantum Bit spends part $\delta_e^* = 0.0636 / 0.5 = 0.1272$ of its killed entropy on compensation the uncertain gap of a priory logic being $\delta_i^* = 0.0636 / 0.7 = 0.09$ part of whole Bit. Memorizing both entropy paths of transition through the gap and the controls' killing the entropy volume encloses both priory and its posteriory certain logic, perform functions Bit-Participator-a primary information observer without any apriori physical law.

**15**. Time interval of conversion during *gap* $\delta_o$, which includes both transition and killing by the impulse, evaluates ratio
$$\delta_t^o = u_{oi} / s_{cev} \cong 0.4375 \times 10^{-15} \sec, \tag{3.9}$$
where $s_{cev}$ (3.5b) is speed of killing this entropy via the finite curved jump during rotation (Sec.II(3)).

At time of transition $\delta_{eio}^t \cong 0.2 \times 10^{-15} \sec$, the killing impulse' part spends time interval
$$\delta_{kio}^t \cong 0.2375 \times 10^{-15} \sec. \tag{3.9a}$$

Since by the end of $\delta_o$, information unit $\mathbf{a}_{io} = u_{oi}$ appears, the real posteriori control produces the finite posteriori speed of generating information, whose maximum limits constant (2.17), while (3.9) estimates the priority speed.

Minimal time interval of reaching a border of generation information unit $\mathbf{a}_{io} \cong \ln 2$:
$$\delta_t^o = \mathbf{a}_{io} / c_{mi} \cong 0.375 \times 10^{-15} \text{ sec} \tag{3.10}$$
also includes the time of transition up to real time-space information dynamics:
$$\delta_{eio}^t \cong 0.2 \times 10^{-15} \text{ sec}. \tag{3.10a}$$

**16**. Unit $\mathbf{a}_{io}$ contains an information equivalent $s_{ev} = i_{ev}$ of energy $e_{ev}$ spent on converting the entropy to information. Energy $e_{ev}$ delivers the certain step-up control action of converting the invariant entropy impulse to equivalent information. This energy compensates for that conserved in the entangled rotation, specifically by angular moment multiplied on angular speed during time (3.10a), which evaluates entropy $s_{evo}$ in Nats. The cost $e_{ev}$ satisfies Landauer principle and compensates cost of Maxwell demon.

**17**. Boltzmann constant $k_{eb}$, measuring ratio of the unit generating radiation energy transferred to temperature of heat that dissipates this energy, is an analogy of the quantum process' ratio $k_{eu}$, which is the information $\mathbf{a}_{io}$ cost for transforming entropy $\mathfrak{u}_{oi}$ and a logical equivalent of Maxwell Demon's energy spent on this conversion. The cost of logic requires to overcome $\delta_o$, including transferring of the entangled entropy and generation of the unit of information during time interval $\delta_t^o$. Since the entangled movement condenses its entropy in the volume of entanglement, this logical cost is a primary result of time shift $\delta_t$ initiating movement, which compensates for both logical and energy costs, delivering the certain step-wise control. The impulse cutting real time interval $\delta_t^o$ covers cost in conversion entropy in equivalent information.

**18.** Encoding each Bit extracts its hidden position by the invariant cut, which erases the Bit of information at cost of energy. Bit is memorized during that impulse time interval in the ending impulse stopping state. Such impulse encoding merges its memory with the time of encoding information, which minimizes that time.
Each impulse time interval encodes invariant unit of information in physical process whose interactive time interval carries energy equivalent of the impulse information cutting from the correlation. That compensates for Maxwell Demon's cost while producing information during the invariant interactive impulse, which implements the minimax origin of information. Since both interactions and its elementary discrete inter-actions ↓↑ define standard information unit Bit, the impulse code is *universal* originates in any natural process generated in interactions.
The specific physical interaction's energy quantity and quality limit the code length via its final encoding bit's information density; and via the energy entropy quality's probability when the information encoding starts. The minimax also brings both stability of the forming physical impulse and appearance a posterior *certain* physical logic of such multiple elementary information units.

**19.** The Byesian probabilistic certainty by overcoming the gap with probability 1, instantaneously reaches the logical certainty, which concludes the elementary path from the units of uncertainty to information unit of certainty and confirms both as a *fact*.

**20.** <span style="color:red">The curved impulse (Sec.II(2), Fig.1a) preserves the invariant entropies defining the invariant curvatures. But with growing density of both entropy and information, the curvatures related to decreasing impulse wide grows accordingly. These curvatures of the information impulses currying energies are sources of the Einstein gravitation, squeezing the impulse and initiating information form of mass-energy (Sec.VII(3)).</span>

## IV. Arising the observer collective information

### 1. Rising a cooperative attraction

**1**. Cooperative rotation and ordering start with entangling entropy increments in the entropy volumes.
Then, the sequential impulse' entropy increments with their volumes involve in collective rotating movement.

**2.** Since each following impulse may start only after the previous impulse cutting time $\delta_{ei}^t$ (3.3b) will triple, time interval between the impulse cutoff actions:
$$\Delta_t = 3\delta_{ei}^t \cong 1.2 \times 10^{-15} \text{ sec} \tag{4.1}$$
imposes limitation on adjoining a pair of the impulse entropies in collective movement; $\Delta_t$ is imaginary time course [10].

**3**. This limitation is applicable also to virtual Observer with its virtual impulse, which can initiate a virtual collective movement of the adjoining entangled volumes that requires an entropic force

$$X_e = \Delta s_o / \Delta l_o, \tag{4.2}$$

where $\Delta s_o \cong 0.25 Nat$ is a minimal entropy increment between the impulse, and $\Delta l_o \cong \Delta_{lo}$ (2.24) is a distance between nearest impulses with that entropy. At these relations, the entropic force is

$$X_e \cong 0.25 / 14.4 \times 10^{-5} \cong 0.1736 \times 10^4 [Nat / m]. \tag{4.3}$$

**4**. Speed of rotating moment $\delta M_e / \delta t$ defined by force (4.3) and velocity (2.22a) in

$$\delta M_e / \delta t = X_e w_o \tag{4.4}$$

characterizes intensively of the rotation which evaluates

$$\delta M_e / \delta t = 0.344 \times 10^6 [Nat / m\sec]. \tag{4.4a}$$

**5**. Since both the rotating moment's speed and entropy force proceed between the impulses, relation (4.4a) describes intensively of attracting rotation intended on capturing next impulse's entropy increment in joint distributed rotating movement, which measures cooperative connection of the impulses entropy increments prior forming next information units. Thus, virtual collective movement (between the impulses' of the entropy increments) may emerge before killing the following impulse's entropy volumes. Perhaps, that movement engages a cooperative transition of the entangled entropy volumes (with starting speed (3.6)) to the cutting gap. It makes possible a collective entanglement which not requires spending energy. Study [26] "demonstrates that the spreading of entanglement is much faster than the energy diffusion in this nonintegrable system". Such a cooperative virtual distribution rotates the involving entangled groups- an ensemble for a joint preparation them to the following killing-cut, producing information units.

Therefore, the distributed rotation primary involves the entropies of the Bayesian linked probabilities, resulting from virtual interactive probes of different frequencies. Then, it involves the entangled entropies, cooperating in the entropy volumes, engaging in the transition movement up to cutting them off on the information units. The rotation persists entanglement connecting the forming information units.

*2. Forming a triple consolidation of information units during cooperative rotation*

**1.** Imaginary entropy in each virtual impulse, according to the minimax, predefines the real information of certain impulse with information $\mathbf{a}_{io}$, which depends on the impulse real starting information speed $\alpha_{io}$ and time interval $t_i$: $\mathbf{a}_{io} = \alpha_{io} t_i$.

Speed $\alpha_{io}$ of generation $\mathbf{a}_{io}$ starts a single real information unit (after killing the entropy volume).

Speed of cutting correlation is an initial source of real information speed, while imaginary information speed arises in the microprocess. Ending value of this speed is speed of transition of the entropy volume:

$$\beta_{io} = c_{ev} . \tag{4.5}$$

Imaginary $\beta_i$ and real $\alpha_i$ speeds are components of the process complex eigenvalues:

$$\operatorname{Re}\lambda_i = \alpha_i, \operatorname{Im}\lambda_i = \beta_i, \lambda_i = \alpha_i \pm j\beta_i,,$$

which determines ratio of their starting imaginary and real components $\beta_{io}, \alpha_{io}$: $\gamma_{io} = \beta_{io} / \alpha_{io}$.

The minimax requests turning to zero the imaginary component by the end of time interval $t_i$ when the unit gets complete information $\mathbf{a}_{io}$ with speed $\alpha_i(t_i)$.

That requirement leads to Eq. connecting $\mathbf{a}_{io}$ and $\gamma_i$ [11]:

$$2\sin(\gamma_i \mathbf{a}_{io}) + \gamma_i \cos(\gamma_i \mathbf{a}_{io}) - \gamma_i \exp(\mathbf{a}_{io}) = 0. \tag{4.5a}$$

From that, at $\mathbf{a}_{io} \cong \ln 2$ follows $\gamma_{io} = \beta_{io} / \alpha_{io} \to (0.4142 - 0.5)$. \hfill (4.5b)

Constrain (4.5) imposes limitation on $\beta_{io}$ by minimal speed of transferring entropy volume (3.6):
$$\beta_{io} = c_{ev} = 0.596 \times 10^{15} \, Nat/\sec. \quad (4.6)$$
Then, from (4.5)-(4.6) follow information speed of starting a single real information (after killing the entropy volume):
$$\alpha_{io} = c_{iv} \cong 2.4143 \times 0.596 \times 10^{15} \, Nat/\sec \cong 1.44 \times 10^{15} \, Nat/\sec, \quad (4.6a)$$
which determines minimal time interval of completion $\mathbf{a}_{io}$:
$$t_i^o \cong 0.48 \times 10^{-15} \sec, \quad (4.6b)$$
satisfying the minimax; estimation (4.6b) is close to $\delta_{ei}^t$ (3.3b).

Invariant of imaginary information $\mathbf{b}_o' = \beta_{io} t_k$ (between the impulses), according to minimax, satisfies condition of turning the real component to zero by the end of time interval $t_k$ at completion this imaginary information. From that condition follows Eq
$$2\cos(\gamma \mathbf{b}_o') - \gamma \sin \gamma(\mathbf{b}_o') - \exp(\mathbf{b}_o') = 0 \quad (4.7)$$
connecting $\mathbf{b}_o'$ with related ratio $\gamma = \gamma_k$.

Solution of (4.7): $\beta_{io} t_k = \pi/6$ at (4.6) identifies $t_k : t_k = \pi/6/0.596 \times 10^{15} = 0.8785 \times 10^{-15} \sec$,

which determines minimal interval of turning real speed (4.6a) to zero; $t_k$ is close to the constraint interval between impulses (4.1).

**2**. Retio (4.5b) evaluates *attractiveness* a real speed by the imaginary speed, or how, at the same fixed time $\delta_{tio}$, imaginary entropy $\delta_{Eio} = \beta_{io} \delta_{tio}$ enables *attracting real information* at forming a single unit:
$$\delta_{Iio} = \alpha_{io} \delta_{tio}. \quad (4.8)$$
According to (4.5b), at a minimal coefficient of attraction:
$$\delta_{Eio} = 0.4142 \delta_{Iio}, \quad (4.8a)$$
a unit of entropy $\delta_{Eio} = 1$ may attract $0.4142 \, \delta_{Iio}$ units of information.

Entropy volume $s_{ve} = 0.0636 Nat$ moving with minimal speed $c_{ev}$ (3.6) may attract potential-information (entropy) $i_v \cong 0.02634 Nat$ of a nearest impulse on minimal time interval $\Delta_t$ with attracting information speed
$$c_{ivo} = i_v/\Delta_t, \; c_{ivo} \cong 0.0548 \times 10^{15} \, Nat/\sec. \quad (4.9)$$
That might increase minimal transition speed $c_{ev}$ up to
$$c_{ev} + c_{ivo} \cong 0.65 \times 10^{15} \, Nat/\sec. \quad (4.9a)$$
With transition speed (4.9a), the rotating entropy volume moves to the cutting gap, engaging other entangled entropy volumes in a joint (collective) rotation with the speed of rotating moment (4.4).

**3**. Impulse, carrying $\cong 0.25 Nat$, cuts random process with entropy volume which evaluates entropy (3.3c):
$$\delta_e \cong 0.4452 Nat. \quad (4.9b)$$
The control cutting the volume should compensate for the interacting entropy increments $0.117 Nat$ (2.16a), which requires information of the control $0.25 + 0.117 = 0.367 Nat$.

This control, cutting the phase volume (3.1a), could bring information
$$s_{evo} = 0.367 \times 1.272 = 0.466824 Nat. \quad (4.10)$$
Additional potential information conveys the entropy of transferring volume $s_{ve} = 0.0636 Nat$, which may decrease the amount of attracting potential information (entropy) $s_v = i_v \cong 0.02634 Nat$ (determined by ratio in (4.8a) on difference

$$\delta s_{ve} = s_{ve} - s_v = 0.0374. \tag{4.10a}$$

Sum $\delta s_{ve} + s_{evo} \cong 0.5 Nat$ (4.10b)

coincides with (3.5a), which determines the amount of information delivered by the impulse cutting the random process. This information compensates for entropy of the virtual probing that delivers the entangled entropy volume whose potential cut might bring total entropy (4.10b), which the real impulse converts to equivalent information thru memorizing.

Assuming that virtual impulse spends the entropy volume part $\delta s_{ve}$ on transition to cutting gap, the real impulse should overcome entropy threshold $s_{evo}$ by the real cut, producing information $0.5 Nat$, which includes information compensation for the virtual entropy volume

$$\delta s_{ve} = \delta s_{iv}. \tag{4.10c}$$

In multi-dimensional virtual process, correlations grow similarly in each dimension under manifold of impulse observation. The correlations, accumulated sequentially in time, increases with growing number of currently observed process' dimensions, entropy volumes, and collective activities. Each impulse cuts the increasing correlation-entropy volume, leading to rising density of the cutting entropy even at the same impulse size, which increases its speed (as the volume related to the cutting impulse wide). Killing the distinct volumes densities converts them in the Bits distinguished by information density, while each Bit accumulates the observation of complimentary events. Between these different Bits, an information gradient of attraction force rises, minimizing the difference, which predisposes the memory.

That connects elementary information Bits in units of information process of information micro-macrodynamics [10, 11, 86].

*3. Conditions of forming optimal triplet with stable cooperative information structure*

**1.** Information of each previous impulse starts attracting next cutting information in the rotating movement during the impulse imaginary time interval (4.1) with entropy force, moment's speed (4.4, 4.4a), and potential attracting information

$$i_v \cong 0.02634 Nat \tag{4.11a}$$

Information ratio in (4.8a) allows evaluate the attracting information brought by each cutting impulse:

$$i_{vo} = 0.5 \times 0.4142 = 0.207 Nat, \tag{4.11b}$$

whose sum with the potential attracting information (4.11a) adds information for the attraction:

$$i_{vf} = i_v + i_{vo} = 0.23334 Nat. \tag{4.11}$$

As a part of information delivered with the impulse, $i_{vf}$ evaluates its *free* information enables attract next cutting information until this information spends on adjoining the following cutting information.

Each step-down cut requests information (1.2b) (of ~1/3 of the impulse):

$$1/3(\ln 2 + 0.05) Nat \cong 0.24766 Nat \approx 0.25 Nat. \tag{4.12}$$

That, concurring with (1.2b), measures the cooperative attraction, which could deliver the free information (4.11).

**2.** Elementary information unit Bit is an impulse, which encloses the equivalent of entropy $\ln 2$ Nat, whose cut converts this entropy to information. Within the impulse's step-down and step-up actions, the first one, cutting entropy of the final probe, memorizes it delivering an information equivalent cost of energy. The impulse step-up action limits the cutting information to equivalent Nat that satisfies the minimum of maximal cutting entropy volume.

The Bit encloses the probing impulses logic, conserving through information cost of energy as a logical equivalent of Maxwell Demon's energy spent on this conversion, and genetates the free information of attraction.

Thus, the information Bit is a self-participator in both converting entropy to equivalent information, which memorizes logic of its entropy prehistory, and in extending a posterior logic in persistence attraction, engaging the multiple impulse's Bits sequence in an information process.

This primary self-participating interaction shares the difference between the entropy of last virtual step-up and real step-down cut actions within the ending impulse actions, which could emerge in natural or artificial processes.

The real interactive impulse carries both real microprocess, attracting next Bit, and the information cost of getting the Bit. That distinguishes the real impulses from the virtual impulses of random activity.

**3.** The information **e**quivalent of the impulse wide $\delta_{ue}^i \cong 0.05 Nat$ limits its size and the extension minimal time interval $\delta_{te} \approx 1.6 \times 10^{-14}$ sec. Potential information speed of attraction within the impulse:

$$c_{ia} = 1/3(\ln 2 + 0.05)/\delta_{te} \cong 0.1548 \times 10^{14} Nat/\sec, \qquad (4.13)$$

is less than both maximal speed $\alpha_{io}$ (4.6a), starting a single real information from imaginary entropy, and the speed between the nearest impulses on time interval $\Delta_t$:

$$c_{ika} \cong 0.0516 \times 10^{14} Nat/\sec. \qquad (4.13a)$$

Maximal speed (4.13a) conveys a flow of the formed information Bits-an information process, which carries the enclosed entropy, memory, energy and logic, enclosing free information (4.11) between the impulses. A single Bit (in Nat), moving with speed (4.13a), spends its free information of $\sim 1/3 Nat$ to attract next Bit on time interval $t_{ika}$:

$$t_{ika} = 1/3\ln 2 / c_{ika} \cong 1/3\ln 2 / 0.0516 \times 10^{-14} = 4.477 \times 10^{-14} \sec. \qquad (4.13b)$$

Minimal distance $\Delta_{lo} \approx 14.4 \times 10^{-15} m$ to next Bit limits a maximal dynamic spatial speed of information attraction:

$$c_{lo} \approx 14.4 \times 10^{-15} m / 4.477 \times 10^{-14} \sec = 3.216 \times 10^{-1} m/\sec. \qquad (4.13c)$$

In this process, the impulses pair adjoins in a doublet which encloses bound free information spent on the attraction.

**4**. According to Efimov's Scenario [27-30], optimal collective dynamics join a triple of its units in the triplet structure, which was early proposed in Borromean Universal three-body relation, including Borromean knot [31] and ring. Ancient Borromean Rings represent symbols for strength in unity. Information triplet, satisfying the minimax, is forming during the cooperative rotation of information units, applied to each eigenvector, as well as to their groups: doublets and triplets.

**5.** *Space-time trajectory solving the minimax variation problem (VP)*
Equations of information macrodynamics (IMD), following from minimax variation principle (VP), describe the averaged attracting microprocesses, assembling Bits in information macroprocess.
In the IMD Hamiltonian dynamics, time-space movement rotates the opposite directional-complimentary conjugated trajectories $+\uparrow SP_o$ and $-\downarrow SP_o$, forming spirals located on conic surfaces **Figs.2, 3**.

The trajectories compile opposite *segments of information process* $\pm SP_i, i = 1,...,n$ up to maximal process dimension $n$. Each $\pm SP_i$ segment averages the microprocesses attracting a Bit with step-up and step-down control. The control selects sections $-SP_i, \pm SP_i$ ending on a border of bridge $\pm \Delta SP_i$, which binds the spiral at each segment border $i$. Each opposite directional segment's border enables attracting half of each process information unit $\pm UP_i$. The impulse joining No-Yes action connects opposite units $-UP_i$ and $+UP_i$ in unit $UP_i$ of Bit through the bridge $\pm \Delta SP_i$. Dynamics of the border stripe describes trajectory of switching controls located on a middle between the opposite spirals (**Fig.3**). Each pair of opposite sections $+SP_i$ and $-SP_i$ forms local circles $\uparrow \odot \downarrow$ with sequentially reverse directions of their movement while total directions along the conjugated trajectories $+\uparrow SP_o$ and $-\downarrow SP_o$ preserve.

The persistent attracting and assembling Bits hold them in information process, whose information integrates and measures information path functional (IPF) on the process trajectories.

Transfer from the VP extremal's maximum to minimum limits dynamic constrain [11] (equalizing the IPF dynamic and diffusion components).

The IPF integrates each $UP_i$ into $UP_{i+1}$ along the space-time trajectory, and when $-SP_i$ transfers to $+SP_i$, the maximum transforms to the minimum on the bridge.

The IMD linear and nonlinear equations describe information flows initiated via the IPF gradients as information forces, which are physical analogies of the thermodynamics flows and forces in irreversible thermodynamics [11, 12].

**6.** *Assembling information units in triplets*

While the ending microprocess binds each pair $\pm UP_i$ in information unit $UP_i$ on the bridge, the macroprocess assembles each three *information units* $UP_i$ in new formed triplet's units $UP_{oi}$ through their ending minimal information speeds having opposite directions.

**Fig.4** illustrates simplified dynamics of assembling tree units $UP_i$ $i=1,2,3$ and adjoining them in triplet $UP_{o1}$ along the segments of space–time trajectory $\pm SP_i$ at changing the opposite information speeds on the trajectory between the segments from maximal $|\mp\alpha_{ito}|^{\max}$ to $|\pm\alpha_{it}|^{\min}, i=1,2,3$, where the bridge of unit $UP3$ connects the units with $UP_{o1}$. (Fig.4 shows symmetrical part of the conjugated dynamics).

Suppose segment $-SP_1\downarrow$ starts moving $-UP1\uparrow$ with maximal speed $|-\alpha_{1t}|$, segment $+SP_1\uparrow$ starts moving $+UP1\uparrow$ with $|+\alpha_{1t}|$, and the local circle rotates in right direction $\uparrow\odot\downarrow$.

The next segment $-SP_2\downarrow$ starts moving $-UP2\downarrow$ with maximal speed $|-\alpha_{2t}|<|-\alpha_{1t}|$ and $+SP_2\uparrow$ starts move $+UP2\uparrow$ with speed $|+\alpha_{2t}|<|+\alpha_{1t}|$. Each of these speed forms second local circle $\downarrow\odot\uparrow$ rotating in left direction with absolute speed $|\alpha_{1t}|$ less those in the previous circle $|\alpha_{1t}|$.

Third segment $-SP_3\downarrow$ starts moving $-UP3\downarrow$ with maximal speed $|-\alpha_{3t}|<|-\alpha_{2t}|$, and $+SP_3\uparrow$ starts move $+UP3\uparrow$ with maximal $|+\alpha_{3t}|<|+\alpha_{2t}|$.

Third local circle rotates in right direction $\uparrow\odot\downarrow$ with absolute speed satisfying relation

$$|\alpha_{3t}|<|-\alpha_{2t}|<|\alpha_{1t}|. \tag{4.14}$$

The opposite directional speeds within each circle attract each $-UP_i$ to $+UP_i$, minimizing the ending speeds down to $|\alpha_{it}|=\alpha_{it}^{\min}$ as these pairs approach, and can bind them in related units $UP_i$ through the starting attracting force (4.2) in sequence

$$\alpha_{1t}^{\min}\to\alpha_{2t}^{\min}\to\alpha_{3t}^{\min}. \tag{4.14a}$$

Current units $UP1, UP2$, made at the growing time-space intervals $t_{o1}<t_{o2}<t_{o3}$, at $t_{o3}\geq t_{o1}+t_{o2}$, are automatically integrated and initially memorized in $UP3$, then are condensed in $UP_{o1}$ while forming the triple knot. Since the IPF is sequentially summing $UP_i, i=1,2,3$ information and memorizing only current $UP_{o1}$, in the time-space course, the previous $UP_i$ are erased as their information integrates $UP_{o1}$. When the condensed information of $UP_{o1}$ integrates sequence $UP1\to UP2\to UP3$ in a primary triplet, it automatically implements the IPF with minimization of total time of building the triplet.

When speeds $+\alpha_{13t}$ and $-\alpha_{13t}$ of each opposite segments' connection move close to speeds $+\alpha_{3t}, -\alpha_{3t}$ by the end of interval $t_3$, according to (4.14, 4.14a), it makes possible connecting conjugated information

$+\alpha_{23t3}t_3$ with $-\alpha_{23t3}t_3$ , (4.14b)

This closes the process of joining each of two complementary units in a third time-space loop.

The triple knot's generated free information, forming the closing loop of the complimentary triplet's processes, which allows self-formation of such joint triplet in the following consecutive attraction.

Entropy of virtual impulse probes through the bridges between the segments is converting to information, which, closing the loop in the cycle, transforms information to information.

Such a triple self-supporting cyclic process indeed requires an initial flow of entropy converted to information.

**6a.** At forming $UP3$, new triplet Bit may appear if the $UP1$ ending speed, minimized by the opposite speed of $UP2$, equalizes with third segment minimal speed $\alpha_{3t}^{\min}$, and the $UP2$ ending speed, also, being minimized in the opposite movement, equalizes with third segment minimal speed $\alpha_{3t}^{\min}$ by the moment of forming $UP3$. Adjoin the two with the third allows forming $UP3$ *during* formation of $UP1$ and $UP2$ joining with a minimal information speed equals to that of the two segments ending speeds.

These speeds minimize the attracting information $\mathbf{a} = 1/3 bit \approx 0.23 Nat$ of each $UP1, UP2, UP3$ whose sum $3\mathbf{a} \cong \mathbf{a}_o$ can join all three in new triplet Bit with information $\mathbf{a}_o$.

Actually, attracting information $\mathbf{a}$ of $UP1$ decreases speed of its starting movement $\alpha_{13t} = \alpha_{1t} - \Delta\alpha_{1t} \to \alpha_{1t}^{\min}$ on such speed increment $\Delta\alpha_{1t}$ which can equalize with $\alpha_{3t}^{\min}$ for $UP3$.

Attracting information $\mathbf{a}$ of $UP2$ decreases speed of its starting movement $-\alpha_{23t} = -\alpha_{2t} - \Delta\alpha_{2t} \to -\alpha_{2t}^{\min}$ on such speed increment $\Delta\alpha_{2t}$ which allows equalize it also with $\alpha_{3t}^{\min}$, where the speed signs correspond the direction of rotation in each local circle. The attracting movement connects these speeds in the triple movement arranging

$+\alpha_{1t}^{\min} \Rightarrow -\alpha_{2t}^{\min} \Leftarrow +\alpha_{3t}^{\min}$ (4.15)

when ending speed $-\alpha_{2t}^{\min}$ emanated from $UP2$ joins equal minimal speeds $+\alpha_{1t}^{\min}, +\alpha_{3t}^{\min}$ forming the end of triple knot which is assembling the triplet unit $UP_{o1}$.

The binding information in the knot memorizes the accumulated information at moving speed

$\alpha_{1t}^{\min} = |\alpha_{2t}^{\min}| = \alpha_{3t}^{\min} = \alpha_{uo1}$ . (4.15a)

An example of forming a triplet space structure shows **Fig.5.** Let's identify the time formation of this space structure.

Assume forming $UP1$ Bit on segment $SP_1$ needs time interval $t_1$ and time interval $\Delta t_{13}$ to attract $UP3$.

Forming second $UP2$ Bit on segment $SP_2$ needs time interval $t_2$ and time interval $\Delta t_{23}$ to attract $UP3$. Forming $UP3$ Bit takes time interval $t_3$ on third segment $SP_3$.

Since all three segments move with decreasing information speeds, it is possible to reach equality

$t_{13} = t_3 = t_{23} + t_{12}$ (4.16)

where $t_{12}$ is time interval between $UP1$ and $UP2$.

Satisfaction of (4.16) allows adjoining all three segments during time interval $t_3$ of forming $UP3$.

Since intervals $t_1, t_2, t_3$ are connected by the same information invariant

$\mathbf{a} = 1/3 bit \approx 0.23 Nat$, it brings joint relation
$$\alpha_{1to}^{max} t_1 = \alpha_{2to}^{max} t_2 = \alpha_{3to}^{max} t_3 = \mathbf{a}_o \cong 3\mathbf{a}. \tag{4.17}$$
The simulated speeds dynamics determine information delivered at each of these intervals [11]:
$$\mathbf{a}^{13} = \alpha_{13t} t_{13} \cong 0.232, \quad \mathbf{a}^{23} = \alpha_{23t} t_{23} \cong 0.1797 \text{ and } \mathbf{a}^{33} = \alpha_{3t} t_3 \cong 0.268. \tag{4.17a}$$
At $t_{13} = t_3$ and $\alpha_{3to} t_3 \cong \mathbf{a}_o, \alpha_{13t} t_{13} \cong 1/3 \mathbf{a}_o \cong \mathbf{a}$, we get $\alpha_{3to} / \alpha_{13t} \cong 3$. (4.17b)

Then to satisfy (4.17) at $\mathbf{a}^{13} = \mathbf{a}^{33} = \mathbf{a}$, the information spent on attraction and assembling triplet unit $UP_{o1}$ should also be $\mathbf{a}$. Therefore, if information spent on attraction $UP3$ is $\alpha_{23t} t_{23}$, the difference:
$$\Delta \mathbf{a}^{23} = \mathbf{a} - \alpha_{23t} t_{23} \cong 0.23 - 0.1797 \cong 0.05 \tag{4.17c}$$
spends on assembling $UP_{o1}$ using the delivered information $2\mathbf{a} = \mathbf{a}^{13} + \mathbf{a}^{33}$.
Following that, relations
$$\Delta \mathbf{a}^{23} = |\alpha_{23t}| \delta t_{23}, |\alpha_{23t}| \cong 1/3 \alpha_{3ot}, \alpha_{3ot} = \mathbf{a}_o / t_3, \tag{4.18a}$$
determine the time interval on assembling $UP_{o1}$:
$$\delta t_{23} \cong 3 t_3 \Delta \mathbf{a}^{23} / \mathbf{a}_o, \tag{4.18b}$$
which evaluates
$$\delta t_{23} \cong 0.214 t_3. \tag{4.18c}$$

Assembling information $\mathbf{a}_o \cong 3\mathbf{a}$ in the knot with speed $-\alpha_{23t}$, determines $-UP_{o1}$ sign.

The assembling loop, connecting speeds (4.15a), builds the triplet $-UP_{o1}$ knot, which binds primary information of $UP1, UP2, UP3$, being spent on the attraction, in information of secondary forming triplet. Attracting free information from triplet $-UP_{o1}$ or $+UP_{o1}$ on forming secondary information unit $UP_{o1}$ minimizes time of building $UP_{o1}$ compared to $UP_i, i=1,2,3$ sequential connections in the macroprocess.

**6b**. The $n$-dimensional process trajectory locate multiple conjugated pairs $+\uparrow SP_{oi}$ and $-\downarrow SP_{oi}$, which could assembles each triple $UP_i, i=1,2,3$ in cooperative $\mp UP_{oi}$ Bit, where the sign of each unit depends on the sign each second segment $SP_{2i}$ which binds each of these units in the triple segment knot.

This attracting movement assembles their three Bits in new formed $\mp UP_{i0}$ Bit forming new loops on higher (second) level which connects the equal speeds of each triple (**Fig.6**).

Particularly, the attractive motion of rotating triples units $-1_o(+UP_{o1}, -UP_{o2}, +UP_{o3})$ can cooperate next form triplets unit $-UP_{4o}$ depending on $-SP_{2i}$. The sign of $+SP_{2i}$ at cooperating opposite triple $+2_o(-UP_{o5}, +UP_{o6}, -UP_{o7})$ forms triplets unit $+UP_{5o}$, other three rotating triple $-3_o(+UP_{o8}, -UP_{o9}, +UP_{o10})$ cooperates in unit $-UP_{6o}$. That closes each of three rotating circles.

Cooperative motion of units $+4_{1o}[-UP_{4o}, +UP_{5o}, -UP_{6o}]$ assembles them in new formed triplet $+UP_{I0}$ Bit composing third levels of triplets. If above units $-1_o, +2_o, -3_o$ have equal attracting speeds by the moment of cooperation, they may join simultaneously in composite unit $+4_{10}$ during building this triple.

Information of units $UP_{o1}, UP_{o2}$, currently formed at the growing time-space intervals $t_{io1} < t_{o2} < t_{io3}$, at $t_{io3} \geq t_{io1} + t_{io2}$, are automatically integrated and memorized in $UP_{oit}$ along with $UP_{o3}$ forming the triple knot $UP_{4ot} = UP_{oit}$. Sequential built triplet knots, memorizing only current $UP_{oit}$, while the previous units information have erased, automatically implement the IPF minimax, minimizing total time of building each composite information unit. The minimax leads to sequential decreasing ending information speed of each $UP_{oi}$, and therefore to decreasing starting information speed next cooperative unit.

The minimax requires the ordered connection of binding speeds at forming triplets which memorize the bound information units, sequentially structures the observer's information dynamics, building and connecting the $UP_{oi}$ and new cooperating triple units.

**6c.** Building the units' dynamic information structure from multiple $\pm SP_{ij}$ segments on $n$-dimensional process trajectory requires assembling each $UP_{ij}$ through two segments, taking from conjugated segments of the trajectory of one dimension, with other opposite directional segment on other conjugated pair, which belongs to other dimension. Building a cooperative form triplet $UP_{oij}$ involves three such complimentary parts from three different process' dimensions analogous to **Figs.3,6**.

In the minimax attracting movement of $\pm UP_{ij}$, three of the time–space segments spend only the time-space interval of a third segment, joining three segments simultaneously, while both the first and second segments attract third segment during their segments intervals accordingly. The opposite moving segments on the macrotrajectory cooperate the symmetrical-complimentary $\pm UP_{oij}$ triplets, enclosing half of each complementary Bits, which through self-joining enable creating complete Bit whose attracting information can assemble other $UP_{oij}$ triplets.

The minimax movement decreases ending information speed of each complementary units of $\pm UP_{oij}$, whose enfolding increases information density of each following Bit.

Therefore, each triple unit contains four Bits including the forth Bit, which enfolds the triplet information in a triple knot. The attracting space-time movement selects, orders and assembles the rotating segments cooperating speeds along the time-space dynamic trajectories.

Indeed. Each primary $\pm UP_{o1}$ from different conjugated pairs emanates three time–space segments $\pm SP_{o1}, \pm SP_{o2}, \mp SP_{o3}$ selected on the minimax trajectory. The symmetry of opposite moving segments on the macrotrajectory actually generates half of each complementary $\pm UP_{o1}$ unit's $\pm BitI4, I = 0,1,....m$ which could self-join, creating complete Bit $|BitI4|$ of $UP_{o1}$, whose attracting information can assemble other triplets units $\pm UP_{Io1}$ and so on.

The triplets' knots create *new class* of information Bits that distinct from the first class information Bits of the assembling units, which were generated via virtual probes with entropies of cutting random process.

Forming the closing loop in processing a complimentary triplet allows self-formation of such joint triplet, while the triple knot generates free information for a next consecutive attraction. The joint triplet with free information produces the cooperative assembling which brings new information in each triple knot.

The self-building continues during the time-space information dynamics, which self-assemble space-time information structure of *information network* (IN) **(Fig.7),** enclosing the growing number of triplets. Each triplet accumulates three Bit's information logic enfolded in the knot's information Bit. Such self-forming structure automatically implements the IPF integration of the process information in a last IN node.

**7.** The *triplet information dynamics* start with information speed of its first segment's eigenvalue

$$\alpha_{io} = c_{iv} \cong 2.4143 \times 0.596 \times 10^{15} \, Nat/\sec \cong 1.44 \times 10^{15} \, Nat/\sec \,, \qquad (4.19)$$

where $\alpha_{io} = \alpha_{1o}$ is potential information speed predicted by the moving entangled entropy volume.

Each triplet self-formation joins two trajectory segments with positive eigenvalues by reversing their unstable eigenvalues and attracting a third segment with negative eigenvalues, whose rotating trajectory moves it up to the two opposite rotating eigenvectors and cooperates all three information segments in a triplet's knot **(Figs 3,4,5).**

Prior to joining, the segment entropy-information, satisfying the minimax, should end with its minimum evaluated by (4.9b) and (4.10). The minimax optimal triplet requires minimal time interval (4.1) spent on equalization of each eigenvalue with the following segment eigenvalue. In forming triplet at the two steps' consolidation, the second segment cooperation may bind $2i_{vf} = 2(i_v + i_{vo}) = 0.4668 Nat$ of both segments attracting information, which equals to that for overcoming entropy threshold $\delta_e \cong 0.4452 Nat$ (4.9b) for third segment. Third segment encloses information of the two triplet segments including information that binds them. Its ending free information attracts next information segment of following triplet, which builds the attracting information units for the subsequent information dynamics.

**8**. Dynamic invariant $\mathbf{a}(\gamma)$ that connects the ending eigenvalues of triplet's segments determines ratios of starting information speeds $\gamma_1^\alpha = \alpha_{io}/\alpha_{i+1o}$ and $\gamma_2^\alpha = \alpha_{i+1o}/\alpha_{i+2o}$ needed to satisfy (4.15a):

$$\gamma_1^\alpha = \frac{\exp(\mathbf{a}(\gamma)\gamma_2^\alpha) - 0.5\exp(\mathbf{a}(\gamma))}{\exp(\mathbf{a}(\gamma)\gamma_2^\alpha/\gamma_1^\alpha) - 0.5\exp(\mathbf{a}(\gamma))}, \gamma_2^\alpha = 1 + \frac{\gamma_1^\alpha - 1}{\gamma_1^\alpha - 2\mathbf{a}(\gamma)(\gamma_1^\alpha - 1)} \quad (4.20)$$

where multiplication $\gamma_1^\alpha \times \gamma_2^\alpha = \gamma_{13}^\alpha$ holds the eigenvalue ratio $\gamma_{13}^\alpha = \alpha_{io}/\alpha_{i+3o}$.

Invariants $\mathbf{a}_i(\gamma) = \alpha_i^t t_i$ and $\mathbf{a}_{io}(\gamma)$, at $\lambda_i^t = \lambda_{io}^t \exp(\lambda_{io}^t t_i)[2 - \exp(\lambda_{io}^t t_i)]^{-1}$, connects by Eq.

$$\mathbf{a} = \mathbf{a}_o \exp(-\mathbf{a}_o)(1+\gamma^2)^{1/2}[4 - 4\exp(-\mathbf{a}_o)\cos\gamma\mathbf{a}_o + \exp(-2\mathbf{a}_o)]^{-1/2}] \quad (4.20a)$$

In dynamics of real eigenvalue: $\alpha_i^t = \alpha_{io}^t \exp(\alpha_{io}^t t_i)[2 - \exp(\alpha_{io}^t t_i)]^{-1}$, invariants $\mathbf{a}_i(\gamma_i), \mathbf{a}_{io}(\gamma_i)$ connect by Eq:

$$\mathbf{a}_i(\gamma_i) = \mathbf{a}_{io}(\gamma_i) \exp\mathbf{a}_{io}(\gamma_i)(2 - \exp\mathbf{a}_{io}(\gamma_i))^{-1} \quad (4.20b)$$

Optimal ratio $\gamma_{io} = 0.4142$ corresponds the minimax with $\mathbf{a}_{io}(\gamma_{io} = 0.4142) \cong 0.73$ and $\mathbf{a}_i(\gamma_{io}) \cong 0.23$.

Each triplet structure identifies firstly minimax invariant $\gamma_i, \mathbf{a}_{io}(\gamma_i)$ then both $\gamma_1^\alpha, \gamma_2^\alpha$.

At known starting eigenvalue $\alpha_{io} = c_{iv} \cong 2.4143 \times 0.596 \times 10^{15} Nat/\sec \cong 1.44 \times 10^{15} Nat/\sec$ and $\alpha_{io} = \alpha_{1o}$, next such speeds in the triplet are $\alpha_{1o}/\gamma_1^\alpha = \alpha_{2o}, \gamma_1^\alpha(\gamma_{io}) = 2.236$, at. $\alpha_{2o} \cong 0.644 \times 10^{15} Nat/\sec$.

Actual starting information speed of first eigenvalue $c_{ika} = \alpha_{1o}, c_{ika} \cong 0.0516 \times 10^{14} Nat/\sec$ leads to second $\alpha_{2o} \cong 0.02345 \times 10^{14} Nat/\sec = 0.2345 \times 10^{13} Nat/\sec$.

Known $\alpha_{2o}, \gamma_{32}^\alpha$ determines third starting eigenvalue's speed:

$\alpha_{3o} \cong \alpha_{2o}/\gamma_2^\alpha = 0.02345/1.6 \times 10^{14} Nat/\sec = 0.1465 \times 10^{13} Nat/\sec$.

Invariants $\gamma_1^\alpha, \gamma_2^\alpha$ determine each triplet's time intervals of information dynamics and related intervals of rotating space movement at

$\alpha_{io}t_{io} = \alpha_{i+1o}t_{i+1o} = \alpha_{i+2o}t_{i+2o} = \mathbf{a}_{io}(\gamma_i)$, $\gamma_1^\alpha = \alpha_{io}/\alpha_{i+1o} = t_{i+1o}/t_{io}$ and $\gamma_2^\alpha = \alpha_{i+1o}/\alpha_{i+2o} = t_{i+2o}/t_{i+1o}$, (4.21)

Particular observations leads to the specific ratios of the triplet's initial eigenvalues $\alpha_{1o}/\alpha_{2o} = \gamma_1^\alpha, \alpha_{2o}/\alpha_{3o} = \gamma_2^\alpha$, satisfying the invariant relations (4.19-4.21).

**9**. Interacting impulses with information measure $\mathbf{a}_{io} = \ln 2 Nat$ each have multiplicative information measure
$$U_m = (\mathbf{a}_{io})^2, \qquad (4.22)$$
which should bind the following double connection in a triplet.
Total information at
$$(\mathbf{a}_{io}(\gamma_{io}))^2 + \mathbf{a}_i(\gamma_{io}) \cong 0.7 \cong \mathbf{a}_{io}(\gamma_{io}) \qquad (4.22a)$$
approaches the delivered information from each impulse.
**10**. Forming a stable triplet, which enables the attraction, limits the maximal ratio
$$4.8 \geq \gamma_1^\alpha \geq 3.45 \;. \qquad (4.23)$$
That determines a boundary of the triplet scale factor $\gamma_1^\alpha$. Approaching $\gamma_1^\alpha = \gamma_2^\alpha \to 1$ leads to repeating the triplet's eigenvalues that limits the related theoretical admissible $|\gamma_i| \in (0.0 - 1.0)$).

Approaching information locality $\mathbf{a}_o(\gamma_i = 1 - o)$ of $\gamma_i = 1$ indicates both a jump of an event with that information to the event information $\mathbf{a}_o(\gamma_i)$ and the time ratio of the following and preceding intervals.

For example, when $\gamma_i$ approaches 1, $\mathbf{a}_o(\gamma_i)$ is changing from $\mathbf{a}_o(\gamma_i = 1 - o) = 0.56867$ to $\mathbf{a}_o(\gamma_i) = 0$, and the corresponding time's ratio of the following and preceding intervals reaches limit ratio $\tau_{i+1}/\tau_i = 1.8254$ with changing sign of the eigenvalues' ratio: $\alpha_{io}/\alpha_{it} \cong -1.9956964$. Eq (4.20) at $\gamma_i = 1$ leads to $\mathbf{a}_o(\gamma_i = 1) = 0$ and information contributions for regular control $\mathbf{a}(\gamma_i = 1) = 0$ and impulse control $\mathbf{a}_o^2(\gamma_i) = 0$.

The above jumps of time or related eigenvalues is a *dynamic indicator* of breaking up the dynamic constraint, which leads to cutting off the model's dynamics from the initial random process with a possibility of getting more uncertainty and rising chaotic diffusion dynamics. Thus, the appearance of an event, carrying $\gamma_i \to 1$, leads to a *chaos and decoupling* of the events chain, whereas the moment of this event's occurrence *predicts* measuring a current event's information $\mathbf{a}_o(\gamma_i) \neq 0$ and using it to compute $\gamma_i$ applying (4.5).

Maximal practically admissible $\gamma_{ia} \to 0.8$ leads to a *minimal stable* triplet with $\gamma_1^\alpha = \gamma_2^\alpha \to 1.65$ that limits acceptable $\gamma_i \to (0 - 0.8)$, which for $\gamma_{io} \to 0$ corresponds $\mathbf{a}_{io}(\gamma_{io}) = 1.1$ bit and $\mathbf{a}_i(\gamma_{io}) = 0.34$ bits.
The triplet dynamics with $\mathbf{a}_i(\gamma \to 0) \cong 0.23$ hold $\gamma_1^\alpha \cong 2.460, \gamma_2^\alpha \cong 1.817$ and $\gamma_{13}^\alpha \cong 4.6$.
Optimal $\gamma_{io} = 0.4142$ hold $\gamma_1^\alpha \cong 2.21, \gamma_2^\alpha \cong 1.76, \gamma_{13}^\alpha \cong 3.89$, and $\gamma_i = 0.8$ brings $\gamma_1^\alpha \cong 1.96, \gamma_2^\alpha \cong 1.68, \gamma_{13}^\alpha \cong 3.3$.

**10a.** Sequence of the model eigenvalues $(\ldots \alpha_{i-1,o}^t, \alpha_{io}^t, \alpha_{i+1,o}^t)$, satisfying the triplet's formation, is *limited* by the boundaries for $\gamma \in (0 \to 1)$, which at $\gamma \to 0$ forms a *geometrical progression* with
$$(\alpha_{i-o}^t)^2 = (\alpha_{io}^t)^2 + \alpha_{i-1,o}^t \alpha_{io}^t, \qquad (4.23a)$$
representing the geometric "gold section" at $\alpha_{io}^t \cong 0.618 \, \alpha_{i-1,o}^t$ and the ratio
$$G = \frac{\alpha_{i+1,o}^t}{\alpha_{io}^t} \cong 0.618; \qquad (4.23b)$$
and at $\gamma \to 1$ the sequence $\alpha_{io}^t, \alpha_{i+1,o}^t, \alpha_{i+2,o}^t, \ldots$ forms the Fibonacci series, where the ratio $\alpha_{i+1,o}^t / \alpha_{i+2,o}^t = \gamma_2^\alpha$ determines the "divine proportion" $PHI \cong 1.618$, satisfying
$$PHI \cong G + 1; \qquad (4.23c)$$

and the eigenvalues' sequence loses its ability to cooperate.
*Indeed*. At $\gamma_i \to 0$, solutions of (420,4.20ab) determine

$\mathbf{a}_o(\gamma_i \to 1) = 0.231$, $\gamma_1^\alpha \cong 2.46$, $\gamma_2^\alpha \cong 4.47$, $\gamma_2^\alpha / \gamma_1^\alpha = \gamma_{23}^\alpha \cong 1.82$.

Ratio $(\gamma_1^\alpha)^{-1} = \alpha_{io}^t / \alpha_{i-1,o}^t \cong (2.46)^{-1} \cong 0.618$ for the above eigenvalues forms a "golden section" (4.23b)

$G = (\gamma_1^\alpha)^{-1} = \alpha_{io}^t / \alpha_{i-1,o}^t \cong (2.46)^{-1} \cong 0.618$ *and* the "divine proportion" $PHI \cong 1.618$ (4.23c).

Above relations hold true for each primary pair of the triplets' eigenvalues sequence, while the third eigenvalue has ratio $\alpha_{i+1,o}^t / \alpha_{io}^t = (\gamma_{23}^\alpha)^{-1} \cong 0.549$. Solution of equation for invariants (4.5,4.20a) at $\mathbf{a}(\gamma_i \to 1) = 0$) brings $\gamma_1^\alpha = \gamma_2^\alpha = 1$ with $\alpha_{i-1,o}^t = \alpha_{io}^t = \alpha_{i+1,o}^t = \alpha_{i+2,o}^t = ,...$ The eigenvalues' sequence loses its ability to cooperate disintegrating into the equivalents and not connected (independent) eigenvalues.

**11.** *The restrictions on the space-time rotating trajectory with growing dimensions* $1,...,i,...,n$.

Each rotating movement presents $n$ three-dimensional parametrical equations of a helix curve located on a conic surface (**Figs.4,5**). A projection of the space trajectory' radius-vector $\bar{r}(\rho, \varphi, \psi^o)$ on the $i$-cone's base is spiral trajectory with radius

$$\rho_i = b_i \sin(\varphi_i \sin \psi_i^o). \qquad (4.24)$$

At angle $\varphi = \pi k / 2$, $k = 1, 2,...$, the trajectory transfers from one cone to another cone trajectory in the cone points $l_i$ which satisfy the extreme condition for the equation (4.24) at the angles

$$\varphi_i(l_i) \sin \psi_i^o = \pi / 2, \qquad (4.24a)$$

where $\psi_i^o$ is angle on the cone vertex, and the base radious is

$$\rho_i = b_i \sin(\pi / 2) = b_i. \qquad (4.24b)$$

The angle at the cone vertex takes the values

$$\sin \psi_i^o = (2k)^{-1}, k=1, 2,..; \text{ at } k=1, \psi_i^o = \pi/6; \qquad (4.25)$$

with the angle on the cone space points $l_i$

$$\varphi_i(l_i(\tau_i)) = k\pi. \qquad (4.25a)$$

The minimax imposes optimal condition on these angles:

$$\varphi_i = \pm 6\pi, \psi_i^o = \pm 0.08343. \qquad (4.25b)$$

Projection of moving vector $l(\bar{r}) = l(\rho, \varphi, \psi^o)$ on the cone base satisfies Eq

$$dl = [(\frac{d\rho}{d\varphi})^2 \sin^{-2} \psi^o + \rho^2]^{1/2} d\varphi. \qquad (4.26)$$

Spiral space angle $\psi$, depends on angle $\psi_i^o$ (4.25) according to Eq

$$tg\psi = \frac{(1 - \sin \psi^o \cos \psi^o + \sin^2 \psi^o)}{(1 \pm \sin \psi^o \cos \psi^o + \sin^2 \psi^o)} \qquad (4.27)$$

which for $\psi_i^o = \pi/6$ brings $\psi = 0.70311$. For the spirals having equal direction at small angle $\psi^o$ satisfies

$$\psi = \pi / 4 - \psi^o. \qquad (4.27a)$$

For the spirals with the opposite directions: $\psi^1 = \pi / 4$.

A relative increment of information volumes $\Delta V_{m,m+1}$ (**Fig.8**) between the volumes of two sequential triplets' $m$ and $(m+1)$: $V_m$, $V_{m+1}$: $\Delta V_{m,m+1}^* = (V_{m+1} / V_m - 1)$ depends on these triplets scale factor $\gamma_{m,m+1}^\alpha$:

$$\Delta V_{m,m+1}^* = (\gamma_{m,m+1}^\alpha)^3 - 1 \qquad (4.28)$$

The triplet absolute volume determines constant
$$V_c = 2\pi c^3 / 3(k\pi)^2 tg\psi^o \qquad (4.29)$$
which depends on angle $\psi_i^o$, initial space speed $c_{io}$:
$$c_{lo} \approx 14.4 \times 10^{-15} m / 4.477 \times 10^{-14} \sec = 3.216 \times 10^{-1} m/\sec, \qquad (4.30)$$
and parameter $k = 1, 2, ... (4.25)$ of consolidation of the volume number in $V_{m+1}$, starting with constant $k = 1$.

Rotation velocity of attraction $\omega_i[l_i(t_{ika})] = 0.1646 \times 10^{-14}$ $radian/\sec$ during $t_{ika}$ (4.12b) (where $\omega_i$ relates to (2.12), (3.3c)), determines space angle $\psi' = \pi/4$ at this moment, which the information dynamics initiate.

The IMD at the moment $t_{ika}$ brings Eqs (4.24-4.30) and above rotating angles, which determine space interval $l_i(\tau_i)$ corresponding to each triplet parameter $\gamma_1^\alpha, \gamma_2^\alpha$, and the rotating volumes.

The information speed of the joint three eigenvalues form first triplet whose vertex delivers information to next triple units that join in next triplet in the rotating movement generating an observer time course and space intervals. Transfer from one cone's trajectory to another one locates on the cone's base, where the location satisfies condition of a local extreme of entropy –information.

The sequential transfer requires to rotate each spiral on a space angle up to adjoin a next optimal trajectory and relocate it in cooperation (**Fig.6**). The rotating on the cones space-time trajectories, cooperating in the triplet, determine its geometrical structure (**Figs.5, 7**) evolving during each triplet formation. Both IMD and its space structure evolve concurrently, producing each other.

The Egs of rotating space trajectory on the cones and the space volume determine observer geometry (**Fig.9**), generated by information dynamics (through rotating velocity and knots of cooperating volumes, transferred to next triplet by the IN scale parameter).

## V. The conditions for building the IN and restriction on its parameters
### *The IN structure and limitations*
**1.** The IN triplet' structure of increasing the enclosed information, satisfying the minimax, requires an observer to bring information of growing density. This depends on the information necessary for both creating new IN's triplet and attaching this triplet to existing IN.

The IN's information density is determined through its current triplet eigenvalues, depending of the IN's scale parameter $\gamma_{13}^\alpha = \alpha_{1o}/\alpha_{3o}$ (as the ratio of the IN's starting information speed $\alpha_{1o}$ to its third segment's information speed $\alpha_{3o}$).

Density of information in the IN's $i$-th node, related to nearest IN's $k$ node, measures scale parameters
$$\gamma_{ik}^\alpha = \alpha_{io}/\alpha_{ko} = \alpha_{i\tau}/\alpha_{k\tau}, \alpha_{i\tau} = 1/3\alpha_{io}, \alpha_{k\tau} = 1/3\alpha_{ko} \qquad (5.1)$$
defined via a ratio of these nodes starting speeds $\alpha_{io}$, or ending information speeds on the knots satisfying (4.18a,b) at $\alpha_{i\tau} = \alpha_{i23t}, \alpha_{k\tau} = \alpha_{k23t}$.

Each of them depends on number of the enclosed triplets:
$$\alpha_{i\tau} = \alpha_{1\tau}/(\gamma_{13}^\alpha)^i, \alpha_{1\tau} = \alpha_{123t}, \alpha_{123t} = \alpha_{1o}/\gamma_2^\alpha, i = 1, ..., k, ..., m, \qquad (5.2)$$
where $m$ is total triplet' number enclosed into that IN node. This scale parameter depends on the eigenvalues current parameter $\gamma_i$ which determines invariant $\mathbf{a}(\gamma_i)$ or $\mathbf{a}_{io}(\gamma_i)$. The eigenvalues parameter $\gamma_{ik}$ identifies observing $\gamma_{ik}^\alpha(\gamma_{ik}) = \alpha_{io}/\alpha_{ko}$, and limited $\gamma_{iko}^\alpha(\gamma_{iko})$ restrains admissible frequency spectrum

$$f_{ik} = (\gamma_{iko}^{\alpha})^{-1} . \quad (5.3)$$

**2.** *Acquisition of the IN current information which delivers the node' interaction with observable information spectrum.*

Assume the IN current $i$ node's speed $\alpha_{i\tau}$ requests the current information frequency which satisfies attracting $k$ node with speed $\alpha_{k\tau}$ and spending on the request information of attraction $\mathbf{a}(\gamma_i)$.

In the considered model, incoming information $\mathbf{a}_\tau(t-s)$ is delivered during time interval $\Delta_t = t-s$ by sequence of impulse control functions, which on each quantum time interval $t = \tau_k$, $k = 0,1,2,....,m$ brings invariant information $\mathbf{a}_o(\gamma_i)$ as a part of $\mathbf{a}_\tau(t-s)$, where $\gamma_i$ is a priory unknown.

An interactive impact of requested information $\mathbf{a}(\gamma_i)$ on delivered $\mathbf{a}_\tau(t-s)$ evaluates Riemann-Stieltjes integral $I_s = \int_{-\infty}^{\infty} f(t-s) dg(s)$ applied to an information function $f(t-s) \to \mathbf{a}_\tau(t-s)$ [12] in the form:

$$I_s = \int_{-\infty}^{\infty} \mathbf{a}_\tau(t-s)\mathbf{a}(\gamma_i)\delta(s)ds = \mathbf{a}_\tau(t)\mathbf{a}(\gamma_i). \quad (5.4)$$

Solution of (5.4) is found for step-function $dg(s) \to \mathbf{a}(\gamma)du(s)$, which holds derivation $du(s)$ forming impulse function $du(s) = \delta(s)ds$. The step control's attracting information $\mathbf{a}(\gamma_i)$ initiates the model's dynamic process of acquiring delivered information $\mathbf{a}_\tau(t)$.

The interactive impulse impact of the requested node information, which needs external information (as cutoff memorized entropy of a data), measures information $\mathbf{a}_o^2(\gamma_i)$:

$$\mathbf{a}(\gamma_i)\mathbf{a}_\tau(t_{k+1}) = \mathbf{a}_o^2(\gamma_i) \quad (5.4a)$$

which *binds* information $I_{ik} = \mathbf{a}(\gamma_i)\mathbf{a}_\tau(t_{k+1})$ according to (4.22).

The binding impulse memorizes the cutting entropy of information microprocess of external spectrum actually prompting memorizing. The impulse may also directly bind the incoming information, as well as information from other observer's INs.

Applying (5.4a) allows finding delivering information $\mathbf{a}_\tau(t_{k+1})$ requested by $\mathbf{a}(\gamma_i)$ with the IN known $\gamma_i$. For example, at $\gamma_i = 0.3, \mathbf{a}_o(\gamma_i) = 0.743688, \mathbf{a}(\gamma_i) = 0.239661, \mathbf{a}_o^2(\gamma_i) = 0.553$ we get $\mathbf{a}_\tau(t_{k+1}) \cong 2.3$ which for number of impulses $k = 3$ during $\Delta_t \cong 3\delta_{ek}^t$ ($\delta_{ek}^t$ is time duration of each $k$) brings $\mathbf{a}_{ok}(\gamma_k) = 0.76924$, which according to (4.5a), determines new $\gamma_k \cong 0.05$. That allows finding the new IN $\gamma_{ik}^{\alpha} \cong 4.81$, which identifies the requested frequency spectrum $f_{ik}$, defined by ratio of information speeds (5.3) $\gamma_{ik}^{\alpha} = \alpha_{io}/\alpha_{ko} = (f_{ik})^{-1}$. This ratio increases compared to that at $\gamma_i = 0.3$. Thus, known $\alpha_{io}$ determines $\alpha_{ko}$ and time interval $t_{k+1} = \tau_{k+1}^1 - (\tau_{k+1}^o + o_{ko})$ starting at the moment $\tau_{k+1}^o + o_{ko}$ and ending at the moment $\tau_{k+1}^1$ of turning off the $k$ control. The interval of acquisition of the delivered information continues during $k = 3$ step-wise controls, which build new IN triplet during the same time.

For this example, at $\gamma_k \cong 0.05$, Eq (4.5a) brings the following results:

$$\delta_{ek}^t = \mathbf{a}_k(\gamma_k)/\alpha_{ko}, \alpha_{ko} = \gamma_{ik}^{\alpha}\alpha_{io}, \gamma_{ik}^{\alpha} \cong 4.81, \mathbf{a}_k(\gamma_k) \cong 0.2564, \alpha_{io} \cong |-2.57|, \delta_{ek}^t \cong 0.1\sec, \Delta_t \cong 0.3\sec \quad (5.5)$$

and spectrum ratio $f_{ik} \cong 0.2$.

The IN control, delivering new information that needs particular node, runs the IN feedback.
Each IN new information acquisition runs also interaction with other IN accessible spectrum within observer, which enables self-creation new information internally.

**3.** Forming an information dynamic cooperative requires rising cooperative information force between the potential cooperating segments of information macrodynamics. Let's find it using gradient

$$X_{ki}^{\alpha} = -\frac{\delta I_{ki}^{\alpha}}{\delta l_{ki}^{\alpha}}, \qquad (5.6)$$

where $\delta l_{ki}^{\alpha}$ is a difference in positions of the segments $k, i$ at $\partial l_{ki}^{\alpha} = \delta t_{ki}^{\alpha} c_{ok}$, $\delta t_{ki}^{\alpha}$ is the related time shift, and $c_{ok}$ is segments' space speed.

An increment of information between two segments $k, i$ at current moment $t_k : \delta I_{ki}(t_k)$ on finite intervals between segments $\partial l_{ki}^{\alpha} = l_k^{\alpha} - l_i^{\alpha} = \delta t_{ki}^{\alpha} c_{ok} = (t_k^{\alpha} - t_i^{\alpha}) c_{ok}$ and a fixed speed $c_{ok}$ depends on the difference of contribution from segments $k : \alpha_{\tau k}(t_k^{\alpha}) t_k^{\alpha}$ and the contribution from segments $i$, collected by moments $t_k$ during time interval $t_{ki} = t_k^{\alpha} - t_i^{\alpha}$ between these segments: $\alpha_{\tau i}(t_i^{\alpha})(t_k^{\alpha} - t_i^{\alpha})$:

$$\delta I_{ki}(t_k) = \alpha_{\tau k}(t_k^{\alpha}) t_k^{\alpha} - \alpha_{\tau i}(t_i^{\alpha})(t_k^{\alpha} - t_i^{\alpha}). \qquad (5.7)$$

The increment $\delta I_{li}(t_{li})$ at $t_{li} = t_l^{\alpha} - t_i^{\alpha}$ by analogy is

$$\delta I_{li}(t_{li}) = \alpha_{\tau l}(t_l^{\alpha}) t_l^{\alpha} - \alpha_{\tau i}(t_i^{\alpha})(t_l^{\alpha} - t_i^{\alpha}).$$

Both increment determine the finite information cooperative forces between these segments

$$\delta X_{ki}^{\alpha} = X_{ki}^{\alpha} \delta l_{ki}^{\alpha} = -\delta I_{ki}(t_k), \delta X_{li}^{\alpha} = X_{li}^{\alpha} \delta l_{li}^{\alpha} = -\delta I_{li}(t_{ki}).$$

The increments relative to first segments are:

$$\delta I_{ki}(t_{ki})/t_i^{\alpha} = \alpha_{\tau k}(t_k^{\alpha}) t_k^{\alpha}/t_i^{\alpha} - \alpha_{\tau i}(t_i^{\alpha})(t_k^{\alpha}/t_i^{\alpha} - 1), \delta I_{li}(t_{li})/t_i^{\alpha} = \alpha_{\tau l}(t_l^{\alpha}) t_l^{\alpha}/t_i^{\alpha} - \alpha_{\tau i}(t_i^{\alpha})(t_l^{\alpha}/t_i^{\alpha} - 1)$$

and their ratio acquires the form

$$\delta I_{kli}^{*} = \delta I_{ki}(t_{ki})/\delta I_{li}(t_{li}) = \frac{\alpha_{\tau k}(t_k^{\alpha}) \gamma_{ki}^{\alpha} - \alpha_{\tau i}(t_i^{\alpha})(\gamma_{ki}^{\alpha} - 1)}{\alpha_{\tau l}(t_l^{\alpha}) \gamma_{li}^{\alpha} - \alpha_{\tau i}(t_i^{\alpha})(\gamma_{li}^{\alpha} - 1)} = \frac{2 - \gamma_{ki}^{\alpha}}{2 - \gamma_{li}^{\alpha}}. \qquad (5.7a)$$

For a triplet with $\gamma_{ki}^{\alpha}(\gamma \to 0) = 2.46, \gamma_{li}^{\alpha}(\gamma \to 0) = 4.794$, realive increments is $\delta I_{kli}^{*} = 0.1645$; for others:

$\gamma_{ki}^{\alpha}(\gamma = 0.5) = 2.215, \gamma_{li}^{\alpha}(\gamma = 0.5) = 3.895, \delta I_{kli}^{*} = 0.113$;

$\gamma_{ki}^{\alpha}(\gamma = 0.8) = 1.969, \gamma_{li}^{\alpha}(\gamma = 0.8) = 3.3, \delta I_{kli}^{*} = -0.0235$.

It's seen that cooperative increment between first pair of segments is less than that for the second pair, depending of the limited segment's ratio of admissible $\gamma$; it decreases with growing $\gamma$, while the last triplet fails to cooperate.

These increments evaluate information cooperative forces at the time interval starting the triple cooperation. By the end of cooperation, each increment, at $t_{ki} \to 0, t_{li} \to 0$, acquires invariant measure $\alpha_{\tau k}(t_k^{\alpha}) t_k^{\alpha} = \alpha_{\tau l}(t_l^{\alpha}) t_l^{\alpha} = \mathbf{a}_{\tau}$ and ratio $\delta I_{kli}^{*} \to 1$.

Information potential of current IN triplet $m_i$, currying information $\delta I_{ik}^{m} = \mathbf{a}_i + \mathbf{a}_{oi}^2 \cong \mathbf{a}_{oi}$, intends to attract $m_k$ triplet, depends on both $\mathbf{a}_{oi}(\gamma)$ and on relative distance

$$(t_i^{m} - t_k^{m})/t_i^{m} = (l_i^{m} - l_k^{m})/l_i^{m} = \partial l_{ik}^{m*} : X_{ik}^{lm} = -\frac{\delta I_{ik}^{m}}{\delta l_{ik}^{\alpha}} = \mathbf{a}_{oi}(\gamma)(\gamma_{ik}^{m} - 1), \qquad (5.8)$$

which measures this potential directly in Nats (bits). The information potential, relative to information of first IN triplet, determines coperative force between these triplets:

$$X_{1k}^{Im1} = (\gamma_{1k}^m - 1). \tag{5.8a}$$

The required relative cooperative information force of the first and second triplets:

$$X_{12}^\alpha \geq [\gamma_{12}^\alpha - 1], \tag{5.8b}$$

at limited values $\gamma_{12}^\alpha \to (4.48 - 3.45)$, restricts the related cooperative forces by inequality

$$X_{12}^\alpha \geq (3.48 - 2.45). \tag{5.9}$$

The quantity of information, needed to provide this information force, is

$$I_{12}(X_{12}^\alpha) = X_{12}^\alpha \mathbf{a}_o(\gamma_{12}^\alpha), \tag{5.9a}$$

where $\mathbf{a}_o(\gamma_{12}^\alpha)$ is invariant, evaluating quantity of information concentrated in a selected triplet by $\mathbf{a}_o(\gamma_{12}^\alpha) \cong 1bit$ at $\gamma_{12}^\alpha = \gamma_{1o}^\alpha$, From that follows

$$I_{12}(X_{12}^\alpha) \geq (3.48 - 2.45) bits. \tag{5.9b}$$

The invariant's quantities $\mathbf{a}_o(\gamma_{io} \to 0)$, $\mathbf{a}(\gamma_{io} \to 0)$ provide *maximal* cooperative force $X_{12}^{am} \cong 3.48$. Minimal quantity of information, needed to form a very first triplet, estimates dynamic invariants

$$I_{o1} \cong 0.75 Nats \cong 1 bits, \tag{5.9c}$$

Therefore, total information, needed to start adjoining next triplet to the IN, estimates

$$I_{o12} = (I_{o1} + I_{12}(X_{12}^\alpha)) \geq (4.48 - 3.45) bit. \tag{5.10}$$

which supports the node cooperation [32] and initiates the IN feedback above. This information equals or exceeds information of the IN current node needed for sequential cooperation the next triplet. Minimal triplets' node force $X_{1m}^\alpha = 2.45$ depends on the ratio of starting information speeds of the nearest nodes, which determines the force scale factor $\gamma_{m1}^\alpha = 3.45$ satisfying the minimax.

The observer, satisfying both minimal information $I_{o1}$ and admissible $I_{12}(X_{12}^\alpha)$ delivering total information (5.10), we call a *minimal* selective observer, which includes the control's carried free information (Sec.3) for the triplet's node. These limitations are the observer boundaries of *admissible* information spectrum, which also apply to multi-dimensional selective observer.

**4.** At satisfaction of cooperative condition (5.10), each following observed information speed, enables creating next IN's level of triplet's hierarchy, delivers the required density of the information spectrum. This leads to two conditions: *necessary*-for creating a triplet with required information density, and *sufficient* -for a cooperative force, needed to adjoin this triplet with an observer's IN. Both these conditions should be satisfied through the observer's ability to *select* information of growing density. At satisfaction of necessary condition, each next information units joins a sequence of triplet's information structures forming the IN, which progressively increases information bound in each following triplet- at satisfaction of sufficient condition, and the IN ending triplet's node conserves all IN information.

The node *location* in the IN spatial-temporal hierarchy determines *quality* of the information bound in IN node, which depends on the node enclosed information density.

**5.** *Extending the IN requires quantity and quality of information (5.8-5.10), which could deliver an external observer, satisfying the requested information emanating from the current IN ending node.*

Let us evaluate the interactive information impact of an external observer, carrying own information, on the IN requested information to form potential new triplet of the current observer.

Such request carries the control information $\mathbf{a}_m \sim 0.25 Nat$ with information speed $\alpha_m^t$ determined by the requested IN node, which encloses information density

$$\gamma_{m+1}^\alpha = (\gamma_{12}^\alpha)^{m+1} = (\gamma_{m=1}^\alpha)(\gamma_{12}^\alpha)^m \times \alpha_{1o}, \tag{5.11}$$

where $\alpha_{1o}$ is information speed on a segment of the IN initial triplet, whose ratio to its third segment speed is $\gamma_{o13}^\alpha = \alpha_{1o}/\alpha_{3o} \cong 3.45 = \gamma_{m=1}^\alpha$, at ratio $\gamma_{12}^\alpha = (\gamma_{m=1}^\alpha)$ which supposedly equals for all $m+1$ triplets; and $\gamma_{m+1}^\alpha$ is scale factor for $m+1$-the triplets of requesting IN's node. The requested density $\gamma_{m+1}^\alpha$ requires relative information frequency

$$f_{1m+1} = (\gamma_{m+1}^\alpha)^{-1}. \tag{5.11a}$$

Minimal IN with single triplet node and potential speed of attraction (4.13a) requests its information density with speed

$$c_{m+1}^\alpha = (3.45)^2 \times 0.1444 \times 10^{14} \approx 11.9 \times 0.1444 \times 10^{14} \approx 1.7187 \times 10^{14} \, Nat/\sec. \tag{5.12}$$

To adjoin the requested information in the IN, the requesting control should carry information $I_{o12}$ (5.9c) with speed (5.12) which requires time interval for transporting this information:

$$t_m \cong 3.45 \times 0.7 \, Nat / 3.564 \times 10^{14} \, Nat/\sec = 0.6776 \times 10^{-14} \sec. \tag{5.12a}$$

This relates to interval of time communication within the observer inner processes, which is less than the impulse wide's time $\delta_{te} \approx 1.6 \times 10^{-14}$ sec that carries the requested information binding $I_{ik}$ (5.4).

The decrease ratio $\delta_{te}/t_m \cong 2.36$ corresponds to increase the initial speed of attraction (4.13a) $0.1444 \times 10^{14} \, Nat/\sec$ in $\sim 12$ times. The increased impulse information density decreases the impulse time wide $\delta_{te}$ to

$$\delta_{tm} = \delta_{te}/12 \cong 0.1333 \times 10^{-14} \sec, \tag{5.12b}$$

which is less than the communication time (5.12a) in $\sim 5$ times.

Impulse with time wide $\delta_{tm}$ (5.12b) transfers the requested information to Observer's external window where it interacts with observing external process via the probing impulses. This information should deliver the step-down cut of external impulse that requires quantity information $0.25 Nat$ -the same as the information which carries the requested control. Communication time (5.12b) to get the needed actual frequency-density (5.11a) from the window observing external process also determines the frequency of the probing impulses.

That requires to increase the initial impulse's attracting information density

$$i_{od} = 0.25 Nat / \delta_{te} = 0.25 Nat / 1.6 \times 10^{-14} \sec = 0.15 \times 10^{14} \, Nat/\sec \tag{5.13}$$

in $\sim 12$ times up to

$$i_{md} \cong 1.8 \times 10^{14} \, Nat/\sec. \tag{5.13a}$$

The observer time' increase follows from preserving information invariant $\mathbf{a}_{om} = \alpha_{1o}\tau_{1o} = \alpha_{m+1}\tau_{m+1}$ along the IN nodes at

$$(\gamma_{12}^\alpha)^{m+1} = \alpha_{1o}/\alpha_{m+1} = M_\tau = \tau_{m+1}/\tau_{1o}, \tag{5.13b}$$

where for $\tau_{1o} = t_{io}, \tau_{m+1} = t_{ko}$ in (5.5), the above densities and the time scale (5.13b) evauates $M_\tau = t_{md}/t_{od} = i_{md}/i_{od} = 11.9 \approx 12$.

The increase of information density corresponds increasing quality of information to be enfolded in the current IN.

Thus, each Observer *owns the time of inner communication,* depending on the requested information, and *time scale,* depending on *density* of accumulated (bound) information.

If new node formation requires $k$ cutting information units, time interval of such cuts $\delta_{eik}^t$ will depend of the IN time scale, increasing proportionally: $\delta_{eik}^t = \gamma_{m=k}^\alpha \delta_{eio}^t, \gamma_{m=k}^\alpha = (\gamma_{12}^\alpha)^k$, which, for $k=10$ increases the $\delta_{eio}^t \cong 0.2 \times 10^{-15}$ sec in $\cong 2.57 \times 10^{17}$ times up to $\delta_{eik}^t \cong 51.4$ sec. That allows memorize a "movie" of moving $k$ information units during the gap, including time-space dynamic of the entangling units. Thus, free information, carrying attracting information force (5.9) with quantity of the force information $I_{o12}$ (5.10) delivers related quality to the forming IN by cutting external information with density (5.13). The cutting information enables forming new IN triplet with $\mathbf{a}_o(\gamma_{12}^\alpha) \cong 1bit \cong \ln 2 Nat$, which should be attached to the current IN. The impulse, carrying this triplet, interacts with existing IN node by impact, which provides information $0.25 Nat$ - the same as at the interaction with an observer external process. The relative information effect of impact estimates ratio $\ln 2 / 0.25 \approx 3$. Taking into account the attracting information $0.231 Nat$ carrying with the triplet Bit, total increase brings $(3\ln 2 + 0.231) Nat \cong 3.573 bit$ which compensates for $I_{12}(X_{12}^\alpha) \geq (3.48 - 2.45) bits$ requested the current IN triplet node. This is minimal threshold for building elementary triplet (5.9b), which enables attract and deliver the requested information to IN node (5.10) that the selective observer can select with the needed frequency of the probing impulses. The triplets are elementary selective objective observers, which unable getting the requested IN level quality of information that ultimately evaluates the IN distinctive cooperative function.

A subjective observer, in addition to the objective observer, *selects* the observing process to acquire needed information according to its *optimal criterion* for *growing the quality*.

The *necessary* and *sufficient conditions* for adjoining information units in the observer's IN sequentially increase quality of the enfolded information.

The identified information threshold separates subjective and objective observers.

Subjective observer enfolds the concurrent information in a temporary build IN's high level logic that requests new information for the running observer's IN and then attaches it to the running IN, currently building its hierarchy.

**6.** *The limitations on the IN's cooperative dynamics and its parameters.*

**6a**. Minimal eigenvalue ratio $\gamma_i = \gamma^*$ limiting the IN dimensions.

Let us find such $\gamma_i = \gamma^*$ - as minimal increment $\gamma^* - \gamma_o = \Delta \gamma_o$ from its minimum $\gamma_o = 0$, whose dynamics with information invariants $\mathbf{a}_{io}(\gamma^*), \mathbf{a}_i(\gamma^*)$ can build a single cooperation-doublet.

Segment information dynamics with $\gamma_o$ transfers for building cooperation with a next segment with $\gamma^*$ the amounts of information consequently

$$\mathbf{a}_{io}(\gamma_o)^2 + \mathbf{a}_i(\gamma_o) = \mathbf{a}_{i^o}(\gamma_o) \quad (5.14), \qquad \mathbf{a}_{io}(\gamma^*)^2 + \mathbf{a}_i(\gamma^*) = \mathbf{a}_{i*}(\gamma^*). \qquad (5.14a)$$

Since each transferred $\mathbf{a}_{i^o}(\gamma_o)$ $\mathbf{a}_{i*}(\gamma^*)$ is deducted from total segment's information $\mathbf{a}_{io}(\gamma_o)$ or $\mathbf{a}_{io}(\gamma^*)$ accordingly, the relative information values

$$\mathbf{a}_{i^o}(\gamma_o) / [\mathbf{a}_{io}(\gamma_o) - \mathbf{a}_{i^o}(\gamma_o)] = m_o(\gamma_o) \qquad (5.15)$$

and $\mathbf{a}_{i*}(\gamma^*) / (\mathbf{a}_{io}(\gamma^*) - \mathbf{a}_{i*}(\gamma^*)) = m(\gamma^*)$ (5.15a)

will be spent on binding cooperative model's dimension $m_o(\gamma_o)$, starting with single information unit at $\gamma_o = 0$, or model's dimension $m(\gamma^*)$, starting with single unit at $\gamma^*$. Each binding reduces the process initial dimension, while the IN dimensions build different starting $\gamma$ and related $\gamma_m^\alpha(\gamma)$.

Finding $\gamma^*$ whose $m(\gamma^*)$ decreases initial dimension $m_o(\gamma_o)$ on 1 requires satisfaction

$$m_o(\gamma_o) - 1 = m(\gamma = \gamma^*). \qquad (5.16)$$

Applying formulas (4.5a), (4.20a) at $\gamma = \gamma_o$ brings $\mathbf{a}_{io}(\gamma_o) = 0.76805$, $\mathbf{a}_i(\gamma_o) = 0.231960953$, which according to (5.15) are able binding initial dimension $m_o = 15.2729035$.

Using (5.16) at unknown $\gamma = \gamma^*$ in form

$$15.2729035 - 1 = m(\gamma^*) = \mathbf{a}_{i*}(\gamma^*) / (\mathbf{a}_{io}(\gamma^*) - \mathbf{a}_{i*}(\gamma^*)) \qquad (5.16b)$$

and applying (4.5a), (4.20a) determine

$\mathbf{a}_{io}(\gamma^*) = 0.762443796$ and $\mathbf{a}_i(\gamma^*) = 0.238566887$, satisfying $\gamma^* = 0.007148$. (5.16c)

From that follows a minimal decreased dimension $m(\gamma^*) = 14.2729035$.

Inequality $\gamma^* > \gamma_o = 0$ imposes *upper limit* on the information invariants in (5.16c). Ratio

$$h_\alpha^o = (\mathbf{a}_{io}(\gamma_o) - \mathbf{a}_{io}(\gamma^*)) / \mathbf{a}_{io}(\gamma_o) = 0.00729922 = 1/137 \qquad (5.17)$$

evaluates minimal elementary uncertainty separating the model nearest dimensions, needed to overcome for binding a single double cooperation at minimal $\gamma^*$.

Number $m_o = 15.2729035 \approx 15$ limits potential maximum of the binding doublets, while $m(\gamma^*) \approx 14$ evaluates their real number after overcoming the barrier of uncertainty (5.17). If each IN is limited by dimension $m(\gamma^*)$, then possible cooperation of such multiple INs is also limited by $n^{max} \cong 14^{14}$ and confirms information (2.18c). Relations $\delta \mathbf{a}_{io}(\gamma^*) = [\mathbf{a}_{io}(\gamma_o) - \mathbf{a}_{io}(\gamma^*)]$ estimates a minimal elementary equivalent entropy-uncertainty separating the model's dimensions and imposes the limit on both.

Ratio of maximal $\gamma = 1$ to minimal admissible $\gamma^* = 0.007148$: $N_o = (\gamma = 1)/\gamma^* \cong 140$ determines maximal number of potential double cooperations up to their destruction at $\gamma = 1$.

Number $N_{o1} = (\gamma = 0.8)/\gamma^* \cong 112$ determines number of the double cooperation on admissible interval of $(\gamma = 0.007148 \to 0.8)$ with different doublet sequence and dissimilar $\gamma$ within this interval.

With a decay of the cooperative model (at $\gamma \to 1$), its current dimension $n_i$ is changed. Potential minimal growth the IN dimension from $n_i$ corresponds to a possibility of adding two cooperating eigenvalues (doublet) to the last triplet's eigenvalue of the $n_i$ dimensional system, forming a new triplet'a system $n_{i+1}$ dimension, and so on. Thus, we get the sequence of feasible triplet systems' dimensions $n_i, n_{i+1}, n_{i+2}$, which satisfy a simple relation $n_{i+1} = n_i + 2$. The current information measures of the starting eigenvalues' sequence, together with the values of $\alpha_{1o}^t$, $\gamma_i^\alpha$, $\gamma$, and $\mathbf{a}_{io}(\gamma)$, $\mathbf{a}_i(\gamma)$ allow us to calculate the IN dimension, restore the complete cooperative dynamics with the IN's space-time hierarchical information structure (**Fig.7**). The space-time node's position, following from the IN's cooperative capability, supports the *self-forming* of observer's information structure (**Fig.3, 5**), whose self-information has the distinctive *quality measure. Thus, information emanated from different IN nodes encloses distinct qualities and the logic remembered in knots.*

*7. Options of creating the IN that encloses both complimentary and the assembled triplets.*

1-Connecting a sequence of the triplets in information time-space dynamic network $IN_o$ (**Fig.7**) where each second triplet knot will accumulate information of both the first one and its own knot, sequentially growing information concentrated in each following knot' node. Each $IN_o$ node differentiates the quantity of concentrated information by its density as measure of quality information for this node- the $IN_o$ level of quality. Final node of this collective $IN_o$ concentrates and memorizes all hierarchical levels of the $IN_o$ collected information, and generates first class of information processing.

2- Connecting the triplet's knots of different $IN_o$ with forming new triplet knot via binging three triplets $UP_{3o}$ from a three parallel working $IN_o$. Such information dynamics concentrate three units of $UP_{3o}$ in new information unit $UP_{31} = \bigcup_{1-3} UP_{3o}$, or nine units $UP_o$. A sequence of such $UP_{31}$, formed from each of three multiple parallel $IN_o$, generates a second class of information processing.

3-Connecting $UP_{31}$ in the enclosed sequence, where each second knot, taken from three knots of the second level of these triple $IN_o$, enfolds both these three knots and the $UP_{31}$ from the first level, producing third kind of triple unit $UP_{32} = \bigcup UP_{31}$. Such information time-space dynamic process concentrates triple $UP_{32}$ knots in new third class of collective network $IN_1$ which triplicates density-quality of its node information at each hiearachical level.

4- A sequence of $UP_{32}$ from each $IN_1$ hierarchical level can organize a forth class of information processes, which involves multiple $IN_1$ as well as multiple $IN_o$.

5-Multiple $IN_1$ triples can be concentrated in a higher triplet's unit $UP_{33} = \bigcup UP_{32}$ whose enclosed sequence organizes new $IN_2$, and so on, with related information process connecting each $IN_2$ level.
Such grouping in the ordered triplets with growing concentration of information leads to ability of enfolding all potentially collected information in a single triplet's knot, memorizing the entire reachable information. Each Bit of different $UP_{ij}, i, j = 1, 2, 3......$ encloses information of distinct density, which depends on both particular observation, quantity and quality of prior collected information.
Thus, a Bit, produced differently, enfolds distinct information, or observations. Even when an observer encodes the same number of Bits, their enclosed information can be dissimilar.
The observed information acquires *specific* quality for each observer IN.

**VI. Encoding information units in the IN code-logic, and observer's computation using this code.**
*The code, program, and information quality of code-logic*
**1.** Each triplet unit generates three symbols from three segments of information dynamics and one impulse-code from the control, composing a minimal *logical code* that encodes this elementary physical information process. The control joins all three in a single triple unit and transfers this triple code to next triple, forming next level of the IN code. The IN triplet's dynamic space-time connection holds information minimum of three triplet's logic structures. Each information unit has its unique position in the time-spaced information dynamics, which defines the exact location of each triple code in the IN. Even though the code impulses are similar for each triplet, their time-space locations allows the *discrimination* of each code and the formed logics, distinction both codes and its units.
The IN code includes digital time intervals, encloses code's geometry, which depends on the digits' collected information. The shortened process intervals in the IN condense the observing information in rotating space-time triplets' knots, whose nodes cooperate in the IN information structure.
The observing process, chosen by the observer's (0-1) probes, following the information logic integration, determines the IN information code units that encode the IN code-logic.
The IN automatically requests higher information values, measured by the IN node hierarchy, via an adaptive feedback's information force, generated by the density of process correlations and defined by the information space curvature.
The IN information geometry holds the node's binding functions and an asymmetry of triplet's structures.
In the DSS information geometry, these binding functions are encoded adapting the requested external information.
The Observer's self-built information space–time networks (IN) hierarchical enfolds multiple observing information triplets encoding the observer logical structure in triplet code. Hence, the information of observing process moves and self –organizes the information geometrical structure creating the information Observer.
The information path functional collects the information units, while the IN performs logical computing operations using the doublet-triplet code of the observer's created program, which is specific for each observer, and therefore is self-encrypting. Such operations, performed with the entangled memorized information units, model a quantum computation [33, 34].

The operations with classical information units, which observer cooperates from quantum information units and runs the units in the IN, model a classical computation.

An observer that unites logic of quantum micro- and macro- information processes enables composing quantum and/or classical computation on different IN levels.

The program holds a distributed space-time processing generated by the observer information dynamics. Information emanated from different IN nodes encloses distinctive quality measure and logic, which encodes the observer IN *genetic code*. Triplet is elementary logical unit holds the IN genetic code which encloses helix geometrical structure (**Fig.3, 6**) analogous to DNA [35, 12].

The genetic code can reproduce the encoded system by decoding the IN final node and specific position of each node within IN structure. *This naturally encodes the impulse interaction with environment.*

**2.** The observed information *specific* quality for each observer IN depends on information density $N_b^{sc}$, defined by the number of information units (bits) that each of this information unit encodes (compresses) from any other source-code.

Since each triplet's bit encodes 3 bits, it information density is $N_b^1 = 3$. A following triplet also encodes 3 bits, but each of its bit encodes 3 bits of the previous triplet's bits. Thus, the information density of such two triplets is equal to $N_b^2 = 9$, and so on. Hence, for the $m$-th triplet we have $N_b^m = 3^m$ bits of this $m$-th triplet which encodes $3^m$ bits from all previous triplet's codes. The IN's final node with $m = n/2$ has $N_b^m = 3^{n/2}$, determined by the process' dimension $n$. The information density, related to the IN's level of its hierarchy, measures also the *value* of information obtained from this level.

For such a code, its information density also measures its valueability. *For example*, an extensive architecture of an ARM chip provides the enhanced code density: it stores a subset of 32-bit instructions as compressed 16-bit instructions and decompresses them back to 32 bits upon execution.

In each particular IN, the triplet elementary logical unit-cells self-organize and compose the observer Logical Structure, satisfying the limitations.

## VII. Information conditions of self-structuring and self-replicating an observer
### 1. Specifics, limitations, and invariance of observer's IN self-replication

**1**. The space-time's conditions (5.9- 5.11), following from the IN's cooperative ability of self-selection and renovation determine *self-structuring* of the observer's IN with its logic.

The invariance of information minimax law for any *information observer* preserves their common regularities of requesting, accepting, proceeding information, and building its information structure. That guarantees objectivity (identity) of basic observer's personal actions with *common information mechanisms* (Secs.II-V), which enable creation of *specific* information structures for each particular observed information, with its particular goal, preferences, energy, material carriers and various implementation not considered here.

The free information in (4.11) is an integral part of observer's interaction with environment, satisfying an *objective* natural information law, but its information content depends on individual observer, which imposes *subjectivity* on requested information and the environment.

*The same observable process may create different observers whose feedback's interacting information might adapt or change the observable process.*

**2.** *The limited the IN time-scale, speed of cooperation, and dimension*

Minimal admissible time interval of impulse acting on observable process is limited by $\delta^o_{t\min} \cong \mathbf{a}_{io}\hat{h} \approx 0.391143\times 10^{-15}$ sec.
Minimal wide of internal impulse $\delta_{te} \approx 1.6\times 10^{-14}$ sec limits ratio $\delta_{te}/\delta^o_{t\min} \cong 41$ which evaluates the limited IN time scale $(\gamma^\alpha_{12})^{m+1} = M_{\tau m} = \tau_{1o}/\tau_{m+1}$ and scale ratio $\gamma^\alpha_{m+1} = (\gamma^\alpha_{12})^{m+1} = 3.45^{m+1} = 41..$ That, for $m+1=3$ triplets brings $M = 41.06$ which for minimal IN two bound triplets enfolding $n=5$ process dimension brings the IN *geometrical* bound scale factor $\sqrt{(\gamma^\alpha_{m1})^n}$ which for this IN holds $\sqrt{(\gamma^\alpha_{m1})^5} \cong 22.1$.

In Physics, three particles' bound stable resonance has been recently observed [28, 30] with predicted scale factor $\cong 22.7$. If an observer enables condense the external information in a decreased wide of its impulse then the number of the IN enclosed triplet grows. This number limits a cooperative speed of IN's last triplet $m$ node, whose ratio to initial triple node information speed evaluates invariant

$$C_{oc} \cong 1/2\mathbf{a}_{io}(\gamma)\mathbf{a}_i^{-1}(\gamma)(\gamma^\alpha_{i=m}-1)(\gamma^\alpha_{i=m})^m . \quad (6.1)$$

Since each pair cooperation requires $1/2\ln 2 Nat$ of information, applied during time of cooperation $\delta_{te}$, the maximal speed of cooperation is

$$c_{oc} = 0.35 Nat/\delta_{te} = 0.21875\times 10^{14} Nat/\sec, \quad (6.2)$$

which is closed to maximal potential speed (4.19):

$$c_{oa} \cong 0.1444\times 10^{14} Nat/\sec \cong 20\times 10^{13} bit/\sec . \quad (6.2a)$$

The ending node cooperative speed $c_{ico} = 1bit/\sec \cong 0.7Nat/\sec$ leads to ratio $c_{iam}/c_{oa} = C_{icm} = 0.2062857\times 10^{14}$.
Applying to (6.1) the optimal minimax invariants:
$C_{ocm} \cong 1/2\ln 2/0.33\ln 2(2.45)(3.45)^m = 1.515\times 2.45(3.45)^m = 3.77(3.45)^m$, at $C_{icm} = C_{ocm}$
determines related maximal $m_o \cong 23.6, n_o \approx 48$.

Maximal cooperative speed $c_{am} \cong 10^6 bit/s$ and a single neuron' low speed $c_{hco} \approx 10 bit/s$ leads to $C_{oc} \cong 10^5$ and $m = 7.3 \cong 7, n = 14$. (6.2b)

At such $m$, a complete IN of human being's information logic can build the IN sections, each with $m\cong 7$ levels, which cooperate in a triplet of future IN, composing all observed integral information (**Figs.6,7**).
At $c_{hco} \approx 10 bit/s$, it requests starting information frequency $B_f \cong 10^6 bit/s \approx 10^{-3} Gbit/s$.

A single IN's maximal information level for human being IN approximates $m_M \cong 7$, and a minimal selective subjective observer is limited by $4 > m_{M1} > 3$ levels. The minimal observer with a single triplet can build the minimal IN with two triplets $m_{Min} = 2$ by adding one more triplet. The observer is able building multiple information Networks, when each three ending nodes of maximal admissible $m_M$ can form a triplet structure with enfolds all three separate local INs, increasing the encoded information in $3m_M$ and then multiply it on $m_M$: $3m_M m_M$ by building new IN starting with this triplet.
That process allows progressively increase both quantity and quality of total encoded information in
$(3m_M m_M)\times(3m_M m_M)\times....... = (3m_M m_M)^{m_M} = N_m$ times of the initial IN's node information $\mathbf{a}_{1o} = \ln 2$:

$$I_m = \ln 2(3m_M m_M)^{m_M} \quad (6.3)$$

with maximal density $I_m^d = \ln 2(\gamma^\alpha_{12})^{N_m}$. (6.3a)

and time scale $M_\tau = (\gamma^\alpha_{12})^{N_m}$ (6.3b)

This a huge quantity and quality of information is limited by maximal $m_M$. Maximal information available from an external random information process is also limited at infinite dimension of the process [10].

*3. Information conditions for structuring a multiple selective observer*

Satisfaction of sufficient condition (5.2-5.4) in multiple IN's interactions determines a multiple selective observer, whose attracting cooperative information force grows from $X_{12}^{\alpha} \geq [(\gamma_{12}^{\alpha}) - 1]$ to

$$X_{12}^{\alpha N_m} \to (\gamma_{12}^{\alpha}[\gamma_m])^{N_m} \text{ at } (\gamma_{12}^{\alpha}[\gamma_m]) \to (\gamma_{12}^{\alpha}[\gamma_m])^{N_m}, \qquad (6.4)$$

accumulating maximal information

$$I_{12}(X_{12}^{\alpha N_m}) = X_{12}^{\alpha N_m} \mathbf{a}_o[(\gamma_{12}^{\alpha})^{N_m}], \quad \gamma_m \to 0. \qquad (6.4a)$$

Multiple communications of numerous observers send a message-demand, as *quality messenger* (qmess), enfolding the sender IN's cooperative force (5.8a, 5.9a), which requires access to other IN observers [14]. This allows the observer-sender generates a collective IN's logic of the multiple observers.

Each observer's IN memorizes its ending node information, while total multi-levels hierarchical IN memorizes information of the whole hierarchy. The observer, requesting maximal quality information (by its intentional information (6.4a)), generates probing impulses, which select the needed density-frequency's real information among imaginary information of virtual probes. That brings to the observer-sender all current IN logical information; while the IN information dynamics enable renovate the existing IN in a process of exchanging the requested information with environment, and rebuild the IN by encoding and re-memorizing the recent information. Since the whole multiple IN information is *limited* as well as a total time of the IN existence, the possibility of the IN self-replication arises.

<u>Comments 6.1.</u> Observer's information units interact in an information channel, producing randomness, entropy that generates a nonrandom information of the channels errors, which corrupts the sender information.

*4. The invariant information and energy conditions of observer's IN self-replication*

The IN node's maximal admissible $m_M$-th level ends with a single dimensional process, which loses ability to enfold new attracting information. Hence, extending that IN does not satisfy more minimax information law, it bringing its instability. Specifically, after a completion of the IN cooperation, the control of last IN ending node initiates one dimensional process $x_n(t_n) = x_n(t_{n-1})(2 - \exp(\alpha_{n-1}^t t_{n-1}))$, which at $t_{n-1} = \ln 2 / \alpha_{n-1}^t$ approaches final state $x_n(t_n) = x_n(T) = 0$ with potential infinite relative phase speed $\dot{x}_n / x_n(t_n) = \alpha_n^t = -\alpha_{n-1}^t \exp(\alpha_{n-1}^t t_n)(2 - \exp(\alpha_{n-1}^t t_n))^{-1} \to \infty$.

Since the model cannot reach zero final state $x_n(t_n) = 0$ with $\dot{x}_n(t_n) = 0$, a periodical process arises as result of alternating the macro movements with the opposite values of each *two* relative phase speeds $\dot{x}_{n+k-1} / x_{n+k-1}(t_{n+k-1}) = \alpha_{n+k-1}^t$, $\dot{x}_{n+k} / x_{n+k}(t_{n+k}) = -\alpha_{n+k}^t$. That leads to instable fluctuations of these speeds at each $t = (t_{n+k-1}, t_{n+k})$ starting the alternations with $\dot{x}_n / x_n(t_n) = \alpha_n^t$, $k = 1, 2, \ldots$ at $\gamma \geq 1$.

The instable fluctuations in three-dimensional process involves the oscillational interactions of three ending node other IN's approaching $\gamma \geq 1$, which generate frequency spectrum of model eigenvalues $\lambda_i^*(t_{n+k})$ in each space dimension $i = 1, 2, 3$.

Formal analysis of this instability associates with nonlinear fluctuations, which can be represented [36] by a superposition of linear fluctuations with the frequency spectrum ($f_1, \ldots, f_m$) proportional to imaginary components of the spectrum eigenvalues ($\beta_1^*, \ldots, \beta_m^*$), where $f_1 = f_{\min}$ and $f_m = f_{\max}$ are the minimal and maximal frequencies of the spectrum accordingly.

In our model, the oscillations under the interactive control generate imaginary eigenvalues $\beta_i^*(t)$:

$\operatorname{Im} \lambda_{n+k}^i(t) = \lambda_{n+k-1}^i [2 - \exp(\lambda_{n+k-1}^i t)]^{-1}$ at each $t = (t_{n+k-1}, t_{n+k})$ for these $i$ dimensions. This leads to relation

$$\operatorname{Im} \lambda_n^i(t_{n+k}) = j\beta_n^i(t_{n+k}) = -j\beta_{n+k-1}^i \frac{\cos(\beta_{n+k-1}^i t) - j\sin(\beta_{n+k-1}^i t)}{2 - \cos(\beta_{n+k-1}^i t) + j\sin(\beta_{n+k-1}^i t)}, \quad (6.5)$$

**at** $\beta_i^* = \beta_{n+k}^i, \beta_i^* \neq 0 \pm \pi k$, where that $\beta_n^i(t_{n+k})$ includes real component

$$\alpha_n^i(t_{n+k}) = -\beta_{n+k-1}^i \frac{2\sin(\beta_{n+k-1}^i t)}{(2 - \cos(\beta_{n+k-1}^i t))^2 + \sin^2(\beta_{n+k-1}^i t)}, \quad (6.5a)$$

at $\alpha_i^* = \alpha_i^*(t_{n+k}) \neq 0$, with the related parameter of dynamics

$$\gamma_i^* = \frac{\beta_n^i(t_{n+k})}{\alpha_n^i(t_{n+k})} = \frac{2\cos(\beta_{n+k-1}^i t) - 1}{2\sin(\beta_{n+k-1}^i t)}. \quad (6.6)$$

At $\gamma = 1$ it corresponds $(\beta_{n+k-1}^i t) \approx 0.423 rad (24.267^o) = 0.134645\pi$.

These fluctuations may couple the nearest dimensions by an interactive double cooperation through overcoming *minimal elementary* uncertainty UR, *separating the model's dimensions' measure via invariant* $h_\alpha^o$, which may border the IN maximal stable level $m_M$.

Suppose, the $k$-th interaction, needed for creation a single element in the double cooperation, conceals information $s_c(\gamma) = \mathbf{a}_{ok}^2(\gamma)$, which should compensate for increment of minimal uncertainty of invariant $h_\alpha^o \mathbf{a}_o(\gamma = 0) = 0.76805/137 = 0.0056$ Nat by information equals $\delta \mathbf{a}_k = \mathbf{a}_o(\gamma = 0) - \mathbf{a}_o(\gamma^*)$.

This invariant evaluates the UR information *border* by the increment of the segment's entropy concentrated in UR, while equality $h_\alpha^o \mathbf{a}_o(\gamma = 0) = \mathbf{a}_{ok}^2(\gamma)$ evaluates minimal interactive increment

$\delta \mathbf{a}_k = \mathbf{a}_{ok}^2(\gamma) = 0.0056$ (6.6a) with minimal information $\mathbf{a}_{ok}(\gamma) = 0.074833148$, (6.6b)

needed for a single interaction in each dimension.

Each $k$-th interaction changes initial $\gamma \geq 1$ on $-\Delta\gamma = -\gamma^*$ bringing minimal information increment in each dimension (6.6b). Minimal information attraction enables cooperate a couple needs three these increments, generating information

$3\mathbf{a}_{ok}(\Delta\gamma) = 0.2244499443 \cong 0.23 = \mathbf{a}_k.$ (6.6c)

Information attraction $\mathbf{a}_k$, generated in each dimensional interaction, can cooperate that interactive information in invariant $\mathbf{a}_{ok}(\gamma) \cong 0.7$ which binds three dimension in single bit thru total nine interactions.

The question is how the interactive fluctuations enable creating a triplet which self-replicates new IN?

Dynamic invariant $\mathbf{a}(\gamma) = \mathbf{a}_k$ of information attraction determines ratios of starting information speeds $\gamma_1^\alpha = \alpha_{io}/\alpha_{i+1o}$ and $\gamma_2^\alpha = \alpha_{i+1o}/\alpha_{i+2o}$ needed to satisfy invariant relations (4.15).

To create new triplet's IN with ratio $\gamma_1^\alpha = \alpha_{io}/\alpha_{i+1o}$, relation (6.6) requires such ratio $\frac{\beta_i^*(t_{n+k})}{\beta_{n-1,o}(t_{n-1,o})} = l_{n-1}^m$ which deliver imaginary invariant $(\beta_{n+k-1}^i t) \to (\pi/3 \pm \pi k), k = 1, 2, ...$ at each $k$ with ending information frequency $\beta_{lo}(t_o) = \beta_i^*(t_{n+k})$ that would generate needed $\alpha_n^i(t_{n+k}) = \alpha_{lo}^m(t_o)$.

In case $\Delta\gamma \to 0 \to \gamma^*$, it can be achieved in (6.6) at $2\cos(\beta_{n+k-1}^i t) \to 1$, or at

$(\beta_{n+k-1}^i t) \to (\pi/3 \pm \pi k), k = 1, 2, ...$, with $\alpha_n^i(t_{n+k}) = \alpha_{lo}^m(t_o) = \lambda_{lo}^m \cong -0.577\beta_{n+k-1}^i$ (6.7)

That determines maximal *frequency's ratio*

$l_{n-1}^m = \beta_{n+k-1}^i/\beta_{n-1,k=o}^i$ (6.7a)

which at $\gamma=1$, $\beta_{n-1,o}(t_{n-1,o})=\alpha_{n-1,o}(t_{n-1,o})$ and $\beta^i_{n+k-1}=\alpha^i_{l,k=3}/0.577$ holds
$l^m_{n-1}=\alpha^i_{l,k=3}/0.577/\alpha^i_{n-1,k=o}$, $\alpha^i_{l,k=3}/\alpha^i_{n-1,k=o}=\gamma^\alpha_1$. It identifies (6.7a) connecting the triplet ratio by invariant
$$l^m_{n-1}=\gamma^\alpha_1/0.577, \tag{6.8}$$
generated by the initial $(n-1)$-dimensional spectrum with an imaginary eigenvalue $\beta_{n-1,o}(t_{n-1,o})$ by the end of the interactive movement. Invariant (6.7a), (6.8) leads to $\gamma^\alpha_1=\alpha^i_{l,k=3}/\alpha^i_{n-1,k=o}=3.89$ and to ratio of initial frequencies $l^{m=1}_{n-1}\cong 6.74$. Next nearest $\alpha^i_{l+1,k=6}/\alpha^i_{lo,k=3}=\gamma^\alpha_2$ needs increasing first ratio in $l^{m=2}_{n-1}\cong 3.8$, and the following $\alpha^i_{l+2,k=9}/\alpha^i_{l+1,k=6}=\gamma^\alpha_2$ needs $l^{m=3}_{n-1}\cong 3.8$. The multiplied ratios
$$l^{m=1-3}_{n-1}=l^{m=1}_{n-1}\times l^{m=2}_{n-1}\times l^{m=3}_{n-1}\cong 97.3256 \tag{6.8a}$$
need to build new triplet, which at $\gamma=1$ not ends with segment, while $l^{m=1-3}_{n-1}$ identifies maximal ratio of spectrum frequencies generated by the instable fluctuations.

Each three information $3\mathbf{a}_{ok}(\Delta\gamma)$ binds pair of nearest spectrum frequencies, starting with pair $\beta_{n-1,o}(t_{n-1,o})$, $\beta^i_{n,k=1}$ which sequentially grows with each $k$ interactive information (6.6a-c), intensifying the increase of frequency. First three pairs bind three dimensions in single Bit $\mathbf{a}_{ok}(\gamma)\cong 0.7$ thru nine interactions of frequencies requiring ratio $l^{m=1}_{n-1}\cong 6.74$, next pair ratio grows in ~2.24 times, each next in 1.26 times.

Thus, a natural source to produce the very first triplet is nonlinear fluctuation of an initial dynamics, involving, as minimum, three such native dynamics dimensions that enclose some memorized information by analogy with ending IN node, which at $\mathbf{a}_o(\gamma=1)=0.58767$, $\mathbf{a}(\gamma=1)=0.29$ brings $\gamma^\alpha_1\cong 2.95$ and needs ratio $l^{m=1}_{n-1}\cong 5.1126$. Natural source of maximal speed frequency is light wavelength whose time interval $t_{lo}\approx 1.33\times 10^{-15}\sec$ determines maximal frequency $f_{max}\cong 0.7518\times 10^{15}\sec^{-1}$ that at each interaction brings information $h^o_\alpha\mathbf{a}_o(\gamma=0)=0.005\,\text{Nat}$ (6.8b) changing the initial frequency.

The required frequency ratio $l^{m=1-3}_{n-1}$ identifies minimal frequency $f_{min}=0.7724586\times 10^{13}\sec^{-1}$.
The triplet information invariant allows finding the equivalent energy invariant for creating such triplet.
Invariants $h^o_\alpha\cong 1/137$ coincides with the Fine Structural constant in Physics [37]:
$$\alpha^o=2\pi\frac{e^2}{4\pi\varepsilon^o hc}, \tag{6.9}$$
where $e$ is the electron charge magnitude's constant, $\varepsilon^o$ is the permittivity of free space constant, $c$ is the speed of light, $h$ is the Plank constant.
The equality $h^o\cong\alpha^o$ between the model's and physical constants allows evaluate the model's structural parameter energy through energy of the Plank constant and other constants in (6.9):
$$h=\frac{e^2}{2\varepsilon^o ch^o}=C_h\alpha_h, \alpha_h=(h^o)^{-1}=inv, \quad \frac{e^2}{2\varepsilon^o c}=C_h=h/(h^o)^{-1}=9.0831\times 10^{-32}J.\sec \tag{6.9a}$$
where $C_h$ is the energy's constant (in [J.s]), which transforms the invariant $\alpha_h$ to $h$.

In the information approach, (6.9a) evaluates energy that conceals the IN bordered the stable level $m_M$. The triplet creation needs nine such interacting increments, which evaluate the triplet energy's equivalent
$$e_{tr}=8.1748\times 10^{-29}J.\sec. \tag{6.9b}$$

Invariant conditions (6.6), (6.9a) enable the model's *cyclic* renovation, initiated by the two mutual attractive processes, which not consolidate by the moment of starting the interactive fluctuation. After the model disintegration, the process can renew itself with the state integration and the transformation of the imaginary into the real information during the dissipative fluctuations bringing energy for a triplet.

Initial interactive process may belong to different IN macromodels (as "parents") generating new IN macrosystem (as a "daughter"), at end of the "parents" process and beginning of the "daughters".
The macrosystem, which is able to continue its life process by renewing the cycle, has to transfer its coding life program into the new generated macrosystems and provide their secured mutual functioning.

A direct source is a joint information of three different IN nodes $\mathbf{a}_{o1}(\gamma_1), \mathbf{a}_{o2}(\gamma_2), \mathbf{a}_{o3}(\gamma_3)$ enable initiate attracting information with three information speeds where one has opposite sign of the two; while the information values cooperating an initial triplet will satisfy the above invariant relations.

Creating a triplet with specific parameters depends on the starting conditions initiating the needed attracting information. To achieve information balance, satisfying the VP and the invariants, each elementary $\mathbf{a}_{oi}(\gamma_i)$ searches for partners for the needed consumption information. A double cooperation conceals information $s_c(\gamma) = \mathbf{a}_o^2(\gamma)$, while a triple cooperation conceals information $s_{cm}(\gamma) = 2\mathbf{a}_o^2(\gamma)$ and it could produce a less free information, while both of them depend on $\gamma$. With more triplets, cooperating in IN, the cooperative information grows, spending free information on joining each following triplet.

Minimal relative invariant $h_\alpha^o = 0.00729927 \cong 1/137$ evaluates a maximal *increment* of the model's dimensions $m_M \cong 14$ (6.9a), and the quantity of the hidden invariant information (6.8b) produces an elementary triple code, enclosed into the hyperbolic structure **Fig.9**, with its cellular geometry.
This hidden *a non-removable uncertainly also enfolds a potential DSS information code.*

Results (6.6-6.9a) impose important *restrictions* on both maximal frequency generating new starting IN and maximal IN dimension which limits a single IN, and whose ending node may initiate this frequency. For example, $\gamma = 1$ corresponds to

$$(\beta_{n+k-1}^i t) \approx 0.423 rad (24.267^o), \text{ with } \beta_i^*(t_{n+k}) \cong -0.6 \beta_{n+k-1}^i. \tag{6.10}$$

Here $\beta_i^*(t_{n+k}) \cong \alpha_{lo}^m(t_o)$ determines maximal frequency $\tilde{\omega}_m^*$ of fluctuation by the end of optimal movement:
$\alpha_{lo}^m(t_o) \cong -0.6\beta_{n+k-1}^i$, where $\alpha_{lo}^m(t_o) = \alpha_{1o}(t_o)(\gamma_{12}^\alpha)^{N_m}$.
Starting speed $\alpha_{1o}(t_o) = 0.002 \sec^{-1}$, $m = 14$, determines $\alpha_{14o}(t_o) = 0.00414 \sec^{-1}$, $\beta_{14} \cong 0.0069 \sec^{-1}$. (6.10a)

The new macromovement starts with this initial frequency. This new born macromodel might continue the consolidation process of its eigenvalues, satisfying the considering restrictions on invariants and cooperative dynamics up to ending the consolidations and arising the periodical movements. This leads to cyclic *micro-macro functioning* when the state integration alternates with the state disintegration and the system decays with possible transformation of an observable virtual process into the real certainty-information. Functional schema of the cyclic evolutionary informational mechanisms **(Fig10)** and the elements have been simulated by the IMD software package [38].

**2. *Limitations on evolution macrodynamics***
**1.** The information speed of evolution dynamics determines the model's average Hamiltonian
$$\hat{H} = Tr[A], A = (\alpha_{it})_{i=1}^n \tag{6.11}$$

where $\alpha_{it}$ is information speeds along a segment dynamics.
The average Hamiltonian for for the segments of IN with $m = n/2$ **t**riplets:

$$\hat{H} = \alpha_1(t_{1o}) \sum_{m=1}^{m=n/2} (\gamma_{13}^{\alpha})^{-m}, \quad \gamma_{13}^{\alpha} = \alpha_{io} / \alpha_{i+3o} = \tau_{i+3o} / \tau_{io} \tag{6.11a}$$

decreases with growing triplet's number $m$ for macromodel's dimension $n$, and regular time course of $t_{1o}$. Each following triplet has less information speed than the previous one, and a total model's information speed is slowing down with growing a time of enfolding the triplet node-an evolution time. Sum of relative derivations of the model's eigenvalues spectrum at the current derivations:

$$D = \sum_{i=1}^{n} \frac{\Delta \alpha_{it}}{\alpha_{it}}, \alpha_{it} = \mathbf{a}(\gamma)\alpha_{io} \tag{6.12}$$

produced by the increment of correlations, is called an *indicator* of model's *diversity*, while the spectrum preserves *stability* at the VP satisfuction. The relative increment of averaged Hamiltonian (6.11):

$$H_{\Delta} = \frac{\Delta \hat{H}}{\hat{H}}, \Delta \hat{H} = \sum_{i=1}^{n} \Delta \alpha_{it} / \alpha_{it} \tag{6.13}$$

at $H_{\Delta}(\Delta \alpha_{io}, \alpha_{io}) = D$ measures the model initial diversity.

A $\max D = D_o$ defines the spectrum's maximal variations, which are limited by model's ability to preserve the spectrum dimension $n$. $D_o = H_{\Delta}^o$ measures maximal admissible increment of the average evolutionary speed $H_{\Delta}^o$ for a *given* macrosystem, whose $n$ dimensions represent a *population* of the macrosystem's interactive elements (subsystems).

A possibility of changing the initial relative $a_{io}^*(\gamma) = \Delta \alpha_{io} / \alpha_{io}$ under the stochastic deviations drives the model's variations. A *maximum* amongst the admissible model's variations evaluates potential of evolution $P_e$:

$$P_e = \max \sum_{i=1}^{i=n} a_{io}^*(\gamma) = \max D_o \to \max H_{\Delta}^o. \tag{6.14}$$

In the model's evolution, a limited $P_e$ counterbalances the admissible eigenvalue's deviations.

*Extending the model population by increasing dimension n and growing the sum in (6.14) raises the potential of evolution.*

The model limits potential of evolution by admissible maximal value $P_{em}$ which preserves each $m$-th triplet invariant:

$$P_{em} = \max \alpha_{mo}^*(\gamma). \tag{6.14a}$$

For triplet's eigenvalue's ratios $\gamma_1^{\alpha} = t_{m2}/t_{m1}, \gamma_2^{\alpha} = t_{m3}/t_{m1}$, segment's time intervals $t_{m1}^t, t_{m2}^t, t_{m3}^t$, $\alpha_{mo}^*(\gamma)$ satisfies:

$$\alpha_{mo}^* = [(\alpha_{m2,o}^t)^2 - (\alpha_{m1,o}^t \alpha_{m3,o}^t)]^{1/2} / \alpha_{m3,o}^t = ((\alpha_{m2,o}^t / (\alpha_{m3,o}^t)^2 - \alpha_{m1,o}^t \alpha_{m3,o}^t / (\alpha_{m,o}^t)^2)^{1/2}$$
$$= (t_{m3}^2 / t_{m2}^2 - t_{m1} t_{m3} / t_{m3}^2)^{1/2} = [(\gamma_1^{\alpha} / \gamma_2^{\alpha})^2 - (\gamma_2^{\alpha})^{-1}]^{1/2} = \varepsilon(\gamma), \tag{6.14b}$$
$$\gamma_1^{\alpha}(\gamma) = inv, \gamma_2^{\alpha}(\gamma) = inv, \alpha_{m1,o}^t t_{m1}^t = \alpha_{m2,o}^t t_{m2}^t = \alpha_{m3,o}^t t_{m3}^t = \mathbf{a}_o(\gamma) = inv.$$

These relations will be compromised, if any $\alpha_{m2,o}^t$ approaches $\alpha_{m1,o}^t$, or $\alpha_{m2,o}^t$ approaches $\alpha_{m3,o}^t$, *because* if for both $|\alpha_{m2o}^t - \alpha_{im1,o}^t| \to 0, |\alpha_{m2o}^t - \alpha_{m3,o}^t| \to 0$, the triplet disappears, and macromodel's dimension $m = n/2$ decreases.

The eigenvalue's distance $\varepsilon(\gamma)$ in (6.14b) is limited by the admissible minimal distance $\varepsilon_{\gamma}(\gamma^*)$ at minimal $\gamma^* = 0.007148$. The triplet's eigenvalues ratios $\gamma_1^{\alpha}(\gamma^*) \cong 2.74, \gamma_2^{\alpha}(\gamma^*) \cong 4.9$ maximize admissible error of deviations for triplet:
$$\varepsilon^2(\gamma^*) \cong 0.1086, \varepsilon(\gamma^*) \cong 0.33. \tag{6.14c}$$
The triplet's *information capacity* to counterbalance maximal admissible deviations (6.14c) defines the triplet's potential:
$$P_e^m = \max |\varepsilon(\gamma)|, \max \varepsilon(\gamma) = \varepsilon(\gamma^*), \tag{6.15}$$

The macromodel potential with $m$ triplets $P_e^n = m \, P_e^m$ limits maximal *acceptable* dimension $n_m$ that sustains cooperation:

$$P_e^n \cong m_n/3, \; m_n = \frac{n_m - 1}{2}, \; P_e^n \cong (n_m - 1)/6n. \tag{6.16}$$

This means, $P_e^n$ admits a maximal decrease of the model's current dimension $n_c$ in $n_m \cong n_c/3$ time. Interval of $\varepsilon(\gamma \in (0.001718 \to 0.8)) \cong 0.33 \to 0.23$ determines the admissible interval of changing of the model potential ratio

$$P_e^* = (P_e^n(\gamma^*) - P_e^n(\gamma = 0.8))/P_e^n(\gamma^*)) \cong 0.3. \tag{6.16a}$$

At minimal $\gamma = 0.1$, model potential $P_e^n \cong 0.275$ leads to ratio $P_{em}^* \cong 0.2$, which accepts changing of the current dimensions $n_{mr} \cong n_c/3.6$ in fewer diapason of $\gamma$ with average potential $P_e^n(\gamma = 0.5)) \cong 0.256$.

Potential $P_e^n$ differs from $P_e$ (6.14), which generally does not support the evolution hierarchy of the process. Relation $P_e^n \leq P_e$ limits variations that restrict both maximal increment of dimension and sustain model's cooperative functions.

A *triplet's robustness* preserves the triplet's invariants under admissible maximal error $\varepsilon_m(\tau_{km})$ in the model time interval $\tau_{km}$ at forming $m$-th triplet-as a threshold for fluctuations, which at current and fixed $\gamma = \gamma_*$ satisfies relation

$$P_{mr} = \max \varepsilon_m(\gamma_*), \varepsilon_m(\tau_{km}) = |\Delta \alpha_m(\tau_{km})/\alpha_m(\tau_{km})| \leq \varepsilon_m(\gamma_*). \tag{6.17}$$

External influences (mutations), affecting the error $\varepsilon_m(\tau_{km}) > \varepsilon_m(\gamma_*)$ within an admissible potential $P_{em}^*$, might change the model's invariants, dynamics and dimension, which could also altered the DSS code (in **Fig.10**).

Let us show that at the equal deviations of the model parameter $\gamma_i^\alpha$: $\pm \Delta \gamma_1^\alpha, \pm \Delta \gamma_2^\alpha$, the model threshold $|\varepsilon(\Delta \gamma)|$ is *asymmetric*. The macromodel with $\gamma = 0.5$ has $\gamma_1^\alpha = 2.21$, $\gamma_2^\alpha = 1.76$ and gets $\varepsilon_o = 0.255$.

Admissible deviations of, $\Delta \gamma_2^\alpha = 0.08$ correspond the macromodel with $\gamma \cong 0.01$, $\gamma_1^\alpha = 0.246$, $\gamma_2^\alpha = 1.82$, which determines $\varepsilon_1 \cong 0.35$ and $\Delta \varepsilon(\Delta \gamma_1^\alpha, \Delta \gamma_2^\alpha) = \varepsilon_1 - \varepsilon_o = 0.095$. At $\Delta \gamma_1^\alpha = -0.25$, $\Delta \gamma_2^\alpha = -0.08$, macromodel with $\gamma_1^\alpha = 1.96$, $\gamma_2^\alpha = 1.68$, determines $\varepsilon_2 \cong 0.167$ and $\Delta \varepsilon(-\Delta \gamma_1^\alpha, -\Delta \gamma_2^\alpha) = \varepsilon_2 - \varepsilon_o = -0.088$.

At the equal deviations, the model potential (6.16) with $\max \varepsilon = \max \varepsilon_1 \cong 0.35$ tends to *increase at decreasing* $\gamma$, and vice versa. An essential *asymmetry* of $|\varepsilon(\Delta \gamma)|$ and therefore the dissimilar $P_{em}$ are the results of the macromodel fundamental quality of *irreversibility*. The feedback self-control works *within* the model's *robustness*, which preservers $\alpha_{it}^*$ at a fixed $\gamma = \gamma_*$. An extended self-control is able to *minimize* the maximal admissible triplet's error $\Delta t_i^*(\Delta \gamma) = 1 - \frac{t_i^2}{t_{i+1} t_{i-1}}$ under the perturbations of $\alpha_{it}^*(\Delta \gamma)$ for *any allowable* $\Delta \gamma \in \Delta \gamma_m$. Such control compensates *acceptable* $\alpha_{it}^*(\Delta \gamma)$, *unrestricted* by the *robustness* potential, but *limited* by $\max |\varepsilon_i(\Delta \gamma_i)|$ for any $\varepsilon_i \leq |\varepsilon_i(\Delta \gamma_i)|$ satisfying (6.15).

Optimal self-control, which automatically *adapts the model* to a changed $\alpha_{it}^*(\Delta \gamma)$ within the admissible $\varepsilon_i \leq |\varepsilon_i(\Delta \gamma_i)|$, we call *adaptive* control [32, 10]. This control actions are limited by adaptive potential $P_a = \max |\varepsilon_i(\Delta \gamma_{io})|, \varepsilon_i \leq |\varepsilon_i(\Delta \gamma_i)|$, whose maximum restricts triplet potential $P_e^m : P_a \leq P_e^m$, while $P_a$ provides both preservation of each triplet structure within $n$-dimensional IN and *adapts* the acceptable variations. The adaptive controls support both *robustness* and adaptive potentials that guarantees the stability preserving the triplets and model's dimension.

## 3. *The IN cooperative complexity*

The IN nested structure holds cooperative complexity (CC), which decreases initial complexity of not cooperative information units. The CC measures *origin* of complexity in the interactive dynamic *process*, cooperating elementary duplet-triplet, whose free information anticipates new information, requests it, and automatically builds the hierarchical IN. Existing complexity focuses on measuring complexity of already formed complex system and processes [39-42]. The CC is an *attribute* of the process's *cooperative* *dynamics*. Bound information of cooperating units is *potential* sources of decreasing the CC, which measures the concealed *information,* accompanied by creation of new phenomena. We study both complexity of information macrodynamic process (MC) and CC, arising in interactive dynamics of chnging information flows from $I_i$ to $I_k$, accompanied with changing their shared volume from $V_i$ to $V_k$ : $\Delta I_{ik} = I_i - I_k, \Delta V_{ik} = V_i - V_k$.

The MC defines an increment of concentration of information in the information flow before and after interaction, measured by the flow's increment per the changed information volume: $MC_{ik} = mes[\Delta I_{ik} / \Delta V_{ik}]$, while the flow's increment is measuring by the increment of entropy speeds: $mes\Delta I_{ik} = \partial \Delta S_{ik} / \partial t$:

$$MC_{ik} = (\partial \Delta S_{ik} / \partial t) / \Delta V_{ik} . \qquad (7.1)$$

This complexity determines an instant entropy's concentration in this volume: $\dfrac{\partial \Delta S_{ik}}{\Delta V_{ik} \partial t}$ (the entropy production), which evaluates the specific information contribution, transferred during the interactive *dynamics* of the information flows. Complexity (7.1) is measured *after* the interaction occured, assuming that both increments of speeds and volumes are known. To evaluate complexity arising *during* interactive dynamics, *information measure of a differential interactive complexity* $MC_{ik}^\delta$, *is introduced, defined by* increment of information flow $-\dfrac{\partial \Delta S_{ik}}{\partial t}$ per small volume increment $\delta V_{ik}^\delta$ (within the shared volume $\Delta V_{ik}$):

$$MC_{ik}^\delta = \frac{\partial H_{ik}}{\partial t} / \frac{\partial \Delta V_{ik}}{\partial t}, \qquad (7.2)$$

where (7.2) defines the ratio of the speeds, measured by the increments of Hamiltonian and volume accordingly. The $MC_{ik}^\delta$ automatically includes both $MC_{ik}$ and its increment $\delta MC_{ik}$. The $MC_{ik}$ measures differential increment of information of interactive elements $i,k$, whose current *information difference* $\Delta S_{ik}$ and shared volume $\Delta V_{ik}$ *-before joining,* would be reduced to increment $\delta S_{ik}$ and volume $\delta V_{ik}^\delta$ accordingly after their cooperation during a macrodynamic process. Applying the IMD, allows measuring both $MC_{ik}$ and $MC_{ik}^\delta$ through the VP information invariants and the direct evaluation these complexities in the bits of information code. Both complexity measures characterize each triplet's dynamics through their eigenvalues, which connect them with related geometrical structure's volume. The complexity (7.2) for a triplet, defined at the moment of three segments eigenvalues' equalization, has a form

$$M_{i,i+1,i+2}^\delta = 3\dot\alpha_{i+2,t} / \dot V_{i,i+1,i+2}, \qquad (7.3)$$

where $\dot\alpha_{i+2,t}|_{t=t_{i+2,t}} = [\alpha_{i+2,o}^2 t_{i+2,t}^2 \exp(\alpha_{i+2,o} t_{i+2,t})(2 - \exp\alpha_{i+2,o} t_{i+2,t})^{-1} - \alpha_{i+2,o}^2 t_{i+2,t}^2 \exp 2(\alpha_{i+2,o} t_{i+2,t})] / t_{i+2,t}^2$

$= [\mathbf{a}_o^2 \exp\mathbf{a}_o(2 - \exp(\mathbf{a}_o))^{-1} - \mathbf{a}_o^2 \exp(2\mathbf{a}_o)] / t_{i+2,t}^2 = (\mathbf{a}_o \mathbf{a} - \mathbf{a}_o^2 \exp(2\mathbf{a}_o)) / t_{i+2,t}^2, \mathbf{a} = \exp\mathbf{a}_o(2 - \exp(\mathbf{a}_o))^{-1}$ (7.3a) *an*d

$\dot V_{i,i+1,i+2} = \delta V_{i,i+1,i+2} / \delta t, \delta V_{i,i+1,i+2} = V_c \delta t^3$, $\delta V_{i,i+1,i+2} / \delta t = V_c \delta t^2 = V_c t_{i+2,t}^2 \delta t^2 / t_{i+2,t}^2 = V_c t_{i+2,t}^2 \varepsilon(\gamma)^2$. (7.3b)

Here space area $\varepsilon^2(\gamma)$ is an invariant at fixed $\gamma$, $V_c = 2\pi c^3 / 3(k\pi)^2 tg\psi^o$ is invarint volume at angle $\psi^o = \pi/6$ on the vertex of each cone (**Fig.2**), $c$ is a speed of rotation of each cone's spiral which produces angle $\pi/2$. We get differential complexity $M_m^\delta = M_{i,i+1,i+2}^\delta$.- for any joint *m*-th triplet-node:

$$M_{i,i+1,i+2}^\delta = 3(\mathbf{a}_o\mathbf{a} - \mathbf{a}_o^2 \exp 2\mathbf{a}_o)/V_c t_{i+2,t}^4 \varepsilon(\gamma)^2, \qquad (7.4)$$

Applying $M_m^\delta$ to a cell with volume $\delta V_{i,i+1,i+2} = \delta V_m$, formed during time interval $\delta t_{i,i+1,i+2} = \delta t_m$, we get $M_m^\delta \delta t_m = 3\dot{\alpha}_m \delta t_m / \delta V_m = 3\Delta\alpha_m / \delta V_m$, where $\Delta\alpha_m$ is an increment of information speed during $\delta t_m$.

The related increment of quantity information at the same $\delta t_m$ is $\Delta\alpha_m \delta t_m = a_m^\Delta = \dot{\alpha}_m \delta t_m^2$, where

$$a_m^\Delta = (\mathbf{a}_o \mathbf{a} - \mathbf{a}_o^2 \exp(2\mathbf{a}_o))\delta t_m^2 / t_{i+2,t}^2, \quad \delta t_m^2 / t_{m,t}^2 = \varepsilon_m^2, t_{m,t}^2 = t_{i+2,t}, \tag{7.5}$$

each invariant $3a_m^\Delta$ measures the quantity of information produced during interaction of three equal eigenvalues within area $\varepsilon_m^2$. Increment of entropy (in 7.1) determines the related volume: $M_m^\delta$, measured by equivalent quantity information, related to a cell volume $\delta V_m$, during the time $\delta t_m$:

$$M_m^\delta \delta t_m^2 = M_m^\Delta = 3(\mathbf{a}_o \mathbf{a} - \mathbf{a}_o^2 \exp(2\mathbf{a}_o))\varepsilon_m^2 / \delta V_m. \tag{7.5a}$$

Information $3a_m^\Delta$ binds the three segments in $\varepsilon_m^2$ prior the two impulse controls assmble them (Sec.IV(6)). By the moment of interaction $\tau_k^{i+2}$, three equal eigenvalues have signs $\alpha_{it}(\tau_k^{i+2}) sign\alpha_{it}(\tau_k^{i+2}) = \alpha_{i+1t}(\tau_k^{i+2}) sign\alpha_{i+1t}(\tau_k^{i+2}) = -\alpha_{i+2t}(\tau_k^{i+2}) sign\alpha_{i+2t}(\tau_k^{i+2})$.

Since negative eigenvalues $\alpha_{it}(\tau_k^{i+2}) sign\alpha_{it}(\tau_k^{i+2}) = \alpha_{i+1t}(\tau_k^{i+2}) sign\alpha_{i+1t}(\tau_k^{i+2})$ are stable and positive eigenvalue $-\alpha_{i+2t}(\tau_k^{i+2}) sign\alpha_{i+2t}(\tau_k^{i+2})$ is unstable, their interaction leads to instability, associated with a choatic attraction, which is localized within zone $\varepsilon_m^2$. The controls, delivering information $2\mathbf{a}_o^2$, *cooperate* these segments (within $\varepsilon_m^2$) by joining them into a single segment' node. Thus, (7.5a) measures *cooperative complexity* of the *interactive* three segments, forming *a single node* of $m$-th triplet.

The cooperative node forms the cell within volume $\delta V_m$, where both the eigenvalues' interaction and cooperation takes place. Since quantity information $2\mathbf{a}_o^2 \cong 1 bit$ of the joint segment from $m$-th triplet's node is transferred to a first segment of following $m+1$-th triplet, the quantity of *binding* information $3a_m^\Delta$ (in (7.5)), being spent on holding $m$-th triplet, is concentrated in the volume $\delta V_m$.

Let $M_{cm}^\Delta = 3a_m^\Delta(\gamma) / [\delta V_m / \varepsilon_m^2]$ evaluates the quantity of information per cell volume $\delta V_m$ related to a cell size area $\varepsilon_m^2$. Then using $M_{cm}^\delta = 3\Delta\alpha_m / \Delta V_m$, $M_{cm}^\delta \delta t_m = 3\Delta\alpha_m \delta t_m / \Delta V_m$, we have $M_{cm}^\Delta = 3a_m^\Delta / \Delta V_m$, which for each $\Delta V_m$ evaluates $M_{cm}^{\Delta V} = 3a_m^\Delta$. At $\gamma = 0.5, \mathbf{a}_o \cong -0.75, \mathbf{a} \cong 0.25$, we get $M_{cmN}^{\Delta V} = 3a_m^\Delta(\gamma = 0.5) \cong -0.897 Nat$ per cell, or $M_{cmb}^{\Delta V}(\gamma = 0.5) \cong -1.29 bit$ per cell-volume that each $m$-th node *conserves* during it formation. Being produced during the considered interaction (that primarily binds these segments), it measures a *cooperative* effect of the interactions, as the node's *inner cooperative complexity*.

Such a relative cooperative complexity does not depend on the actual cell volume and the number of nodes that the cell enfolds, and the $M_{cm}^{\Delta V}$ invariant quantity is not transferred along the IN nodes' hierarchy. Actually at $\delta V_m / \varepsilon_m^2 = V_c t_m^2$ and a fixed invariant $\varepsilon_m^2$ and volume $V_c$, increment $\delta V_m$ grows with assembling more nodes. Since that, complexity for any cell's volume $M_{cm}^{\Delta V} = inv(\gamma)$ (according to 7.5)) decreases with assembling more cooperating nodes within this volume. With growing the size of a cooperative, the cooperative complexity per its volume decrses in the ratio $M_{m+1}^\Delta / M_m^\Delta = t_m^2 / t_{m+1}^2 = (\gamma_m^\alpha)^{-2}$ while each following $M_{m+1}^\Delta$ enfolds complexity of the previous $M_m^\Delta$. Absolute value of interval $\delta t_m = t_m \varepsilon$ grows with increasing $t_{m+1}/t_m = \gamma_2^\alpha$, which leads to $\delta t_{m+1}/\delta t_m = \gamma_2^\alpha$ and $M_m^\Delta = 3a_m^\Delta/\delta t_m \delta V_m, \delta V_m = 3V_c \delta t_m^2$, $M_m^\Delta = a_m^\Delta/V_c \delta t_m^3 = a_m^\Delta/V_c \varepsilon_m^3 t_m^3$, while $M_m^\delta = a_m^\Delta/V_c \delta t_m^4 = a_m^\Delta/V_c \varepsilon_m^4 t_m^4$. This confirms the previous relations.

The ratio of the nearest triplet's complexities (7.4) is $M_{m+1}^\delta / M_m^\delta = t_m^4 / t_{m+1}^4$ at

$$(\mathbf{a}_o \mathbf{a} - \mathbf{a}_o^2 \exp(2\mathbf{a}_o))/V_c \varepsilon(\gamma)^2 = A_M(inv(\gamma)). \qquad (7.5b)$$

At $t_m^4/t_{m+1}^4 = (\alpha_{m+1}/\alpha_m)^4 = (\gamma_{m+1})^{-4}$, satisfaction of (7.5b) with $\gamma_{m+1} = \gamma_2(\gamma) = inv_o(\gamma)$, we get

$$M_{m+1}^\delta / M_m^\delta = \gamma_{m+1}^{-4}, \qquad (7.6)$$

which for $\gamma_2(\gamma = 0.5) = 3.89$ takes values $M_{m+1}^\delta / M_m^\delta \cong 0.00437$.

Complexity $M_{m+1}^\delta$, measuring $m+1$ node, also enfolds and condenses the complexity of a previous node. By moment $\tau_m$ of $m$-th triplet's cooperation, its three eigenvalues equalize: $\alpha_{3\tau}^m = \alpha_{2\tau}^m = \alpha_{1\tau}^m$, and at the *moment* of triplet's formation $\tau_m + o$, the cooperative eigenvalues $\alpha_m$ enfolds the joint triplet eigenvalues:

$$\alpha_3^m(\tau_m + o) = 3\alpha_{3\tau}^m = \alpha_m, \qquad (7.7)$$

In the IN, $m$-th triplet's first eigenvalue $\alpha_{1\tau 1}^m$ *equals* to last eigenvalue of $(m-1)$-th triplet $\alpha_{m-1}$: $\alpha_{1\tau 1}^m = \alpha_{m-1}$, where $\alpha_{1\tau 1}^m$ enfolds all three eigenvalues of previous $(m-1)$-triplets, while $m$-triplet holds

$$\alpha_{3\tau}^m / \alpha_{1\tau 1}^m = (\gamma_m^\alpha)^{-1}. \qquad (7.7a)$$

Substituting (7.7a) to (7.7) we have $\alpha_m = 3\alpha_{1\tau 1}^m(\gamma_m^\alpha)^{-1}$, and with $\alpha_{m-1}$ we get we get ratio

$$\alpha_m / \alpha_{m-1}^m = 3(\gamma_m^\alpha)^{-1}. \qquad (7.7b)$$

The sustained cooperation of the IN eigenvalues requires $\gamma_m^\alpha(\gamma = 0.5) \cong 3.9$, which brings the ratio (7.7b) to $\alpha_m/\alpha_{m-1}^m \cong (1.3)^{-1}$. Decreasing the eigenvalues of the cooperated triplets along the IN encloses the increased information density, which condenses more $MC_{ik}^\delta$ complexity. Specifically, at $M_{m+1}^\delta / M_m^\delta = (\alpha_{m+1}^4/\alpha_m^4)/\dot V_{m+1}/\dot V_m$ and $\dot V_{m+1}/\dot V_m = \alpha_m^2/\alpha_{m+1}^2$, $\alpha_{m+1}/\alpha_m = (1/3\gamma_{m+1})^{-1}$ ratio $M_{m+1}^\delta / M_m^\delta = (\alpha_{m+1}^4/\alpha_m^4)/\dot V_{m+1}/\dot V_m = (\alpha_{m+1}^6/\alpha_m^6) = (1/3\gamma_{m+1})^{-6}$ brings decreasing $M_{m+1}^\delta / M_m^\delta \cong 0.203$. Comparing (7.6) with the difference of relative complexities: $\Delta M_m^\delta / M_m^\delta = (M_m^\delta - M_{m+1}^\delta)/M_m^\delta = (1-\gamma_2^4)$, we get $\Delta M_m^\delta / M_m^\delta \cong |0.996|$ at $\gamma_2(\gamma = 0.5) = 3.89$, indicating that the difference decreases insignificantly. Relative sum of these complexities: $\Delta M_{m\Sigma}^\delta / M_m^\delta = (M_m^\delta + M_{m+1}^\delta)/M_m^\delta = (1+\gamma_2^4), \Delta M_{m\Sigma}^\delta / M_m^\delta \cong 1.0044$ also grows insignificantly.

Comparing these complexities with complexities of double cooperation within a triplet, we have $M_{12}^\delta / M_1^\delta = (\alpha_{12}\delta t_{12}/\delta V_{12})/(\alpha_1 \delta t_1/\delta V_1) \cong 2(\alpha_2/\delta V_{12})/(\alpha_1/\delta V_1)$, which at $\delta t_{12} \cong \delta t_1$, $\alpha_2/\alpha_1 = (\gamma_2^\alpha)^{-1}, \delta V_{12}/\delta V_1 = (\gamma_2^\alpha)^{-3}$, lead to $M_{12}^\delta / M_1^\delta \cong 2(\gamma_2^\alpha)^{-4}$, while for triplet: $M_{123}^\delta / M_1^\delta \cong 3(\gamma_3^\alpha)^{-4}$. (7.8)

At $\gamma_1^\alpha = 2.215$, $\gamma_2(\gamma = 0.5) = 3.89$, $M_{12}^\delta / M_1^\delta \cong 0.083$, $M_{123}^\delta / M_1^\delta \cong 0.013$. During a triple cooperation, the complexity decreases more than that in douple cooperation within a triplet. The $MC_{ik}^\delta$ (7.6) between the nearest triplets decreases much faster then that in the cooperation within a triplet. At the cooperation, each following nodes' complexity wraps and absorbs complexity of previous node, binding these node units and concerving the bound information. Decrease the IN cooperative complexity indicates that more cooperations have occurred, while at negative eigenvalues jump (SecIV(10)), the complexity grows with decoupling nodes and rasing choitic movement. The MC for each extremal segment and their ratios:

$M_i^d = \alpha_{it}/\Delta V_{it}, M_{i+1}^d = \alpha_{i+1,t}/\Delta V_{i+1,t}, M_{i+2}^d = \alpha_{i+2,t}/\Delta V_{i+2,t}, M_{i+1}/M_i = (\alpha_{i+1,t}/\alpha_{it})/(\Delta V_{i+1,t}/\Delta V_{it}),$

$M_{i+2}/M_i = (\alpha_{i+2,t}/\alpha_{it})/(\Delta V_{i+2,t}/\Delta V_{it})$ **at** $(\alpha_{i+1,t}/\alpha_{it}) = \gamma_1^{-1}, (\alpha_{i+2,t}/\alpha_{it}) = \gamma_2^{-1}, (\Delta V_{i+1,t}/\Delta V_{it}) = (1-\gamma_1^3),$

$(\Delta V_{i+2,t}/\Delta V_{it}) = (1-\gamma_2^3)$, we get $M_{i+1}^d/M_i^d = \gamma_1^{-1}(1-\gamma_1^3)^{-1}$ and $M_{i+1}^d/M_i^d = \gamma_1^{-1}(1-\gamma_1^3)^{-1}$. For $m$ th triple:

$(M_i^d + M_{i+1}^d + M_{i+2}^d)/M_i^d = \Delta M_{m\Sigma}^d/M_m^d = 1 + \gamma_1^{-1}(1-\gamma_1^3)^{-1} + \gamma_2^{-1}(1-\gamma_2^3)^{-1}$, (7.8a)

and $\Delta M_{m\Sigma}^d/M_m^d \cong 1.05$ at $\gamma_2(\gamma = 0.5) = 3.89, \gamma_1(\gamma = 0.5) = 2.215$.

Invariants relaltions (7.3a) bring invariant forms of these complexities:

$M_i^d = \mathbf{a}/t_i\Delta V_{it}, M_{i+1}^d = \mathbf{a}/t_{i+1}\Delta V_{i+1,t}, M_{i+2}^d = \mathbf{a}/t_{i+2}\Delta V_{i+2,t},$ (7.9)

Comparing the summary MC (7.8a), related to complexity of the triplet's fist segment, with triplet's *cooperative* complexity (7.8) (related to that for the same first segment), at $\gamma_2 > 3$, we have

$1 + \gamma_1^{-1}(1-\gamma_1^3)^{-1} + \gamma_2^{-1}(1-\gamma_2^3)^{-1} \gg 3(\gamma_2)^{-4},$ (7.9a)

for which, at $\gamma_2(\gamma = 0.5) = 3.89, \gamma_1(\gamma = 0.5) = 2.215$ we obtain $1.05 > 0.013$.

The results indicate the essential differeence of both types of complexities: $M_i^\delta$ measures the unit's information intensity, determined by quantity of information intendent to spend on cooperation with other units. When the cooperation occurs, the intensity is deminished, being compensated by information that binds these units and concerves the bound information. A collective unit holds a less information intensity than it was prior to cooperation measured by a summary of each of unit complexities. With more units in the collective, each complexity of attached unit $M_{i,i+1,i+2}^\delta, i = 1,....,m$ tends to decrease. The growing cooperatives intend to spend less information for attracting other units, accepting assembled units with decreasing the information speeds under the minimax. A total (integral) relative *MC*-complexity for entire IN with $m$ triplets approximates sum of (7.8a): $MC_m^\Sigma \cong m$, which grows with adding each new triplet.

The IN integral $MC_{ik}^\delta$ relative *cooperative* complexity $MC_m^{\delta\Sigma} \cong \sum_1^m [3(\gamma_2^{-4})]^m = (1-[3(\gamma_2^{-4})]]^m/[1-[3(\gamma_2^{-4})]$

decreases with adding each new triplet, and at $m \to \infty, \gamma_2(\gamma = 0.5) = 3.89, \gamma_1(\gamma = 0.5) = 2.215$ it holds $MC_m^{\delta\Sigma} \cong 1.013$. As total $MC_m^{\delta\Sigma}$ grows, the complexity of each following cooperation provides a diminishing contribution to complexity of the IN cooperative, and with growing number of such inits, the sum approaches zero. The $MC_m^\Sigma$, defined for a non-cooperating triplet's segment, in $m$ times higher than the IN's cooperative complexy $MC_m^{\delta\Sigma}$ as the triplet's number grows.

The evolution dynamics with adaptive self-controls keep the $MC_m^{\delta\Sigma}$ decrease at each IN update.

*4. Cooperative information mass and information space curvature*

Information of triplet's eigenvectors $\alpha_m = 3\alpha_{i+2}$ cooperating in ajoint volume increment $v_m$, holds this information in form $M_{vm} = \alpha_m v_m$, which we call *information cooperative mass* of this volume.

Applying triplet's Hamilotian $\alpha_m = H_m$ and differential volume $v_m = \delta V_m/\delta t = \dot{V}_m$, determines information mass of diferential volume $M_{vm} = H_m \dot{V}_m$. Connection of entropy derivation $\partial \Delta S_m/\partial t = -H_m$ with entropy's divergence $\partial \Delta S_m / \partial t = c_m div \Delta S_m$ for the same volume $v_m$, linear speed $c_m$ at cooperation of $m$-the triplet, defines the information mass through this divergence:
$$M_{vm} = -(c_m div \Delta S_m)\dot{V}_m. \tag{7.10}$$
The ratio of the information mass for a nearest triplets:
$$M_{vm}/M_{vm+1} = \alpha_m/\alpha_{m+1}(v_m/v_{m+1}), \ v_m/v_{m+1} = \alpha_{m+1}^2/\alpha_m^2 = (1/3\gamma_{m+1}^\alpha)^2, \ M_{vm}/M_{vm+1} = 1/3\gamma_{m+1}^\alpha, \gamma_{m+1}^\alpha \cong 3.9 \tag{7.10a}$$
grows in 1.3 times with adding each following IN's triplet.
The thriplet cooperative complexity (7.3) for the same volume is
$$M_m^\delta = 3\dot{H}_m/\dot{V}_m. \tag{7.11}$$
Multiplication information mass $M_{vm}$ on the complexity (7.11) leads to
$$M_{vm}M_m^\delta = 3H_m\dot{H}_m, \text{ which for } H_m = \alpha_m, \dot{H}_m = \dot{\alpha}_m, \text{ brings } M_m^\delta M_{vm} = 3\alpha_m\dot{\alpha}_m. \tag{7.12}$$
Information curvature $K_m^\alpha$ defines an increment of information speed on an instant of geodesic line $ds$ in Riemann space, which is relative to the information speed on this instant [11].
Information curvature $K_\alpha^m$ at cooperation of three triplet's eigenvectors decsribes a curving phase space at locality within volume $v_m$. This curvature connects to classical Gaussian curvature in a Riemann space [43], defined via fundamental metric tensor $\sqrt{g}$ and the phase space metric $ds = v_m dt$:
$$K_m^\alpha = (\sqrt{g})^{-1} d(\sqrt{g})/ds = (\sqrt{g})^{-1} d(\sqrt{g})/v_m dt, \tag{7.13}$$
where $g$ describes a closeness of the space vectors. For the eigenvectors in information phase space, metrical tensor $\sqrt{g}$ is expressed [11,44] thru information of triplet eigenvectors $\alpha_m$ localized in a space which generates an increment of tensor $\sqrt{g}$ for the triple cooperation:
$$\sqrt{g} = (\alpha_m)^{-3}. \tag{7.13a}$$
Substitution to (7.13) determines triplet curvature
$$K_\alpha^m = -3\dot{\alpha}_m/\alpha_m v_m, \tag{7.14}$$
which connects to (7.12) via
$$K_\alpha^m = -M_m^\delta \dot{H}_m. \tag{7.14a}$$
According to [43], multiplication of physical mass on $\sqrt{g}$ determines the mass density. Multpuling this tensor, expressed thru eigenvector at cooperation (7.13a), on information mass $M_{vm}$, leads to mass *density* $M^*_{vm}$:
$$M^*_{vm} = (\alpha_m)^{-3}\alpha_m v_m = (\alpha_m)^{-2} v_m. \tag{7.15}$$
In simulated IN hierarchy (Figs.7,8), cooperating eigenvalues $\alpha_m$ decrease with growing number of triplets $m \to n/2$, which increases $M^*_{vm}(m)$. That mass estimates information speed of creating-encoding information unit with energy $e_{ev}$ covering entropy cost of conversion to information $s_{ev} = i_{ev}$ (Sec.III(10)).
Epressing (7.14) at $MC_m = \alpha_m/v_m$ in form $K_\alpha^m = -3\dot{\alpha}_m\alpha_m v_m v_m / v_m\alpha_m\alpha_m v_m v_m$ leads to
$$K_\alpha^m = -M^*_{vm}, M_m^\delta MC_m = -M^*_{vm} = MC_m^{\delta e} = M_m^\delta MC_m \tag{7.16}$$
where $MC_m^{\delta e}$ is an effective triplet complexity.
The information curvature is a joint result of cooperation and the memorized information mass-a cooperated information mass, which generates the effective complexity. The cooperation decreases uncertainty and increases information mass at forming each triplet. The cooperation, accompanied by decreases of triplet's eigenvalues and complexity, declines the curvalure of the cooperated IN's structure. The negative curvature (7.15) characterizes a topology of the space area where the cooperation occurs.
*The information mass emerges as a curved information space per cooperative information complexity.*
Multiplication information mass density of the curvature: $M^*_{vm} K_\alpha^m = -3\dot{H}_m H_m^{-1} = -3 d^* H_m^*$ determines relative decrease of increment of energy $d^* H_m^*$ that forming this mass requires.

Simple estimation curvature by inverse radious $r_m$ of cooperating triplet node:
$$|K_{\alpha E}^m| = (r_m)^{-1}, r_m = \varepsilon(\gamma) = [(\gamma_1^\alpha/\gamma_2^\alpha)^2 - (\gamma_2^\alpha)^{-1}]^{1/2} \quad (7.17)$$
connects the estimated curvature with triplet invariants (4.20), (6.14b), and at $\varepsilon_m(\gamma^*) \cong 0.33$ estimates the growing curvature of the IN node knot for each cooperating triplet.

To evaluate maximal speed $c_{mo}$ in elementary single cooperation $\mathbf{a}$ we use $\partial\Delta S_m/\partial t = |\mathbf{a}|/t_m = c_m div\Delta S_m$ leading to $c_{mo} = [t_{mo} div\Delta S_m/|\mathbf{a}|]^{-1}$, where $t_{mo}$ estimates minimal admissible time interval $t_{mo} \cong 1.33 \times 10^{-15}$ sec of light wavelenght $l_{mo} = 4 \times 10^{-7} m$.

Structural invariant of minimal uncertainty estimates normalized divirgence: $div^*\Delta S_m = div\Delta S_m/|\mathbf{a}| \approx 1/137$.

Resulting maximal information speed $c_{mo} \approx 1.03 \times 10^{17}$ Nat/sec restricts the cooperative speed and minimal information curvature at other equal conditions. Information mass (7.10):
$$M_{vm} = -|\mathbf{a}|(c_m div\Delta S_m/|\mathbf{a}|)v_m = -|\mathbf{a}|v_m h_\alpha^o \quad (7.17a)$$
encloses the impulse information $\mathbf{a}$, volume $v_m$, and triplet structural invariants $h_\alpha^o \cong 1/137$.

The elementary cooperation binds *space* information $div^*\Delta S_i$ which limits maximal speed of incoming $i$-information, imposing *information* connection on the time and space. Unbound information code's unit has not such limtation. Ratio of speed $c_{mo}$ to speed of light $c_o$:
$$c_{mo}/c_o \approx 0.343 \times 10^9 Nat/m = 0.343 gigaNat/m \quad (7.18)$$
(measured in a light's wavelength meter) limits a maximal information space speed.

In this case, each wavelength of speed of ligth delivers $\cong 137$ Nats during $t_{mo} \cong 1.33 \times 10^{-15}$ sec.

The material mass-energy that satisfies the law of preservation energy (following the known Einstein equation), distinguishes from the information mass (7.10), (7.15).
The law satisfies when the triplet acquires energy (6.9) as the mass-energy carying the information mass. This triplet tinformation curvature and the effective cooperative complexity also hold physical energy.

## 5. *The Observers Intelligence*

The observer's *selective* actions evaluate the current information cooperative force initiated by free information attracting new high-quality information. *S*uch quality delivers a high density-frequency of related observing information through the selective mechanism. These actions engage acceleration of the observer's information processing, coordinated with the new selection, quick memorizing and encoding each node information with its logic and space-time structure, which minimizes the spending information and complexity.

The observer's optimal *multiple choices,* needed to implement the minimax self-directed strategy, evaluates the cooperative force emanated from the IN integrated node (6.4). Time interval of the control action predicts that required to select the information which carries the requested information density.

The information dynamics concurrently renovate the existing IN in a process of exchanging the requested information with environment, rebuilding the IN by encoding and re-memorizing its total recent information.

The observer's information process, carrying energy, memory and logic of the collected hidden information, conveys the intentional cooperative actions, modelling selective cognitive dynamic efforts that build and organize the observer IN's information space-time dynamic structure.

The self-built structure, under self-synchronized feedback, drives self-organization of the IN and *evolution macrodynamics* with ability of its self-creation. The free information, arising in each evolving IN, builds the Observer specific time–space information logical structure, enclosing its *conscience as intentional ability* to request, predict, and integrate the explicit information in the observer IN highest level. These processes, integrating by the IN space-time structure, express the observer *cognition*, which starts with creation elementary information unit with memories at microlevel.

Results [33] confirm that cognition arises at quantum level as "a kind of entanglement in time"…"in process of measurement", where… "cognitive variables are represented in such a way that they don't really have values (only potentialities) until you measure them and memorize", even "without the need to invoke neurophysiologic variables", while "perfect knowledge of a cognitive variable at one point in time requires there to be some uncertainty about it at other times". The coordinated selection, involving verification, synchronization, and concentration of the observed information, necessary to build its logical structure of growing maximum of accumulated information, unites the observer's organized *intelligence action*, which evaluates the amount of quality of information spent on this action *integrated in the IN node*. This functional organization integrates cognition and conscience of the interacting observers, their levels of knowledge that evaluates the amount of quality information memorized in observer IN highest hierarchical level, which measures the *Observer Information Intelligence.*

The self-directed strategy develops multiple logical operations of a self-programming computation which enhances collective logic, knowledge, and organization of diverse intelligent observers.

The intelligent actions and the intelligence of different observers connect their level of knowledge, build and organizes the observers IN's information space-time dynamic structure.

Increasing the INs enfolds growing information density that expands the intelligence, which concurrently memorizes and transmits itself over the time course in an observing time scale.

The intelligence, growing with its time-space region, increases the observer life span, which limits a vanished memory of the multiple final IN ending node.

Since whole multiple IN information is *limited* as well as a total time of the IN existence, the IN self-replication arises, which enhances the collective's intelligence, extends and develops them, expanding the intellect's growth. The self-organized, evolving IN's time-space distributed structure models *artificial intellect*. The invariance of information minimax law for any *information observer* preserves their common regularities of accepting, proceeding information and building its information structure.

That guarantees objectivity (identity) of basic observer's individual actions' *common information mechanisms,* which enable creation of *specific* information structures for each particular observed information, with individual goal, preferences, energy, material carriers and various implementations.

**VIII. The selected examples and reviews of the scientific investigations in different area of natural sciences supporting the theoretical information results**

*1. General Physics*

**1**.The physicists [45] demonstrate a first direct observation of the so-called vacuum fluctuations by using short light pulses while employing highly precise optical measurement techniques, proving no the absolute nothingness. The positive (red) and negative (blue) regions are randomly distributed in space and they change constantly at high speed. Vacuum is filled with finite fluctuations of the electromagnetic field, representing the quantum ground state of light and radio waves in the quantum light field. The found access to elementary time scales is shorter than the investigated oscillation period of the light waves.

It confirms the approach initial assumption of an initial random probability field, and an observer of this field should have a shorter time scale for its observation (Sec.V (5)).

**2**. Gluons in Standard Model of Particle Physics exists only virtually mediating strong forces at interactions [46]; each carries combination of colors charges; whopping colors and holding two colors own at pair interactions; the increasing interaction forces conserve their shape like a string. The illustration [46] looks similar to our virtual processing.

**3.** In sub-Plank process [47], quantum sates, confined to phase space volume and characterized by `the classical action', develop sub-Plank structure on the scale of shifting-displacing the state positions to orthogonal, distinguishable from the unshifted original. The orthogonality factor moves classical Plank uncertainty in random direction, which reduces limit of a sensitivity to perturbations. It relates to origin of the structure of virtual observer (Sec. I).

**4.** Measurement the probability distributions for mapping quantum paths between the quantum states [48] "reveals the rich interplay between measurement dynamics, typically associated with wave function collapse, and unitary evolution of the quantum state as described by the Schrödinger equation". The wave function collapse only in final measurement.
The measurement starts with a time distributed ensemble trajectories whose rotation in the waveguide cavity produces a space coordinate to the ensemble.
**5.** Study the non-equilibrium statistical mechanics of Hamiltonian systems under topological constraints [49] (in the form of adiabatic or Casimir invariants affecting canonical phase space) reveals the correct measure of entropy, built on the distorted invariant measure, which is consistent with the second law of thermodynamics.
The decreasing entropy and negative entropy production arises in arbitrary a priori variables of the non-covariant nature of differential entropy, associated with time evolution of the uncertainty.
Applying Jaynes' entropy functional to invariant entropy measure requires Euler's rotation with angular momentum identifying appearance of the Cartesian coordinate which satisfies the topological invariant.
These results agree with the applied EF functional, invariant measures of impulse's entropy, and appearance of space coordinate in the rotation preserving the impulse measure (Secs.I-II).
**6**. Breit–Wheeler process of pair production [49a] is the simplest mechanism by which pure light can be potentially transformed into matter. It demonstrates the possibility of transforming the virtual probes in reality.

*2. Neural Dynamics. Integrating an observing information in neurodynamics*
In [50] we have analyzed the selected multiple examples and reviews of different neurodynamic processes.
Here we add some recent results, studying the following publications.
**1**. According to [51], "Recent discussions in cognitive science and the philosophy of mind have defended a theory according to which we live in a virtual world .., generated by our brain…"; "this model is perceived as if it was external and perception-independent, even though it is neither of the two. The view of the mind, brain, and world, entailed by this theory has some peculiar consequences"… up virtual brain thoughts.
Experimental results [50] show that Bayesian probabilistic inference governs special attentional belief updating though trials, and that directional influence explains changes in cortical coupling connectivity of the frontal eye fields which modulate the "Bayes optimal updates". The frequency of oscillations which "strongly modulated by attention" *causes shifts attention* between locations.Neurons increasingly discriminate task-relevant stimuli with learning, modify sensory and non-sensory representations and adjust its processing preferring the rewarded stimulus. This causal stimulus-response, reflecting anticipation choices, predicts the features of observer formalism. Brain learns distinction between what is important and what is not, discriminating between images and optimizing stimulus processing in anticipation of reward depending on its importance and relevance.
**2**. Existence of DSS' triple code confirms [52], uncovered that a neuron communicates by a *trinary code*, utilizing not only zeros and ones of the binary code but also minus ones. The experiment [53] provides "evidence for the analog magnitude code of the triple -code model not only for Arabic digits but represents "semantic knowledge for numerical quantities..."
Results [54] demonstrate decreasing entropy in brain neurodynamic for measured frequency spectral densities with growing neurodynamic organization. Influence of rhythms on visual selection report results [55].
**3**. Importance of decisional uncertainty in learning focuses results [56], where stimulus-is an impulse, decision–getting information, greater distance-more probability-closer to information, comparing that to correct choice.
Evidence from experiments [57] show that "specific region of the brain appears essential for resolving the uncertainty that can build up as we progress through an everyday sequence of tasks, a key node in a network preventing errors in keeping on track.
Study [58] shows how learning enhances sensory and multiple non-sensory representations in primary visual cortex neuron.
**4**. Paper [59] describes how to build a mini-brain which is not performing any cogitation, but produces "electrical signals and forms own neural connections -- synapses -- making them readily producible test beds for neuroscience research".
**5**. Author [60] proposes that all cells are comprised of series of highly sophisticated "little engines" or nanomachines carrying out life's vital functions. The nanomachines have incorporated into a single complex cell, which is a descendant of a three-stage combination of earlier cells. The built complex signaling networks (Quorum Sensing) allowed one microbe to live inside and communicate with its host, forming a binary organism. Third entity, a bacterium that could photosynthesize, gained the ability to synchronize its mechanism with the binary organism. This "trinity organism" became the photosynthetic ancestor of every plant on earth that have driven life since its origin. "Resulting complex nanomachine forms a 'Borromean photosynthetic triplet'.

**6**.Analysis of all selected multiple examples and reviews of different neurodynamic processes, substantiates that our approach' functional regularities create united information mechanism, whose integral logic self-operates this mechanism, transforming multiple interacting uncertainties to physical reality-matter, human information and cognition, which originate the observer information intellect. The *information mechanism* enables specific *predictions* of individual and collective functional neuron information activities in time and space. Neurons' microprocesses retrieve and external information, including spike actions, related to the impulses, which generate the inner macrodynamics. The identified cooperative communications among neurons assemble and integrate their logical information structures in the time-space hierarchy of information network (IN), revealing the dynamics of IN creation, its geometrical information structure, triplet code, and limitations. The found information forces hold a neuron's communication, whose information is generated automatically in the neuronal interactions. Multiple cooperative networks assemble and integrate logical hierarchical structures, which model information brain processing in self-communications. The information mechanism's self-operating integral logic reveals: the information quantities required for attention, portioned extraction, its speed, including the needed internal information dynamics with the time intervals; the information quality for each observer's accumulated information by the specific location within the IN hierarchical logic; the information needed for the verification with the digital code, generated by the observer's neurons and their cooperative space logic; the internal cooperative dynamics build information network (IN) with hierarchical logic of information units, which integrates the observer required information in temporary build IN's high level logic that requests new information enclosing in the running observer's IN ; the IN nodes enfold and memorize its logic in self-forming cooperative information dynamical and geometrical structures with a limited boundary, shaped by the IN information geometry; the IN hierarchical locations of the nodes provide measuring quality of information, while the IN ending node memorizes whole IN information; the IN operations with the sequentially enclosed and memorized information units perform logical computing using the doublet-triplet code; the cooperative force between the IN hierarchical levels selects the requested information as the observer's dynamic efforts for multiple choices, needed to implement the minimax self-directed optimal strategy; the information quantity and quality, required for sequential creation of the hierarchical INs values, organize brain cognitive and *intelligence action* , finally leading to multi-cooperative brain processing and extension of intelligence.

## 3. Self-organizing dynamic motion of elementary micro- and macrosystems

**1**. *The experiments and computer simulations of collective motion exhibit systems ranging from flocks of animals to self-propelled microorganisms.*

The cell migration established similarities between these systems, which illustrates following specific results.

The emergent correlations attribute to spontaneous cell-coupling, dynamic self-ordering, and self-assembling in the persistence coherent angular motion of collective rotation within a circular areas [61]. The persistence of coherent angular motion increases with the cell number and exhibits a geometric rearrangement of cells to the configuration containing a central cell. Cell density is kept constant with increasing the cell number. The emerging collective rotational motion consists of two to eight cells confined in a circular micropatterns. The experimentally observed gradual transition with increasing system size from predominantly erratic motion of small cell groups to directionally persistent migration in larger assemblies, underlining the role of internal cell polarity in the emergence of collective behavior. For each nucleus the angular position was evaluated respectively to the circle center, and angular velocity is normalized and averaged over the individual angular velocities of the N-cell system.

Circle size increases in such a way that the average area per cell is constant at approximately 830 μm2. Probability distribution of the mean angular velocity for systems containing two to eight cells is fitted by a single Gaussian and their mixture is two Gaussians. For all N cells, the probability distribution displays symmetry breaking into clockwise and counterclockwise rotations. Both directionalities are almost equally represented, with a small bias towards clockwise rotation. Average mean squared displacement indicates ballistic angular motion for all cell numbers, while the averaged the displaced intervals of nucleus exhibited diffusive behavior. Both experiments and simulations showed consistently that the persistence of the coherent state increases with the number of confined cells for small cell numbers but then drops abruptly in a system containing five cells. This is attributed to a geometric rearrangement of cells to a configuration with a central only weakly polarized cell.
It reveals the decisive role of the interplay between the local arrangement of neighboring cells and the internal cell polarization in collective migration. The similarities suggest universal principles underlying pattern formation, such as interactions rules [62-66], the systems' generic symmetries [67-69].

**2.** *Confinement stabilizes a bacterial suspension into a spiral vortex*

**1.** 'Enhanced ordering of interacting filaments by molecular motors [70] demonstrate the emergence of collective motion in a high-density concentrated filaments propelled by immobilized molecular motors in a planar geometry".
At a critical density, the filaments self-organize to form coherently moving structures with persistent density modulations.
The experiment allows backtracking of the assembly and disassembly pathways to the underlying local interactions.
The identified weak and local alignment interactions essential for the observed formation of patterns and their dynamics.
The presented minimal polar-pattern-forming system provide new insight into emerging order in the broad class of bacteria and their colonies [71-74], and self-propelled particles [75-79].

**2.** Confining surfaces play crucial roles in dynamics, transport, and order in many physical systems [80-83].
Studying [84] the flow and orientation order within small droplets of a dense bacterial suspension reveals the influence of global confinement and surface curvature on collective motion. The observing competition between radial confinement, self-propulsion, interactions, other induces a steady single-vortex state, in which cells align in inward spiraling patterns accompanied by a thin counter rotating boundary layer.

**3.** The cited experiments validate: "spontaneous cell-coupling, dynamic self-ordering, self-assembling in the persistence coherent angular motion of collective rotation within a circular areas", displacement in angular motion with diffusive behavior of the displaced intervals, emergence of collective order confined on a curved surface, others (Secs.II-IV).

**4.** According to recent discovery [85], "the protein stable shapes adopted by a few proteins contained some parts that were trapped in the act of changing shape, the changes relate to how proteins convert from one observable shape to another". From the process of RNA translation of DNA triplets to enzymes and aminoacids, all proteins start as linear chains of building blocks and then quickly fold to their proper shape, going through many high-energy transitions to proteins multiple biological functions.

**Conclusion**

The information formalism describes the origin and evolution dynamic of information, creation of information observer from hidden information of cutting random process. The approach provides information path from an observing process uncertainty to certainty of an information observer. The process Bayesian probabilities reveal entropy of the observing process impulses, which integrates an entropy functional and transforms to observer's information units. The uncertainty-certainty gap on the edge of reality connects the entangled quantum conjugated entropies *with* classical physics including relativity theory for an observer, and after overcoming the gap with information and thermodynamics. The classical physics proceeding within the gap have probabilistic casualty approaching the gap, local deterministic casualty on edge of the gap (2.15), and information causally $P=1$ at overcoming the gap.

Information path functional (IPF) collects and integrates the units in information process, whose dynamics include both microdynamics of creation the information units and macrodynamics of their processing with highest probability on trajectories compared to the classical dynamics. The information macrodynamic processes model information equivalents of any physical processes, carrying the related energy.
Information potentially encodes different observation in various physical-chemical substances.

This paper mathematically validates each step of the path by defining numerical values of information threshold and constrains, which should be overcame from origin both virtual and information observers enables information self-structuring, evolution, intelligence, and self-recreation. Consecutive overcoming these thresholds requires gradually increase quaintly and quality information which compensate the observer spending along the path, which specifies limitation, constraints, and computer simulation.

Information from observations generates information physical process, builds the observer information network, collects, and orders, memorizes information, and successive repeats these processes up to arising complexity, self-replication and evolving intelligence.
That allows modelling human perception of observation and brain information processing.
The path of growing intelligence [15] is a *tendency*, which however will not achieve the complete certainty while minimizing the uncertainty, which makes fact-truth *incomplete* [40].
The introduced mathematically proven conditions of the origin of information and emergence of an information Life substantiate, extend and distinguish the results from many conventional approaches, and/ or trivial assumptions. It shows the path of building Universe.

The selected experimental results demonstrate, *support and confirm the paper formal results extended in* [86].

The developed computer-based methodology and software [11] were practically used for systems modeling, identification and control of a diversity of information interactions in some physical (technological) and non-physical (economical, information-biological) objects. Specifically, the approach has been applied to intelligent systems [87, 88], biological processes [89, 90], collective economic and social systems [91-94], and was implemented in the processes of solidifications, casting technologies [95], and electro-technologies [11].

# Figures

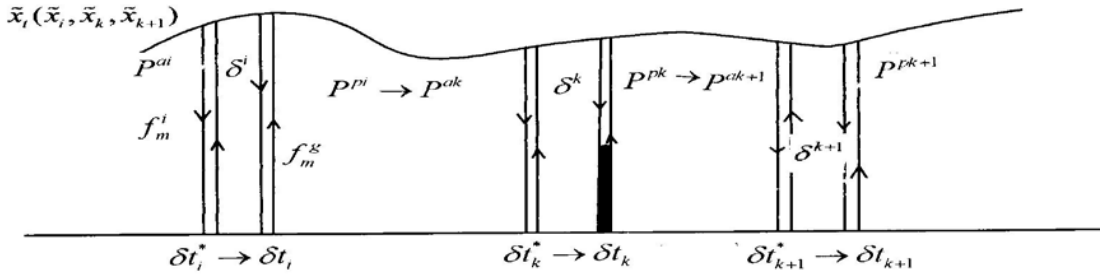

**Fig. A.** Schematic illustration of interactive impulses arising in observable (virtual) process $\tilde{x}_t$ as observer's series of probing action: $\delta^i, \delta^k, \delta^{k+1}$ whose frequencies $(f_m^i,...f_m^g)$ reveal observer's a priori–a posteriori probabilities $P^{ai} \to P^{pi} \to P^{ak} \to P^{pk} \to P^{ak+1} \to P^{pk+1}$ during time intervals $\delta t_i^* \to \delta t_i$, $\delta t_k^* \to \delta t_k$, $\delta t_{k+1}^* \to \delta t_{k+1}$, where each symbol $\to$ indicates the transfer from observable (virtual) time to the observing (certain) time intervals during the probing impulses; within the impulse $\delta^k$ (for the observable process' dimension $\tilde{x}_k$) starts a certain step-up control by index $\uparrow$ (Yes) (shown in bold) with uncertain gap-delay $\delta_o$ ( in the text). The Bayes probability's connection increases each a posteriori correlation and sequentially reduces the relative entropy along the process.

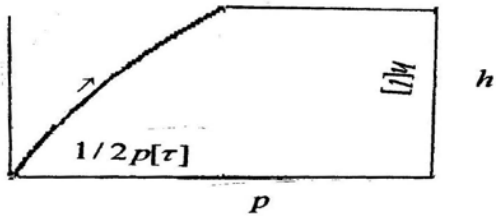

**Fig.1.** Illustration of origin the impulse space coordinate measure $h[l]$ at curving time coordinate measure $1/2p[\tau]$ in rotation.

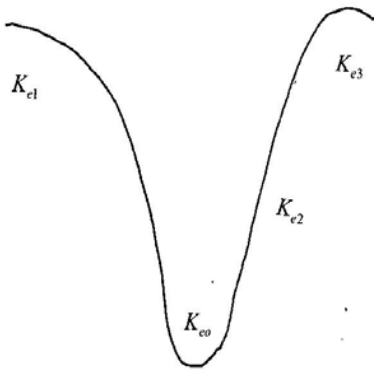

**Fig.1a.** Curving impulse with curvature $K_{e1}$ of the impulse step-down part, curvature $K_{eo}$ of the cutting part, curvature $K_{e2}$ of impulse transferred part, and curvature $K_{e3}$ of the final part cutting all impulse entropy.

**Fig.2. Forming a space -time spiral trajectory with radius** $\rho = b\sin(\varphi \sin \beta)$ **on the conic surface at the points D, D1, D2, D3, D4 with the spatial discrete interval DD1=**$\mu$**, which corresponds to the angle** $\varphi = \pi k /2$, $k = 1,2,...$ **of the radius vector's** $\rho(\varphi,\mu)$ **projection of on the cone's base (O1, O2, O3, O4) with the vertex angle** $\beta = \psi^o$ **( in the text).**

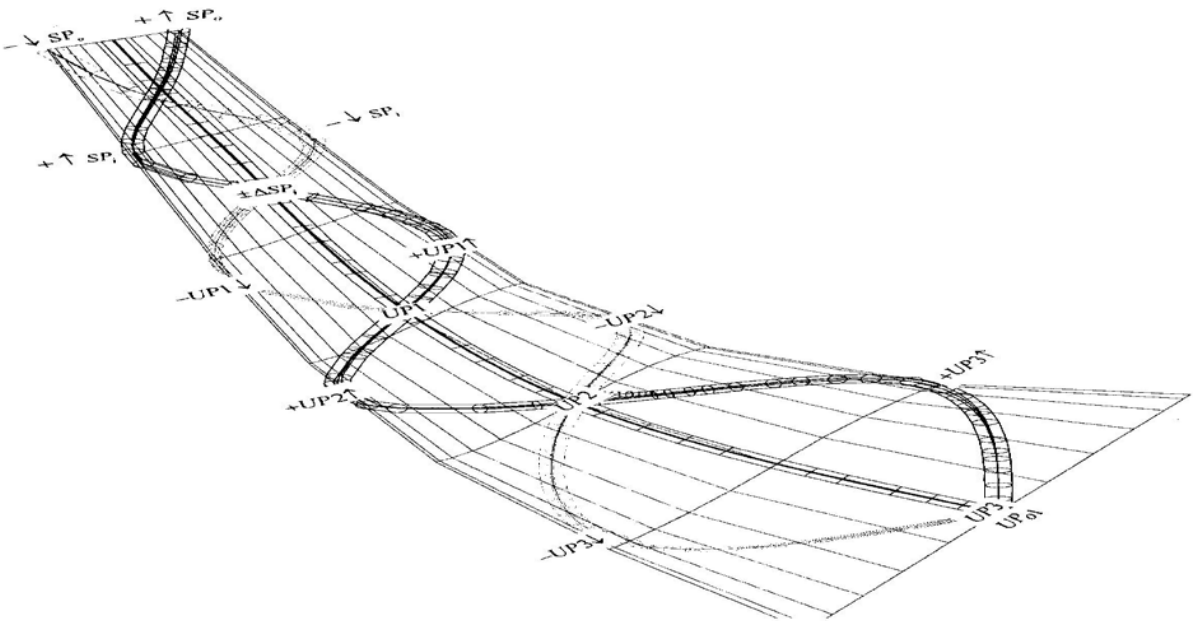

**Fig. 3. Time-space opposite directional-complimentary conjugated trajectories** $+\uparrow SP_o$ **and** $-\downarrow SP_o$**, forming spirals located on conic surfaces (analogous to Fig.2). Trajectory of bridges** $\pm \Delta SP_i$ **binds the contributions of process information unit** $\pm UP_i$ **through the impulse, joining No-Yes actions, which model a line of switching controls.**

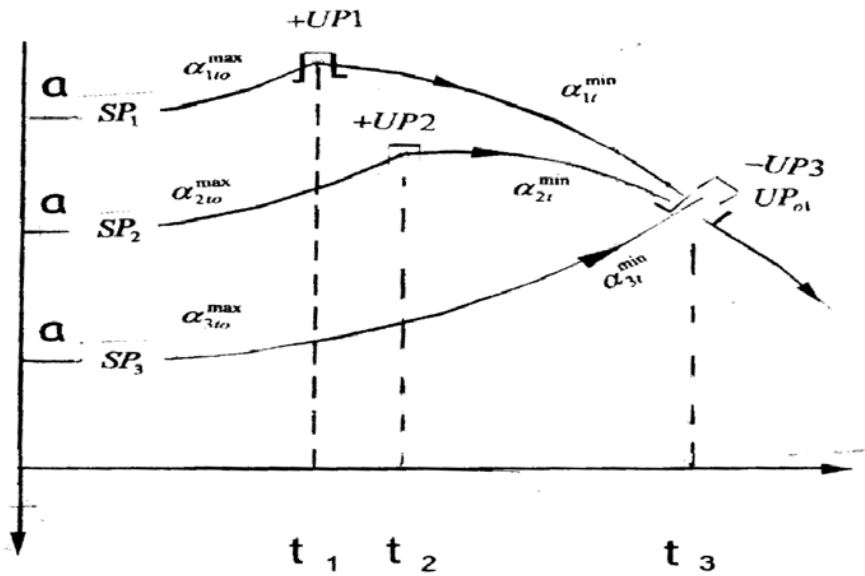

**Fig.4. Illustrative dynamics of assembling units** $+UP1, +UP2, -UP3$ **on the space time trajectory and adjoining them to** $UP_{o1}$ **along the sections of space-time trajectory** $SP_1, SP_2, SP_3$ **(Fig.3) at changing information speeds from** $\alpha_{1to}^{max}$, $\alpha_{2to}^{max}$, $\alpha_{3to}^{max}$ **to** $\alpha_{1t}^{min}, \alpha_{2t}^{min}, \alpha_{3t}^{min}$ **accordingly;** $a$ **is dynamic information invariant.**

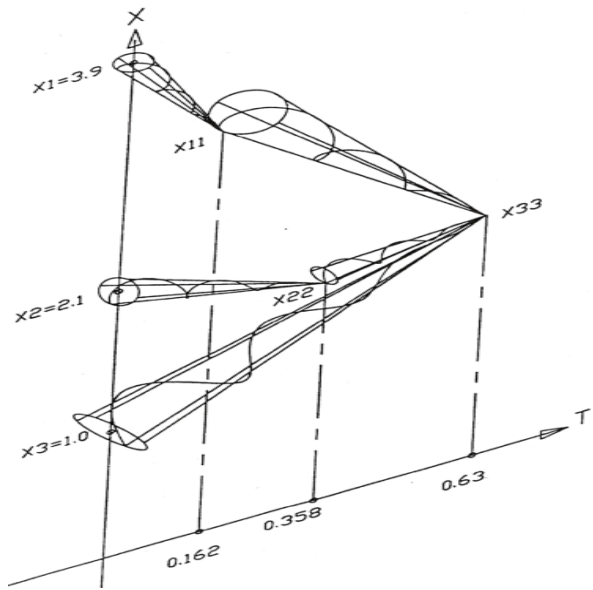

**Fig. 5. Simulation of forming a triplet's space structure, composed from time-space spirals Fig.3 according to Fig.4. The indicated time interval measure real times** $\Delta t_{13}, \Delta t_{13}, t_3$ **during the simulation at the shown cones diameters.**

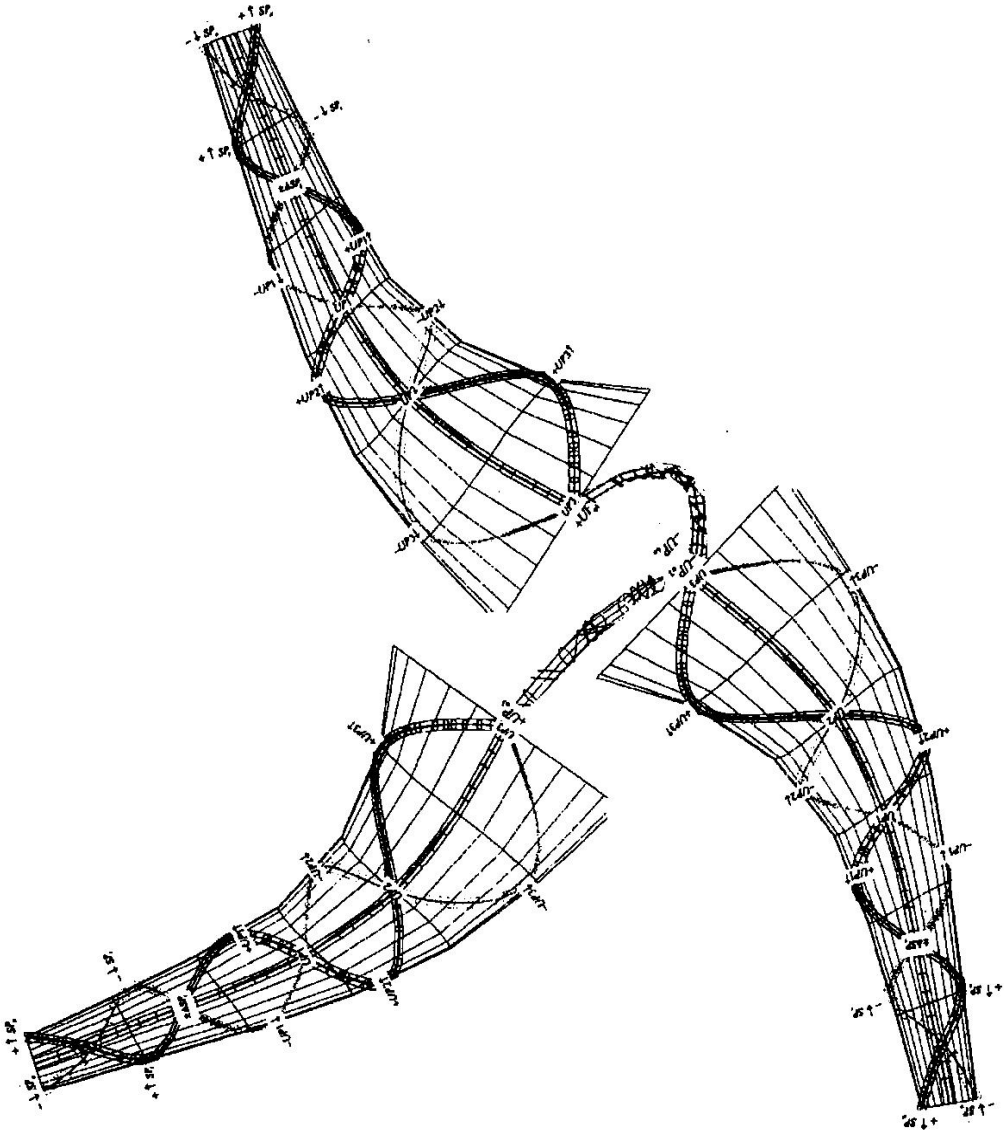

**Fig.6.** Assembling three formed units $\mp UP_{i0}$ in a higher (second) triplet level connecting the units' equal speeds (Fig.4).

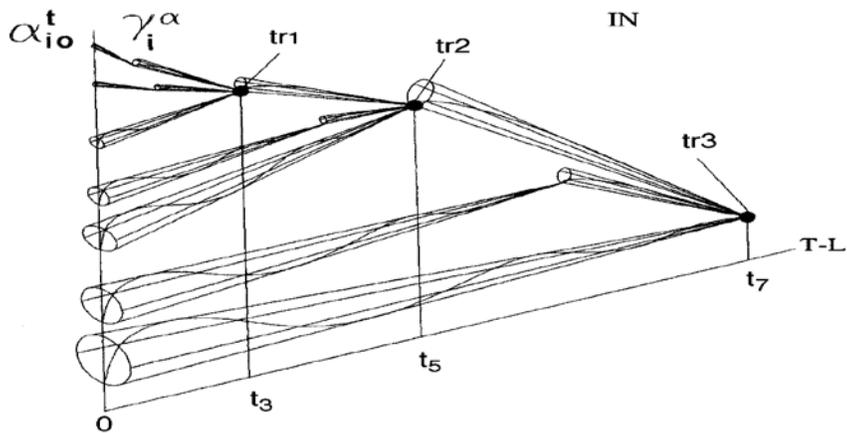

**Fig.7.** The IN information geometrical structure of the hierarchy of the spiral space-time dynamics (Figs.3,5,6) of the triplet nodes (tr1, tr2, tr3, ..); $\{\alpha_{io}^t\}$ is a ranged string of the initial eigenvalues, cooperating around the $(t_1, t_2, t_3)$ locations of T-L time-space.

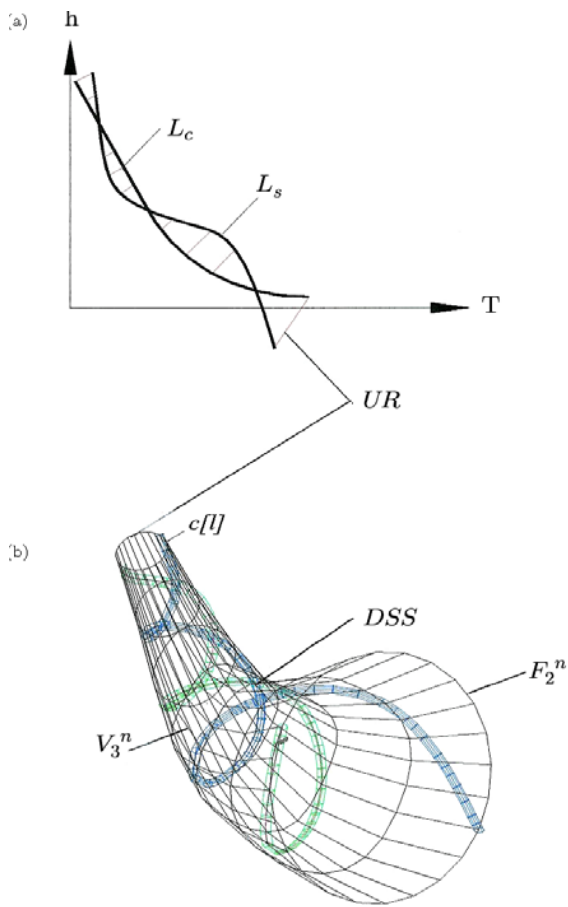

**Fig.8.** Simulation of forming the double spiral cone's structure (*DSS*) with cells (c[l]), arising along switching control line $Lc$ (Fig.3), uncertainty zone UR, surrounding the $Lc$-hyperbola in the form of the $Ls$-line, the spiral space geometry surface $F_2^n$ enfolding volume $V_3^n$. This structure is a space model of each spirals on Fig. 3, 5, 6, 7.

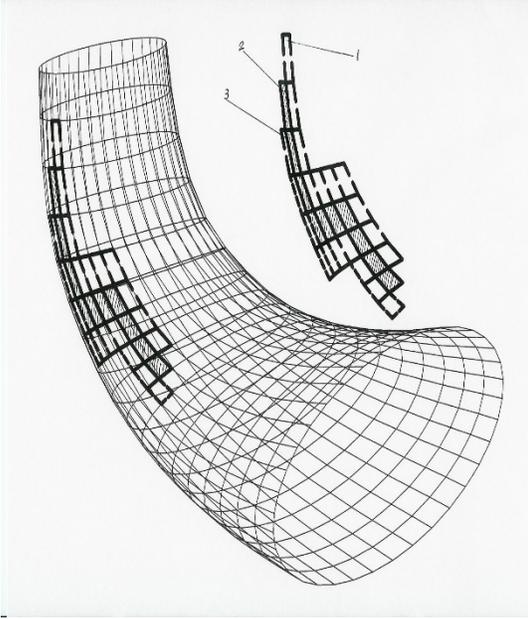

**Fig. 9. Structure of the cellular geometry, formed by the cells of the DSS triplet's code, with a portion of the surface cells (1-2-3), illustrating the space formation. This structure geometry integrates information contributions modelling in the Figs.3, 8.**

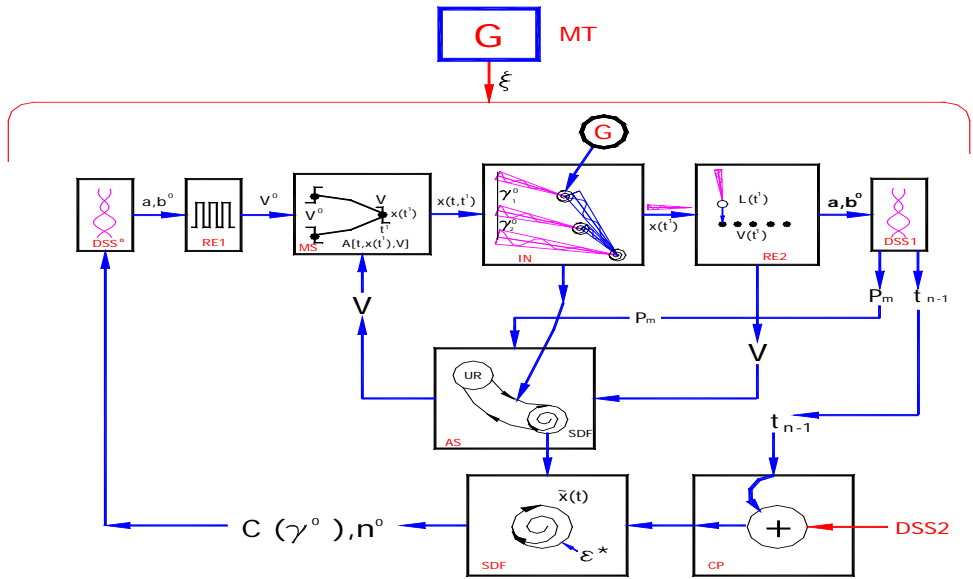

**Fig.10. Functional schema of the cyclic evolutionary informational mechanisms:**
System macrodynamics MS, governed by an inherited double spiral structure $DSS^o$; mechanism of mutations MT delivering external perturbations $\xi$ ; control replication mechanism RE transforming $DSS^o$ code into the initial controls $v_o$; the IN cooperation and macromodel renovation, generating renovated $DSS1$; adaptative and self-organizing mechanisms AS, which generate G and potential MT; replication control mechanism RE2, selecting macrostates $x(\tau)$ and forming current control $v(\tau)$ by duplication of $x(\tau)$; coupling mechanism $C(\gamma^o), n^o$ that carry both parents' $DSS1$ and $DSS2$ invariants; stochastic dissipative fluctuations SDF forming new invariants $(\gamma^o, n^o)$ that define a new $DSS_o^o$ and initiate new MS , IN, MT; repeating cycle AS with UR , SDF after coupling and transferring the inherited invariants to new generated macrosystem.